\documentclass[longauth]{aa}  
\pdfoutput=1
\usepackage{graphicx}
\usepackage{txfonts}
\usepackage{natbib}

\bibpunct{(}{)}{;}{a}{}{,}
\begin{document} 

   \title{Chemodynamics of the Milky Way. \\I.~The first year of APOGEE data}

   \author{F. Anders\inst{1,2}, C. Chiappini\inst{1,3}, B. X. Santiago\inst{3,4}, H. J. Rocha-Pinto\inst{3,5}, L. Girardi\inst{3,6}, L. N. da Costa\inst{3,7}, M. A. G. Maia\inst{3,7}, M. Steinmetz\inst{1}, I. Minchev\inst{1}, M. Schultheis\inst{8}, C. Boeche\inst{9}, A. Miglio\inst{10}, J. Montalbán\inst{11}, D. P. Schneider\inst{12, 13},T. C. Beers\inst{14,15}, K. Cunha\inst{7,16}, C. Allende Prieto\inst{17}, E. Balbinot\inst{3,4}, D. Bizyaev\inst{18}, D. E. Brauer\inst{1}, J. Brinkmann\inst{18}, P. M. Frinchaboy\inst{19}, A. E. Garc\'{i}a P\'erez\inst{20}, M. R. Hayden\inst{21}, F. R. Hearty\inst{20, 12}, J. Holtzman\inst{21}, J. Johnson\inst{22}, K. Kinemuchi\inst{18}, S. R. Majewski\inst{20}, E. Malanushenko\inst{18}, V. Malanushenko\inst{18}, D. L. Nidever\inst{23}, R. W. O'Connell\inst{20}, K. Pan\inst{18}, A. C. Robin\inst{24}, R. P. Schiavon\inst{25}, M. Shetrone\inst{26}, M. F. Skrutskie\inst{20}, V. V. Smith\inst{14}, K. Stassun\inst{27}, G. Zasowski\inst{28} }
   
   \authorrunning{Anders, Chiappini et al.}      
   
     \institute{Leibniz-Institut f\"ur Astrophysik Potsdam (AIP), An der Sternwarte 16, 14482 Potsdam, Germany\\
              \email{fanders@aip.de, cristina.chiappini@aip.de}
         \and{Technische Universität Dresden, Institut für Kern- und Teilchenphysik, Zellescher Weg 16, 01069 Dresden, Germany}
	\and{Laborat\'orio Interinstitucional de e-Astronomia, - LIneA, Rua Gal. Jos\'e Cristino 77, Rio de Janeiro, RJ - 20921-400, Brazil}
	\and{Instituto de F\'\i sica, Universidade Federal do Rio Grande do Sul, Caixa Postal 15051, Porto Alegre, RS - 91501-970, Brazil}
	\and{Universidade Federal do Rio de Janeiro, Observat\'orio do Valongo, Ladeira do Pedro Ant\^onio 43, 20080-090 Rio de Janeiro, Brazil}
	\and{Osservatorio Astronomico di Padova -- INAF, Vicolo dell'Osservatorio 5, I-35122 Padova, Italy}
	\and{Observat\'orio Nacional, Rua Gal. Jos\'e Cristino 77, Rio de Janeiro, RJ - 209
21-400, Brazil}
	\and{Observatoire de la Cote d'Azur, Laboratoire Lagrange, CNRS UMR 7923, B.P. 4229, 06304 Nice Cedex, France}
	\and{Astronomisches Rechen-Institut, Zentrum für Astronomie der Universität Heidelberg, Mönchhofstr. 12-14, 69120 Heidelberg, Germany}
	\and{School of Physics and Astronomy, University of Birmingham, Edgbaston, Birmingham, B15 2TT, United Kingdom}
	\and{Institut d'Astrophysique et de Géophysique, Allée du 6 août, 17 - Bât. B5c, B-4000 Liège 1 (Sart-Tilman), Belgium}
	\and{Department of Astronomy and Astrophysics, The Pennsylvania State University, University Park, PA 16802, USA}
	\and{Institute for Gravitation and the Cosmos, The Pennsylvania State University,
	\and{National Optical Astronomy Observatory, 915 N. Cherry Ave., Tucson, AZ 85719, USA} 
	\and{JINA: Joint Institute for Nuclear Astrophysics} 
	\and{Steward Observatory, University of Arizona, Tucson, AZ, 85721, USA}
	\and{Instituto de Astrofisica de Canarias,  C/ Vía Láctea, s/n, 38205, La Laguna, Tenerife, Spain} 
	\and{Apache Point Observatory and New Mexico State University, P.O. Box 59, Sunspot, NM, 88349-0059, USA}
	\and{Department of Physics \& Astronomy, Texas Christian University (TCU), P.O. Box 298840, Fort Worth, TX 76129, USA}
	\and{Department of Astronomy, University of Virginia, P.O. Box 400325, Charlottesville, VA 22904-4325, USA}
	\and{New Mexico State University, Box 30001 / Department 4500, 1320 Frenger St., Las Cruces, NM 88003, USA}
	\and{The Ohio State University, Department of Astronomy, 4055 McPherson Laboratory, 140 West 18th Ave.,
Columbus, OH 43210-1173, USA}
	\and{Department of Astronomy, University of Michigan, 1022 Dennison, 500 Church St., Ann Arbor, MI 48109, USA}
	\and{Institut Utinam, CNRS UMR6213, Universit\'e de Franche-Comt\'e, OSU THETA de Franche-Comt\'e-Bourgognen, Besan\c{c}on, France}
	\and{Astrophysics Research Institute, IC2, Liverpool Science Park, Liverpool John Moores University, 146 Brownlow Hill, Liverpool, L3 5RF, United Kingdom}
   University Park, PA 16802}
	\and{McDonald Observatory, The University of Texas at Austin, Austin, TX 78712, USA}
	\and{Vanderbilt University, Dept. of Physics \& Astronomy, VU Station B 1807, Nashville, TN 37235, USA}
	\and{Johns Hopkins University, Department of Physics and Astronomy, 3701 San Martin Drive, Baltimore, MD 21210, USA}
	}

   \date{Received \today; accepted ...}

  \abstract
   {The Apache Point Observatory Galactic Evolution Experiment (APOGEE) features the first multi-object high-resolution fiber spectrograph in the Near-infrared (NIR) ever built, thus making the survey unique in its capabilities: APOGEE is able to peer through the dust that obscures stars in the Galactic disc and bulge in the optical wavelength range. Here we explore the APOGEE data included as part of the Sloan Digital Sky Survey's 10th data release (SDSS DR10).}
   {The goal of this paper is to a) investigate the chemo-kinematic properties of the Milky Way disc by exploring the first year of APOGEE data, and b) to compare our results to smaller optical high-resolution samples in the literature, as well as results from lower resolution surveys such as the Geneva-Copenhagen Survey (GCS) and the RAdial Velocity Experiment (RAVE).}
   {
We select a high-quality (HQ) sample in terms of chemistry (amounting to around 20.000 stars) and, after computing distances and orbital parameters for this sample, we employ a number of useful subsets to formulate constraints on Galactic chemical and chemodynamical evolution processes in the Solar neighbourhood and beyond (e.g., metallicity distributions -- MDFs, [$\alpha$/Fe] vs. [Fe/H] diagrams, and abundance gradients).}
   {Our red giant sample spans distances as large as 10 kpc from the Sun. Given our chemical quality requirements, most of the stars are located between 1 and 6 kpc from the Sun, increasing by at least a factor of eight the studied volume with respect to the most recent chemodynamical studies based on the two largest samples obtained from RAVE and the Sloan Extension for Galactic Understanding and Exploration (SEGUE). We find remarkable agreement between the MDF of the recently published local (d $<$ 100 pc) high-resolution high-S/N HARPS sample and our local HQ sample (d $<$ 1 kpc). The \emph{local} MDF peaks slightly below solar metallicity, and exhibits an extended tail towards [Fe/H] $= -$1, whereas a sharper cutoff is seen at larger metallicities (the APOGEE sample shows a slight overabundance of stars with metallicities larger than $\simeq+$0.3 w.r.t. the HARPS sample). Both samples also compare extremely well in an [$\alpha$/Fe] vs. [Fe/H] diagram. The APOGEE data also confirm the existence of a gap in the abundance diagram. When expanding our sample to cover three different Galactocentric distance bins (inner disc, solar vicinity and outer disc), we find the \emph{high-[$\alpha$/Fe]} stars to be rare towards the outer zones (implying a shorter scale-length of the thick disc with respect to the thin disc) as previously suggested in the literature. Finally, we measure the gradients in [Fe/H] and [$\alpha$/Fe], and their respective MDFs, over a range of 6 $ < $ R $ <$ 11 kpc in Galactocentric distance, and a 0 $ < $ z $< $ 3 kpc range of distance from the Galactic plane. We find a good agreement with the gradients traced by the GCS and RAVE dwarf samples. For stars with 1.5 $<$ z $<$ 3 kpc (not present in the previous samples), we find a positive metallicity gradient and a negative gradient in [$\alpha$/Fe].
 }
{}

   \keywords{Galaxy: general -- Galaxy: abundances -- Galaxy: disk -- Galaxy: evolution -- Galaxy: stellar content --  Stars: abundances
               }

   \maketitle

%

\section{Introduction}

Our Galaxy and its companions are the only systems for which large numbers of individual stars can be resolved and analysed spectroscopically. These stars carry a fossil record of the processes involved in the formation and evolution of the Milky Way. By measuring the chemical abundances in the stellar atmospheres, we have access to the gas composition at the time and place of the star's birth. Combining these chemical fossil imprints with the current kinematical properties of a large number of stars (covering large portions of our Galaxy), one can then infer the main processes at play during the formation and evolution of the Milky Way. This method, sometimes referred to as \emph{Galactic Archaeology} or \emph{Near-Field Cosmology}, has proven to be extremely powerful in helping to answer questions related not only to the Milky Way formation but also to stellar evolution, the origin and evolution of chemical elements, and cosmology (\citealt{Pagel2009}, \citealt{Matteucci2001, Matteucci2012}, \citealt{Freeman2002}, \citealt{Gilmore2012a}, \citealt{Rix2013}). 

From the Galactic Archaeology viewpoint, one of the most important issues is the determination and relative quantification of processes shaping the galaxy disc structure and constraining its assembly history. This explains the unprecedented efforts now in place to obtain detailed chemical and kinematical information for a large number of stars in our Galaxy. A suite of vast stellar astrometric, photometric and spectroscopic surveys has been designed to map the Milky Way and answer questions related to its formation. With the data provided by medium- and low-resolution surveys such as RAVE \citep{Steinmetz2006}, LAMOST/LEGUE \citep{Zhao2006, Newberg2012} and SEGUE \citep{Yanny2009}, together with information coming from high-resolution surveys such as Gaia-ESO (GES, \citealt{Gilmore2012}); HERMES/GALAH \citep{Zucker2012} and APOGEE \citep{AllendePrieto2008}, it will be possible to draw a new detailed picture of our Galaxy, providing an ideal testbench for galaxy formation models. Most importantly, the recently launched Gaia satellite (\citealt{Perryman2001}, \url{http://www.rssd.esa.int/Gaia}) and its spectroscopic follow-up missions will revolutionize not only our understanding of the Milky Way, but the whole field of Near-Field Cosmology\footnote{Primary task of ESA's astrometric mission {\it Gaia} is to measure the parallaxes and proper motions of up to one billion (mostly disc) stars with unprecedented accuracy ($\sigma(\pi)\sim20 \mu$as and $\sigma(\mu)\sim20 \mu$as at magnitude $G\sim15$ -- providing a distance accuracy of 1–2\% at 1 kpc; see \citealt{Turon2005}), but it also provides medium-resolution spectra in the CaII triplet region (the $848\dots874$ nm wavelength range) for stars brighter than 17th magnitude, obtaining high precision radial velocities ($\sigma(v_{\mathrm{los}})\sim10$ km/s; \citep{Katz2004}), in addition to low-resolution optical spectra providing well-determined stellar parameters. Thus, Gaia will be able to probe the kinematics of the disc out to several kpc in all directions \citep{Bailer-Jones2009}.}. The combination of these datasets with complementary information coming from asteroseismology \citep{Miglio2013a} data will be an important asset.

The big challenge ahead of us is to build theoretical models able to make predictions to be compared with these huge datasets. The only way to understand the high-dimensional problem of the formation and evolution of a late-type barred spiral galaxy like the MW in a cosmological context is through sophisticated simulations combining chemical and dynamical evolution (see detailed discussion in \citealt{Minchev2013}). Constraining these models has become a primary task of current and future surveys.

In this first of a forthcoming series of papers, we focus on finding new and tighter chemodynamical constraints on models of our Galaxy using data from the Apache Point Observatory Galactic Evolution Experiment (APOGEE; \citealt{AllendePrieto2008}, Majewski et al. 2014, in prep.), one of four experiments operating in the third epoch of the Sloan Digital Sky Survey (SDSS-III; \citealt{Eisenstein2011}), using the 2.5m Sloan telescope \citep{Gunn2006} at Apache Point Observatory (APO). We define a subsample of APOGEE data from the recent data release
(DR10; \citealt{Ahn2013}) for which full kinematical information was obtained for red giant stars spanning distances as large as 10 kpc from the Sun (although most of the high quality data in our sample is confined to distances below 5 kpc). A complementary paper
(Hayden et al. 2013) presents the spatial distribution of mean metallicities for the full DR10 sample, which extends to even larger distances, but without kinematical information.  
Future work will further develop the analyses of these samples, including comparisons
with predictions from star count models like TRILEGAL \citep{Girardi2005, Girardi2012}, chemical evolution models for the Galactic disc and (semi-)cosmological chemodynamical simulations of the MW, such
as the recent model of \citet{Minchev2013}.

In Section~\ref{sample} we describe how our APOGEE high-quality sample (HQ) was selected, both in terms of chemistry and kinematics, carefully discussing what \emph{minimal quality requirements} are necessary to define samples to be used for detailed chemodynamical studies. Section~\ref{kinematics} focusses on the kinematical parameters: we present our computed distances, the adopted proper motions and the computed orbital parameters (along with their uncertainties).
By pruning our sample to include stars with best-determined chemical \emph{and} orbital parameters, we construct what we refer to as the \emph{Gold sample}. In Section~\ref{results}, we first discuss a \emph{local} (Solar vicinity) sample (with d $< $ 1 kpc), and compare it with the high-resolution, very-high $S/N$ HARPS sample of \citet{Adibekyan2011}. We then extend our discussion to further regions outside of the Solar neighborhood. Section~\ref{conclusions} summarizes our main results and discusses some future prospects.

\section{Observations and Sample Selection}
\label{sample}
APOGEE delivers high-resolution ($R\sim 22,500$) high signal-to-noise ($S/N\sim 100$ pixel$^{-1}$) spectra of primarily red giant stars in the H band ($\lambda=1.51-1.69 \mu$m), enabling the determination of precise ($\sim100$ m/s) radial velocities as well as stellar parameters and chemical abundances of up to 15 elements. 
In addition, APOGEE has already proven to be useful in various other fields as well, such as the determination of the Galactic rotation curve \citep{Bovy2012a}, detection of (sub-)\-stellar companions (Nidever et al. 2014, in prep.), spectral variability of hot stars (Chojnowski et al. 2014, in prep.), dark matter distribution in the Sgr dSph galaxy \citep{Majewski2013}, characterisation of diffuse interstellar absorption bands (Zasowski et al. 2014, in prep.) or open star clusters (\citealt{Frinchaboy2013}; Covey et al. 2014, in prep.).\\

APOGEE's final goal is to measure accurate and precise radial velocities, stellar parameters and chemical abundances for around 100,000 red giants candidates.
APOGEE's target selection is a key part of the survey, because it has to be assured that the sample is minimally biased and homogeneous to draw robust conclusions about the underlying stellar populations (see \citealt{Zasowski2013} for details). Here we will explore chemodynamical constraints already produced from the first year of APOGEE data.\\

The database of APOGEE spectra released in SDSS DR10 forms the largest catalogue of high-resolution IR spectra ever obtained. For more than 57,000 stars observed by APOGEE before July 2012, stellar parameters and chemical abundances have been determined by the APOGEE Stellar Parameters and Chemical Abundances Pipeline (ASPCAP; \citealt{Ahn2013}, García Pérez et al. 2014, in prep.). We use these data to assemble a sample of red giant stars with high-quality chemical abundances that will be employed to probe the chemodynamical properties of the Galactic disc.
In this Section we describe the selection criteria and the calibration relations applied to the DR10 catalogue, leading to our \frqq HQ Sample\flqq. A summary of the applied cuts is given in Table \ref{selectionsummary}.

\subsection{Photometry}
\label{photometry}

Although the APOGEE targeting strategy for the main survey was chosen to ensure high quality data, consistency in the input catalogue and a straightforward selection function, this is not always true for stars selected for ancillary science programs, among them giant stars in the {\it Kepler} \citep{Gilliland2010} and CoRoT \citep{Baglin2006} fields. Hence, the NIR magnitudes and errors for the final sample were taken directly from the 2MASS Point Source Catalogue \citep{Cutri2003}, requiring the original quality criteria for the main survey described in \citet[][see their Table 3 for details]{Zasowski2013} and, as some of the ancillary targets\footnote{The main group of ancillary targets in our final sample are the asteroseimic targets from {\it Kepler} and CoRoT. Known cluster members and probable candidates have not been used in the final analysis, due to the additional selection biases this might introduce.} were not strictly selected on the basis of 2MASS astrometry, also requiring positional consistency. 

The mid-IR data used for the estimation of interstellar extinction was adopted from the WISE \citep{Wright2010} and Spitzer-IRAC photometry \citep{Benjamin2005} contained in the APOGEE targeting (requiring only that the uncertainties of the corresponding $[4.5 \mu]$ magnitude be $\le 0.1$ mag), as well as the actual extinction values $A(K_s)$ calculated with the RJCE method \citep{Majewski2011,Nidever2012a}, as described in \citet{Zasowski2013}.

\begin{table*}[t]
  \centering
\caption{Summary Table for the selection of the APOGEE HQ Giant Sample}
\begin{tabular}{l l p{5cm}}
\hline
Parameter & Requirement & Notes\\
\hline\hline
$S/N$ & $>70$/pixel & \\
$\sigma(v_{\mathrm{los}})$ & $\le 1$ km/s& no RV-identified binaries\\
APOGEE\_STARFLAG bits & $\notin\{0,1,3\}$ & no commissioning data or obviously bad spectra\\
APOGEE\_TARGET1 bits & $\notin\{10,15,16, 18, 19,23,24\}$ & avoid, e.g., extended objects, M31 clusters, M dwarfs\\
APOGEE\_TARGET2 bits & $\notin\{4,9,10,13,15,16,17\}$ & avoid, e.g., sky fibres, telluric standards, known cluster members\\
ASPCAP $\chi^2$ & $<25 $ & \\
$T_{\mathrm{eff}}$ & $\in \{3800~\mathrm{K},5200~\mathrm{K}\}$ & avoid too low temperatures\\
$\log g$ & $\in \{0.5~\mathrm{dex}, 3.8~\mathrm{dex}\}$ & select red giant stars\\
$\mathrm{[M/H]}$ & $\in \{-1.0, 0.45\}$ & avoid low metallicities\\
\hline
\end{tabular}
 \label{selectionsummary}
\end{table*}

\subsection{APOGEE data reduction}
\label{aspcaphow}
APOGEE's reduction pipeline delivers 1D flux-calibrated spectra corrected for telluric absorption and sky emission, along with precise ($\delta(v_{\mathrm{los}})\lesssim 0.2$ km/s) and accurate (zero-point accuracy $\approx0.26\pm0.22$ km/s) heliocentric velocities \citep{Nidever2012b}, and data-quality flags that are also included in the higher level catalogues. In particular, we use the data-quality flags, the signal-to-noise ratio (S/N) and the visit-to-visit scatter of the heliocentric velocities $\sigma(v_{\mathrm{los}})$ to clean our sample (see Table \ref{selectionsummary} for a summary).

ASPCAP works in two steps: first, the main stellar parameters are estimated from synthetic template fit to the entire APOGEE spectrum provided by the APOGEE reduction pipeline (see \citealt{Ahn2013} and Nidever et al. 2014 (in prep.) for details). Next, these values are used to fit various small spectral windows containing line features from individual elements to derive their abundances. Before DR10, the pipeline development was focussed on the first step, so that only the set of overall stellar parameters are reported in DR10. Because molecular features (CN, CO, and OH) can be very prominent in cool stellar atmospheres, a global fit needs to allow for variations in at least seven parameters: effective temperature $T_{\mathrm{eff}}$, surface gravity $\log g$, microturbulence $\xi_t$, overall metal abundance [M/H], and relative $\alpha$-element (including oxygen) [$\alpha$/M], carbon [C/M], and nitrogen [N/M] abundances\footnote{The Solar abundance values are adopted from \citet{Asplund2005}. [M/H] is defined as the overall logarithmic metal abundance with respect to the Solar abundance ratio pattern. [X/M] denotes the deviation of an element X from the corresponding Solar abundance ratio, $\mathrm{[X/M] = [X/H] - [M/H]}$. The $\alpha$-elements considered by ASPCAP are O, Ne, Mg, Si, S, Ca, and Ti.\label{f1}}. As the microturbulence is currently approximated as a fixed function of log $g$ to save computing time, six independent parameters are released from the DR10 ASPCAP run. 

\subsection{Spectra quality, signal-to-noise ratio and radial velocities}
Various tests have shown that ASPCAP requires at least a S/N of 50/pixel, but optimally 100/pixel, to deliver robust chemical abundances \citep{AllendePrieto2008, Eisenstein2011, Ahn2013}. In the present work, we adopt a signal-to-noise ratio cut of 70. Our choice is a trade-off to yield a clean, yet statistically significant, sample. 

The radial velocities are taken from the ASPCAP files, and their uncertainties calculated as the quadratic sum of the visit-to-visit scatter and the median visit error in $v_{\mathrm{los}}$ (usually the visit-to-visit scatter dominates). To eliminate likely binaries, it is required that $\sigma(v_{\mathrm{los}})<1$km/s.

\subsubsection{ASPCAP convergence}
 ASPCAP finds the best-fit stellar model atmosphere based on a $\chi^2$ minimisation of the cross-correlation between the observed spectrum and a grid of synthetic model spectra \citep{Meszaros2012, Ahn2013}. However, for a number of stars the algorithm does not yet find a satisfactory match in the set of synthetic spectra, due to a variety of reasons. The most common case in DR10 is that a star has a much cooler atmosphere than even the coolest grid models currently available; this occurs for the extremely luminous M (super-)giants. In some cases the ASPCAP algorithms also fail to find the absolute minimum in the \frqq$\chi^2$ landscape\flqq~of the model grids, and thus the best-fitting synthetic atmosphere. As such cases must be avoided, it is necessary to:

\begin{itemize}
\item Eliminate stars whose ASPCAP parameters lie too near the edges of the current grids of synthetic spectra.
\item Set an upper limit on the (reduced) $\chi^2$ of the ASPCAP fit to avoid poorly converged results.
\end{itemize}
Both these considerations have entered into our sample selection; in this work we require $\chi^2<25$.

\begin{figure*}[th]\centering
 \includegraphics[width=0.4\textwidth]{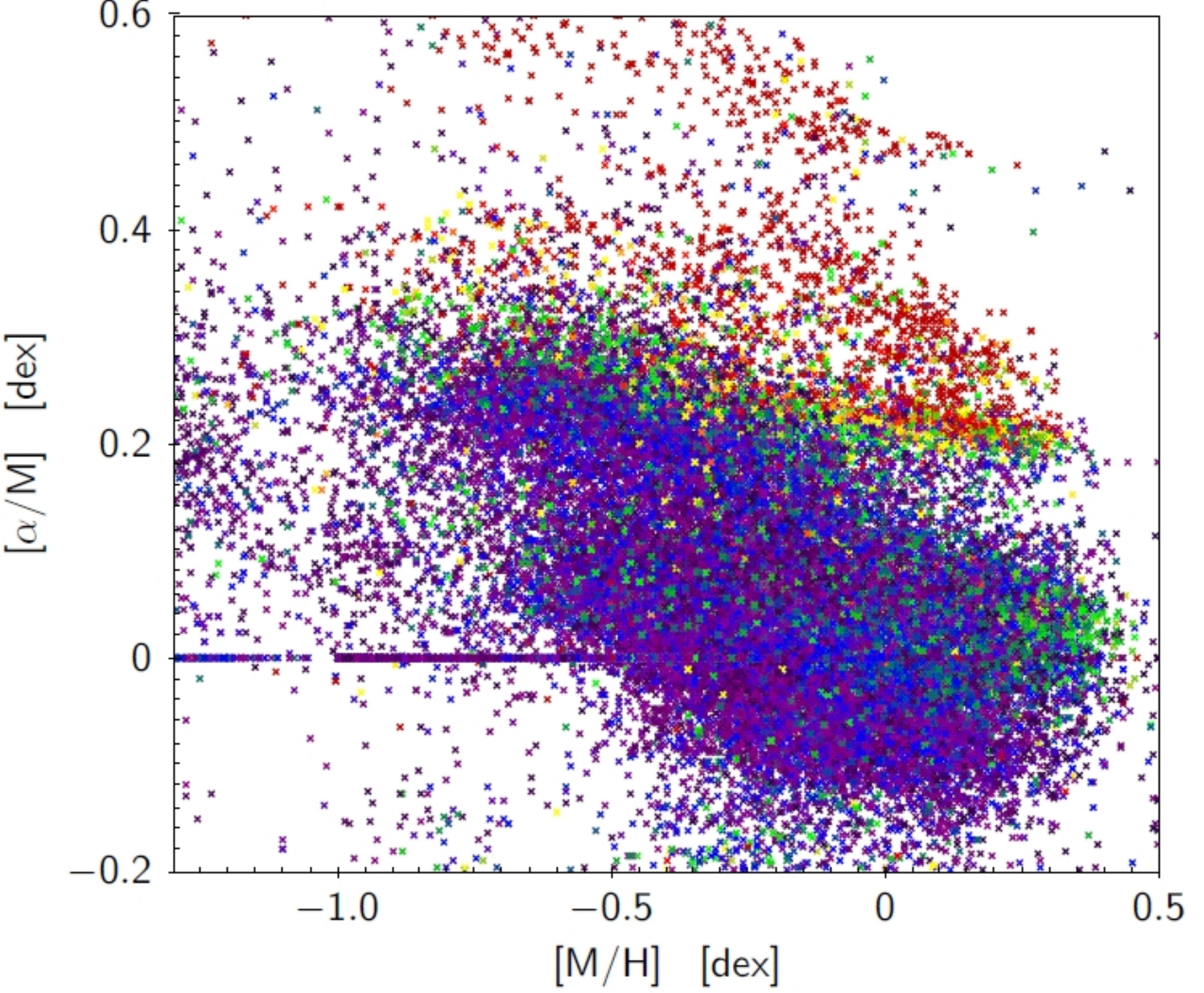}
 \includegraphics[width=0.43\textwidth]{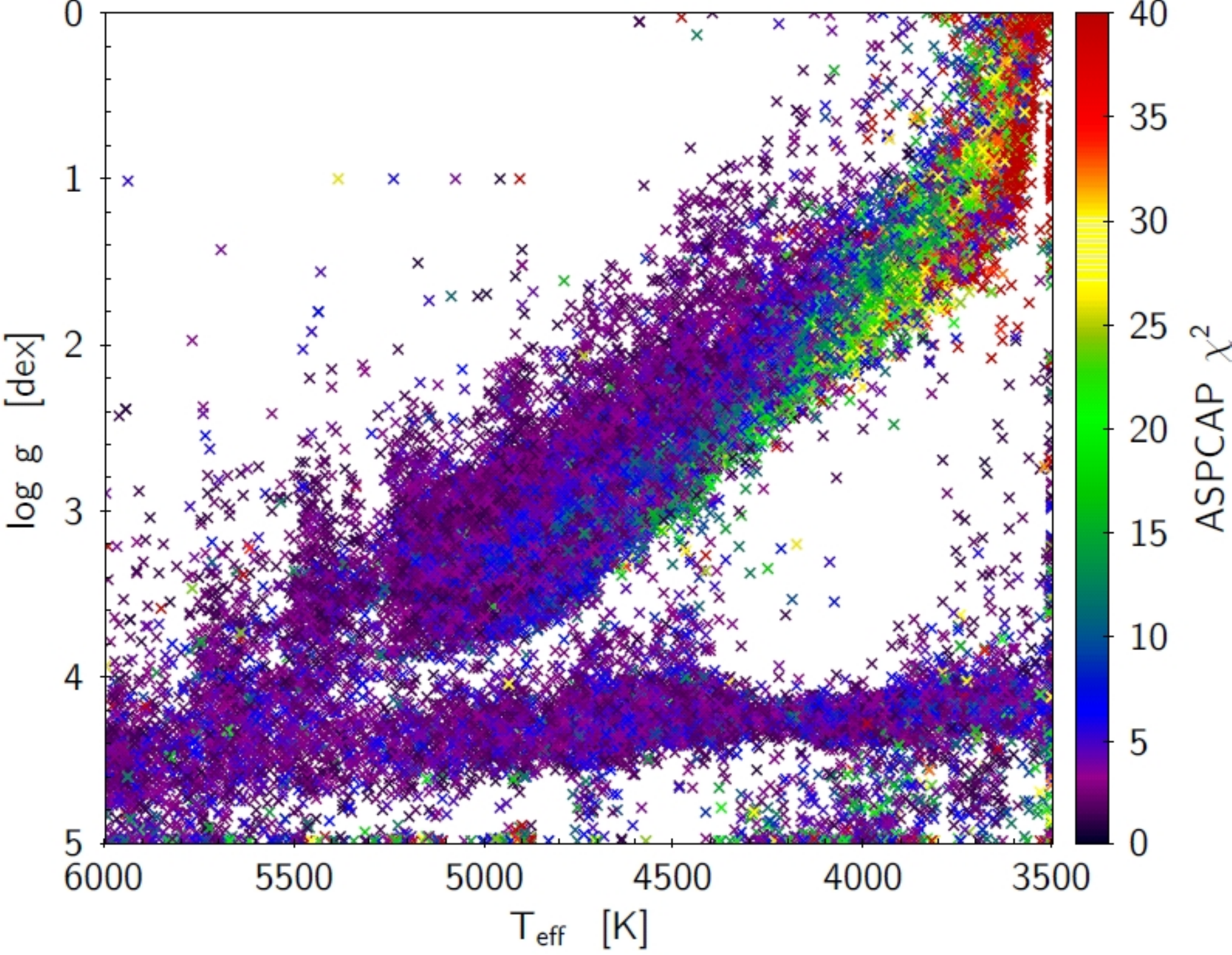}
\caption{Two 2D slices through the 6-dimensional hypercube of ASPCAP parameter space, colour-coded by $\chi^2$. Left panel: [$\alpha$/M] vs. [M/H], the so-called \frqq chemical plane\flqq. 
Some artificial features introduced by ASPCAP are also visible (the region of unphysical, poorly converged best-fit models appearing in red; the line at [$\alpha$/M]$=0.0$ corresponding to the A and F dwarfs forced to Solar $\alpha$-abundances; see Section~\ref{aspcaphow} for details). Right panel: The ASPCAP {\it Kiel} diagram ($T_{\mathrm{eff}}$ vs. $\log g$). Giant stars lie on the diagonal branch, while main sequence stars are aligned in the horizontal sequence in the lower part of the diagram. The latter behaviour is somewhat unphysical -- cooler main sequence stars should have higher surface gravities -- and shows that the pipeline is not optimised for dwarf stars yet.}
\label{chi2}
\end{figure*}

While there is a clear trend of the ASPCAP fit $\chi^2$ with temperature, this fact alone does not mean that cooler stars have more uncertain parameters. In fact, this trend is expected because the spectra of cool stars become considerably more \frqq crowded\flqq~due to the numerous molecular features, and are harder to fit by automated software. But loosening the overall $\chi^2$ criterion for cool stars by allowing, e.g., $\chi^2<40$ for $T_{\mathrm{eff}}<4200$ K, shows that high $\chi^2$ is indeed correlated with issues in the [C/M] and [$\alpha$/M] parameters in the cool regime (see left panel of Fig. \ref{chi2}). We have thus maintained the same $\chi^2$ limit for all temperatures. We are aware that this choice induces a small bias against the most metal-rich part of the upper giant branch. This point should be kept in mind when interpreting our results in Section~\ref{results}.

\subsubsection{ASPCAP parameters}

Most importantly, the giant stars for the HQ sample are selected from the ASPCAP {\it Kiel} diagram ($T_{\mathrm{eff}}$ vs. $\log g$, right panel of Figure \ref{chi2}) based on a generous cut of the giant branch, resulting in a $T_{\mathrm{eff}}$ upper limit of 5200 K and an (uncalibrated) $\log g$ upper limit of 3.8 dex (see below).
ASPCAP DR10 metallicities are generally well-behaved and reliable in the metallicity regime of the Galactic disc ($-1.5\lesssim\mathrm{[M/H]}\lesssim+0.4$, \citealt{Meszaros2013}) with small systematic shifts at the metal-rich end as well as larger shifts in the very low-metallicity regime. In this study we applied a more conservative cut in the metal-poor regime ([M/H]$=-1.0$), which was based on tests with previous ASPCAP versions. To cover the entire metallicity regime of the thin disc and still avoid the ASPCAP grid edge at [M/H] $=+0.5$, we cut the metal-rich end at [M/H] $=+0.45$. As has also been shown by \citet{Meszaros2013}, $\alpha$-element abundances derived by ASPCAP match the results from cluster literature fairly well for $-0.5<\mathrm{[M/H]}<+0.1$; outside this metallicity range some systematic dependencies on the other fit parameters are seen. The applied calibrations and adopted uncertainties for these parameters are discussed in the next Sections.

\subsection{Calibrations}

\subsubsection{Effective temperature}
DR10 effective temperatures derived by ASPCAP are fairly reliable over a wide parameter range, showing a good agreement with independently-derived temperatures from high-resolution spectroscopy (deviating on average by $8\pm161$ K), and a good agreement with effective temperatures derived with the IR flux method using the relations of \citet[][]{GonzalezHernandez2009}, modulo a zero-point shift of 113 K (see \citealt{Meszaros2013} for details). 

Whereas \citet{Meszaros2013} decided to correct for this shift, we currently use the uncorrected DR10 temperatures because of the good agreement with high-resolution optical spectroscopy. It is known that systematic differences between the photometric and spectroscopic temperature scales exist: spectroscopic \frqq excitation temperatures\flqq~often yield lower values than colour--temperature calibrations by a few hundred Kelvins \citep[e.g.,][]{Johnson2002}. \\

\subsubsection{Surface gravity}
\label{loggcalib}

Whereas ASPCAP effective temperatures are currently considered to be remarkably accurate when compared to surveys of similar size, the pipeline still has considerable difficulties in providing reliable estimates for surface gravities; log $g$ offsets of order $0.3-0.5$ dex are documented (\citealt{Meszaros2013}).

\begin{figure}[!h]\centering
 \includegraphics[width=0.5\textwidth]{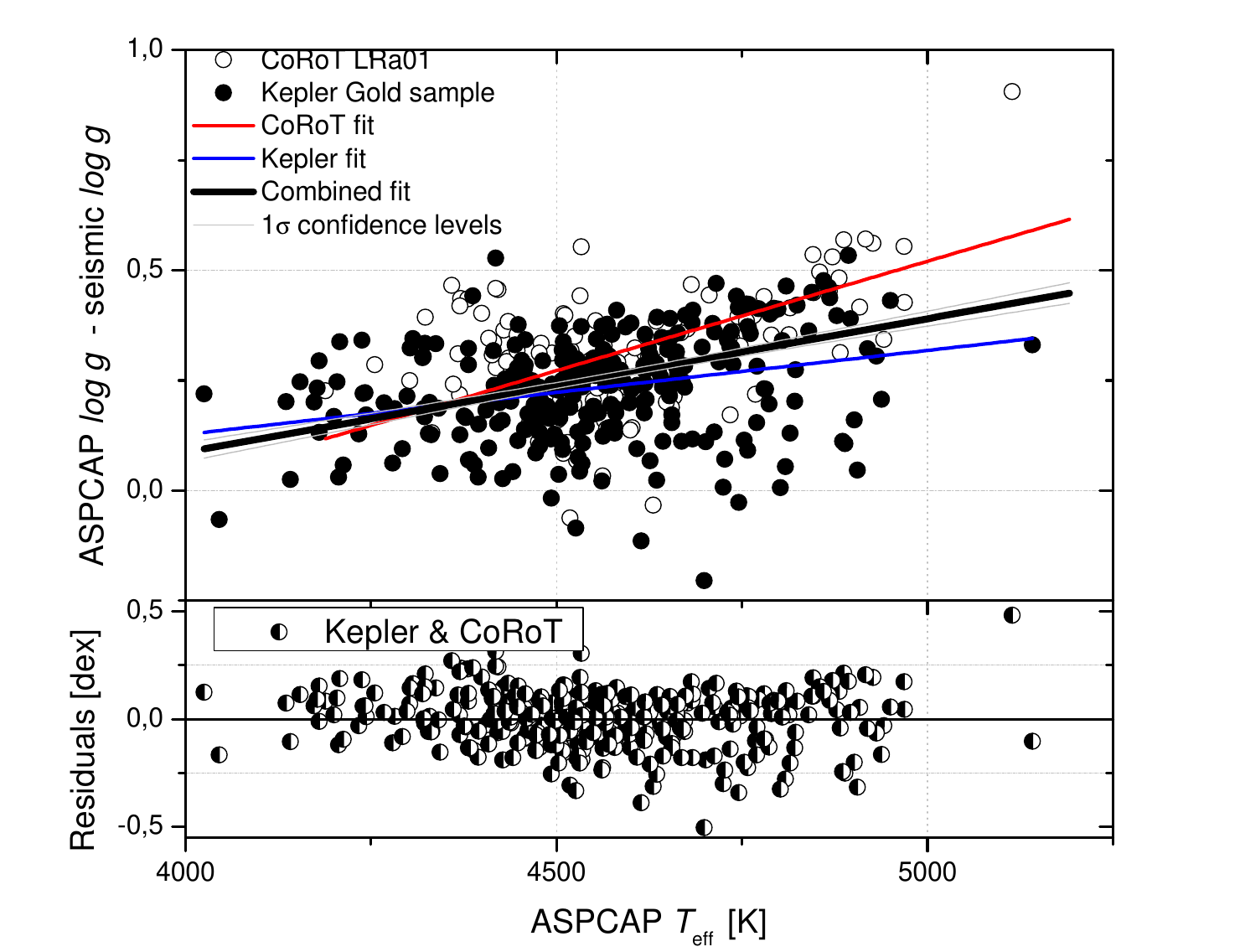}
\caption{Illustration of the applied $\log g$ calibration using asteroseismology data. ASPCAP DR10 $\log g$ is higher with respect to the seismic values by on average $\sim0.25$ dex, with the discrepancy increasing with increasing effective temperature. A linear fit using only CoRoT data (115 stars, open circles) is given by the red line, a fit using only Kepler data (279 stars, black circles) is indicated by the blue line. The fit obtained by combining the two datasets is illustrated by the thick black line. The lower panel shows the residuals, revealing some remaining possible systematics.}
\label{keprot}
\end{figure}

In the present work, we correct for these systematics by calibrating log $g$ using asteroseismic data from 279 Kepler stars contained in the APOKASC\footnote{The collaboration between {\it Kepler} and APOGEE (where KASC stands for the Kepler Asteroseismic Science Consortium).} catalogue (Epstein et al. 2014, in prep.), as well as 115 stars observed by the CoRoT satellite that have been followed up by APOGEE \citep[CoRoT field LRa01, data published in][]{Miglio2013a,Miglio2013b}. As shown in Figure \ref{keprot}, the following linear correction as a function of temperature was applied for $T_{\mathrm{eff}}>4000\text{ K}$:\footnote{As shown in \citet{Meszaros2013}, a pure asteroseismic analysis suggests that the uncorrected DR10 gravities are overestimated in the full metallicity range, whereas a comparison with the cluster isochrones suggest that the DR10 surface gravities are nearly correct, hence implying a dependency of the gravity correction on metallicity only in the metal-poor regime. We instead provide a pure asteroseismic calibration based on an extended sample, also including the CoRoT targets, which is appropriate for the metallicity range considered in the present work (with [M/H] > $-$1).}

\begin{align*}
\mathrm{log~} g_{\mathrm{calib}} & = \mathrm{log~} g_{\mathrm{ASPCAP}}+1.13-3.03\cdot10^{-4}\cdot T_{\mathrm{eff}}. 
\end{align*}

For temperatures between 3800 K $< T_{\mathrm{eff}} <$ 4000 K, no correction was applied.

\subsubsection{Metallicity}

For our analysis, we use the calibration described in \citet{Meszaros2013}, derived using a sample of well-studied open and globular clusters covering a wide range of metallicities ($\mathrm{[Fe/H]}\in\{-2.3,+0.4\}$).

\subsubsection{$\alpha$-element abundance}
Several tests suggest that APOGEE DR10 $\alpha$-element abundances are still to be treated with caution, but can in principle be used in scientific analyses \citep{Ahn2013}. While \frqq $\alpha$\flqq~in principle tracks the elemental abundances of O, Ne, Mg, Si, S, Ca and Ti, the spectral features corresponding to these elements are very sensitive to changes in the effective temperature (in cooler atmospheres, [$\alpha/$M] mainly tracks O and Ti, whereas in warmer atmospheres Ca, Mg and Si features are more important), so that any trends seen with $\alpha$-element abundance should be checked in narrower temperature bins. For cooler metal-poor stars, the lack of Fe lines seems to be the primary source of ambiguity for the overall metal and oxygen abundance. The sytematic trends seen at the metal-rich end still remain poorly understood. 

\subsection{Uncertainties}
\subsubsection{Adopted errors}
The initial ASPCAP parameter error estimates are based on the the random contributions to the errors as derived by inverting the FERRE $\chi^2$ curvature matrix, following the favoured prescription of \citet{Press1992}. However, these values are too small to represent reliable random uncertainties by roughly a factor of 15 when compared to the scatter observed in the calibration clusters (García Pérez et al. 2014, in prep.). For DR10, it has therefore been decided to follow the conservative (though somewhat artificial) uncertainty treatment of \citet{AllendePrieto2006}. The final error on each parameter is calculated as the larger of a) the individual FERRE errors times 15, and b) the general scatter of the clusters as given by \citet{Meszaros2013}:

\begin{equation}
\begin{split}
\Delta T_{\mathrm{eff}}&=(83.8 - 39.8\cdot\mathrm{[M/H]})~\mathrm{K}\\
\Delta \log g&=0.2~\mathrm{dex}\\
\Delta\mathrm{[M/H]}&=(0.055 - 0.036\cdot\mathrm{[M/H]})~\mathrm{dex}\\
\Delta\mathrm{[\alpha/M]}&=0.08~\mathrm{dex}
\end{split}
\end{equation}

We have adopted this prescription for this work, which delivers at least reliable upper limits to the uncertainties.

\subsubsection{Binarity}

It has long been established that a high percentage of the local F- and G-dwarf population lives in multiple stellar systems (e.g., \citealt{Duquennoy1991} and \citealt{Duquennoy1991a} estimate a multiplicity fraction of 65\%, while recent estimates by \citet{Fuhrmann2011} suggest a value of 50\% for Solar-type stars). This underlines the importance of understanding how unresolved companions affect stellar parameter estimates. \citet{Schlesinger2010} used the SEGUE Stellar Parameter Pipeline to estimate the effects of potential contamination by the light from a binary companion on their high-S/N sample of $\sim20,000$ G-K dwarf stars observed by SEGUE, and find that 11$\pm$2\% of the latter is expected to be significantly affected in its temperature or metallicity determination by an undetected companion, resulting most importantly in a systematic shift to cooler temperatures.

Although we cannot provide quantitative estimates of binarity effects on ASPCAP's stellar parameter estimates yet, the affected sample percentage should be even smaller than in SEGUE, for two reasons. First, giant stars are quite luminous, so that the light of the primary is likely to dominate the resulting spectrum. Secondly, APOGEE's split multi-epoch observations permit accurate detections of temporal radial-velocity variations, so that by requiring the radial-velocity scatter $\sigma(v_{\mathrm{los}})$ to be small we already eliminate a significant fraction of the multiple systems (which on the other hand means introducing another bias into our sample).

\subsection{Adopted subsamples}\label{subsamples}

 We have defined, for the first time, a high-quality chemical sample extending at least 4 to 6 kpc beyond the solar circle. This dataset is crucial for constraining chemodynamical models outside the solar region, something urgently needed in the field and so far addressed with SEGUE \& RAVE -- low- and medium-resolution samples heavily biased to high Galactic latitudes. We will use the chemical high-quality (\frqq HQ\flqq) sample to study the inner and outer parts of the disc. 

We further define four high-quality (sub-)\-samples with different characteristics (see Table \ref{samplesummary} for details):
\begin{itemize}
\item An (extended) Solar-vicinity sample of APOGEE red giants confined to a sphere of radius 1 kpc around the Sun, for comparison with previous high-resolution studies, in particular the recent HARPS FGK dwarf sample of \citet{Adibekyan2011}.
\item The HQ$^k$ sample -- a subsample of the HQ sample with fully-determined 6-D phase space coordinates, i.e., valid distance determinations and proper motions (see Sect. \ref{kinematics}). The superscript {\it k} stands for `kinematics'.
\item A chemodynamical disc sample with as precise kinematical information as possible -- not as local as existing high-resolution samples in the literature, but extending to 1--2 kpc in distance. We will define an APOGEE \frqq Gold Sample\flqq~which meets these criteria, by imposing quality limits on distance and proper motion error.
\end{itemize}

While the first two samples are free from any further biases that might be introduced by the proper motion catalogue, the other two samples might possess some biases. In addition, in the case of the extended sample, biases are expected towards the inner Galactic regions mainly due to a sparse coverage of the stellar disc (additional biases affecting the APOGEE DR10 sample as a whole are discussed in \citealt{Hayden2013}). In a forthcoming paper we intend to simulate our sample with a population synthesis model to be able to quantify better the impact of those biases on our results. The present paper mainly focuses on observables that are less affected by potential observational biases.

\begin{table*}[t]
  \centering
\caption{Definitions and sizes of useful subsamples of the HQ sample. }
\begin{tabular}{l l r}
\hline
Name & Requirements & Number of stars\\
\hline\hline
HQ sample & see Table 1 &  21,288\\
HQ sample with reliable $\alpha$-element abundances & $4000~\mathrm{K}<T_{\mathrm{eff}}<5000$ K & 18,855\\
HQ sample with valid distance determination  & distance code \citep{Santiago2013} converges & 21,105\\
HQ sample with (valid) UCAC-4 proper motions  & PM criteria (see Sect. \ref{pm}) are fulfilled & 17,882\\
\hline
HQ$^k$ sample & valid proper motions  \& distances & 17,758\\
Local HQ sample & $d<1$ kpc & 1,975\\
Local HQ$^k$ sample & $d<1$ kpc $\land$ HQ$^k$& 1,654\\
Gold sample & $\sigma(\mu)<4.0$ mas/yr $\land~ \sigma(d)/d)<20\%$ & 3,984\\
\hline
\end{tabular}
 \label{samplesummary}
\end{table*}

\section{Kinematics}\label{kinematics}
\label{kinematics}

To perform a thorough chemodynamical analysis of a stellar survey, it is necessary to measure and interpret the motion of the stars inside the Galaxy and to calculate their orbital parameters.\footnote{In turn, stellar motions and their statistics can in principle also be used to determine the form of the Milky Way potential. The usefulness of APOGEE in this context was recently demonstrated by \citet{Bovy2012a}.} Here, we particularly aim at finding correlations between chemical-abundance patterns and orbital properties. To obtain the full 6-dimensional phase space coordinates of the stars in the HQ sample, the 2MASS astrometry and APOGEE line-of-sight velocities must be complemented by information on stellar distances and proper motions.\\

\subsection{Distances} \label{dist}

The development of sophisticated spectrophotometric parallax methods has been undertaken by many different groups in the past several years \citep[e.g.,][]{RochaPinto2003, AllendePrieto2006, Breddels2010, Zwitter2010, Burnett2010, Burnett2011}. For APOGEE stars, preliminary distance estimates from various groups exist (Hayden et al. 2014, in prep.; Santiago et al. 2014; Schultheis et al. 2014, subm.). We have computed our distances based on the Bayesian approach of \citet{AllendePrieto2006}, which was further developed by us \cite[see][]{Santiago2013} to compute SDSS distances both for APOGEE (giants) and SEGUE (dwarfs). In this section, the general features of the method are briefly described; for a detailed description, the reader is referred to \citet{Santiago2013}.\\

 The goal of isochrone-based distance codes is to find 
stellar models that fit as many spectrophotometric observables as possible (magnitudes, colours, stellar parameters, abundances), and are most likely to be close to the \frqq true\flqq~one. In the Bayesian method adopted in \citet{Santiago2013}, an efficient use is being made of all the available uncertainties and several simple priors (stellar density distribution, initial mass function, uniform star formation history with different cut-offs for the different stellar components, metallicity distributions). A general framework for spectrophotometric distances using Bayesian methods is provided by, e.g., \citet{Burnett2010}.

In brief, one can write the probability of finding the \frqq true\flqq~parameter set for a star ${\bf x}=(l,b,s,M,\tau,{\rm [M/H]})$ when observing the quantities ${\bf y}= (T_{\mathrm{eff}}, \mathrm{log}~g, \rm{[M/H]_{obs}}$, magnitudes, colours, $l_{\mathrm{obs}}, b_{\mathrm{obs}}, \dots$) via Bayes' theorem as 
\begin{equation}\label{bayes}
p({\bf x} | \overline{\bf y},{\boldsymbol \sigma_{\bf y}}, S)\propto P(S|\overline{\bf y},{\bf x},{\boldsymbol \sigma_{\bf y}})\cdot p(\overline{\bf y}|{\bf x},{\boldsymbol \sigma_{\bf y}})\cdot p({\boldsymbol \sigma_{\bf y}}|{\bf x})\cdot p({\bf x})
\end{equation}
Here, $(l,b)$ are the position angles in the Heliocentric Galactic coordinate frame, $s$ the distance from the Sun, $M$ the initial stellar mass, $\tau$ its age and [M/H] the overall metallicity. Quantities with subscript `obs' stand for the corresponding observed values.\\
The actual measured values of the observed parameters {\bf y} and their uncertainties are denoted as $\overline{\bf y}$ and $\boldsymbol \sigma_{\bf y}$, respectively, whereas the property $S$ stands for the fact that the star belongs to our sample. The four factors in eq. \ref{bayes} are
\begin{enumerate}
\item The selection function (SF) of the sample, $P(S|\overline{\bf y},{\bf x},{\boldsymbol \sigma_{\bf y}})$.
\item The likelihood $p(\overline{\bf y}|{\bf x},{\boldsymbol \sigma_{\bf y}})$ that, given the true values ${\bf x}$ and the measurement uncertainties ${\boldsymbol \sigma_{\bf y}}$, the set $\overline{\bf y}$ is measured.
\item The probability $p({\boldsymbol \sigma_{\bf y}}|{\bf x})$ to observe the quoted errors given the variable set ${\bf x}$.
\item A number of multiplicative priors subsumed under the expression $p({\bf x})$.
\end{enumerate}

Each of these terms has to be modeled separately, which in the case of large stellar surveys usually proves a challenging task. However, some of the (sub-)terms peak more sharply than others, thus dominating the full probability distribution function (pdf) in eq. \ref{bayes}. The statistically relevant set of `true' parameters ${\bf x}$ and its uncertainties can then be calculated by computing the moments of this pdf. In particular, a distance estimate $s^\ast$ is computed by marginalizing the pdf over the other parameters and then computing the mean, mode or median of the one-dimensional probability distribution. 

For our APOGEE sample, we adopt the following assumptions for the four terms in eq. \ref{bayes}: 
\begin{enumerate}
\item The dependency of the pdf on the selection function is assumed to be slowly-varying, which may be the main caveat of our current method. However, the sharp magnitude and colour limits in the selection function are already being accounted for by the likelihood term, and we include a term to deal with the Malmquist bias in the priors (see below). In the future, the full selection function or at least a field dependent magnitude distribution will be included in this term: $P(S|\overline{\bf y},{\bf x},{\boldsymbol \sigma_{\bf y}})\propto p(l,b,H)$, representing the distortion of the underlying distribution introduced by APOGEE's targeting scheme.
\item The likelihood $p(\overline{\bf y}|{\bf x},{\boldsymbol \sigma_{\bf y}})$ is modelled by a multivariate Gaussian, meaning that all parameters are assumed to have independent Gaussian errors. We use the photometric uncertainties from 2MASS and the spectroscopic uncertainties as quoted in Section \ref{sample}.
\item The term $p({\boldsymbol \sigma_{\bf y}}|{\bf x})$ is set to unity for simplicity, as the dependence of the full pdf on variations of ${\boldsymbol \sigma_{\bf y}}$ with ${\bf x}$ will be sufficiently weak.
\item As priors on ${\bf x}$ we assume a Chabrier-type initial mass function $p(M)$ \citep{Chabrier2001}, and assume different density and metallicity distributions as well as star-formation histories (SFH) for the Galactic components Bulge, Thin Disc, Thick Disc and Halo, following \citet{Burnett2011}. In addition, we correct for the Malmquist selection bias resulting from the fact that more luminous stars are preferentially detected by magnitude-limited surveys \citep{Malmquist1936}. We account for this effect by including a term $p(M_{\mathrm{abs}})\propto 10^{0.6 M_{\mathrm{abs}}}$. 
\end{enumerate}

 Whereas the first three assumptions are fairly straightforward and well-accepted, the discussion of how restrictive the priors of the underlying {\bf x} distribution should be is still ongoing. \citet{Burnett2010} argue that the approach of starting from simple uniform priors to not overload the modeling with prejudices is difficult to defend, because the justification to prefer, e.g., a uniform age distribution over a uniform distribution in log(age) is not clear. A rigorous calibration of these priors using a combination of asteroseismology and high-resolution spectroscopy is urgently needed in the field and an ongoing project of the SDSS-III/Brazilian Participation Group.

\subsubsection{Differences to other approaches, encountered difficulties and recent upgrades}

Despite the fact that our method is similar to many other approaches used in the field, we wish to stress some refinements, namely:

\begin{itemize}
\item In principle, a number of measures (e.g., the mean, the median and the mode) could be used for finding the \frqq best\flqq~distance to a star from the full probability distribution (eq. \ref{bayes}). As the mode is an unstable quantity when the pdf is rather flat or multi-peaked, and the median is sometimes expensive to compute, we here use the mean, and the second moments of the pdf to obtain an estimate of the uncertainties. Alternatively, we define a different and more extensive prescription for the uncertainties, which is a major advantage of our code, and is described in Section \ref{disterr}.\\
\item The main difficulties in estimating distances for our dataset are the heavy interstellar extinction in the Galactic plane and the not-yet fully-understood systematic uncertainties in the log $g$ parameter, which impacts any spectrophotometric distance estimate\footnote{In fact, the latter issue is true for every currently operating spectroscopic survey.}. Unlike for most of the stars in GCS, RAVE and SEGUE, interstellar reddening is a dominant factor for our APOGEE sample, influencing primarily the NIR photometry. We have accounted for this effect by using RJCE-dereddened magnitudes and colours (see Section~\ref{photometry}).\\
\item
Differing from other groups, the surface gravity parameter was calibrated using only asteroseismology data, as described in Section~\ref{loggcalib}.
\end{itemize}

We have used the newly computed PARSEC isochrones \citep{Bressan2012}, which have a much more detailed grid of theoretical isochrones for the 2MASS $JHK_s$ photometric system than the ones previously available. As the adopted isochrones do not take [$\alpha$/Fe] enhancement into account, we adopted an ad-hoc approach to include the $\alpha$-abundance in the overall metallicity $Z$ of the scaled-solar Padova models using the approximation\footnote{For our APOGEE sample, the relation translates to $\mathrm{[Z/H]}\approx\mathrm{[M/H]_{calib}+[\alpha/M]}$. This approximation is still justified because ASP\-CAP's [M/H] which -- when uncalibrated -- tracks the overall metal abundance (as explained in footnote \ref{f1}), was calibrated on literature {\it iron} values, so that we can use $\mathrm{[M/H]_{calib}}$ as a proxy for [Fe/H].} $\mathrm{[Z/H]}\approx\mathrm{[Fe/H]+[\alpha/Fe]}$. 

Ideally, one would want to use self-consistent stellar models with variable $\alpha$-element content, thus adding an [$\alpha$/Fe] dimension to the isochrone set. New BaSTI \citep{Cassisi2006} and PARSEC models are now being computed with consistent alpha-enhanced compositions, which will solve this problem in the near future. At present, the available sets are still too limited and heterogeneous to be used for producing isochrones over a wide range of ages and metallicities.

\subsubsection{Uncertainties}\label{disterr}
Reliable estimates for the uncertainties of the computed distances are quite complicated to evaluate. Changing a model prior, changing a term in the selection function, or dropping one of the observed parameters can, in some cases, change the weighted mean absolute magnitude and thus the distance by a significant amount. In \citet{Santiago2013}, we estimate uncertainties in two different ways. First, we calculate an \frqq internal\flqq~ uncertainty by taking the second moment of the pdf in equation \ref{bayes}. To assess how sensitive the derived distances are to changes in the choice of the matching parameters, we also define an alternative \frqq external\flqq~uncertainty, based on distance estimates from different subsets of the observables ${\bf y}=\{\log g, T_{\mathrm{eff}}$, [Z/H], $J-H, H-K_s\}$

Various tests have been performed on possible measures of distance uncertainty. An {\it internal} measure of the variation of the pdf (Eq. \ref{bayes}) could be its confidence intervals, standard deviation or the difference between the mode and the mean of the pdf. It has been shown that both the maximum difference of the distances using different sub-datasets and the pdf's standard deviation yield similar and robust error estimates \citep{Santiago2013}.

In the following, we will generally use the \frqq internal\flqq~distance uncertainties. The distance uncertainty distribution for the APOGEE HQ sample is shown in the right panel of Fig.~\ref{disthisto}. 

\subsubsection{Resulting distances}

\begin{figure*}[!t]\centering
 \includegraphics[width=0.49\textwidth]{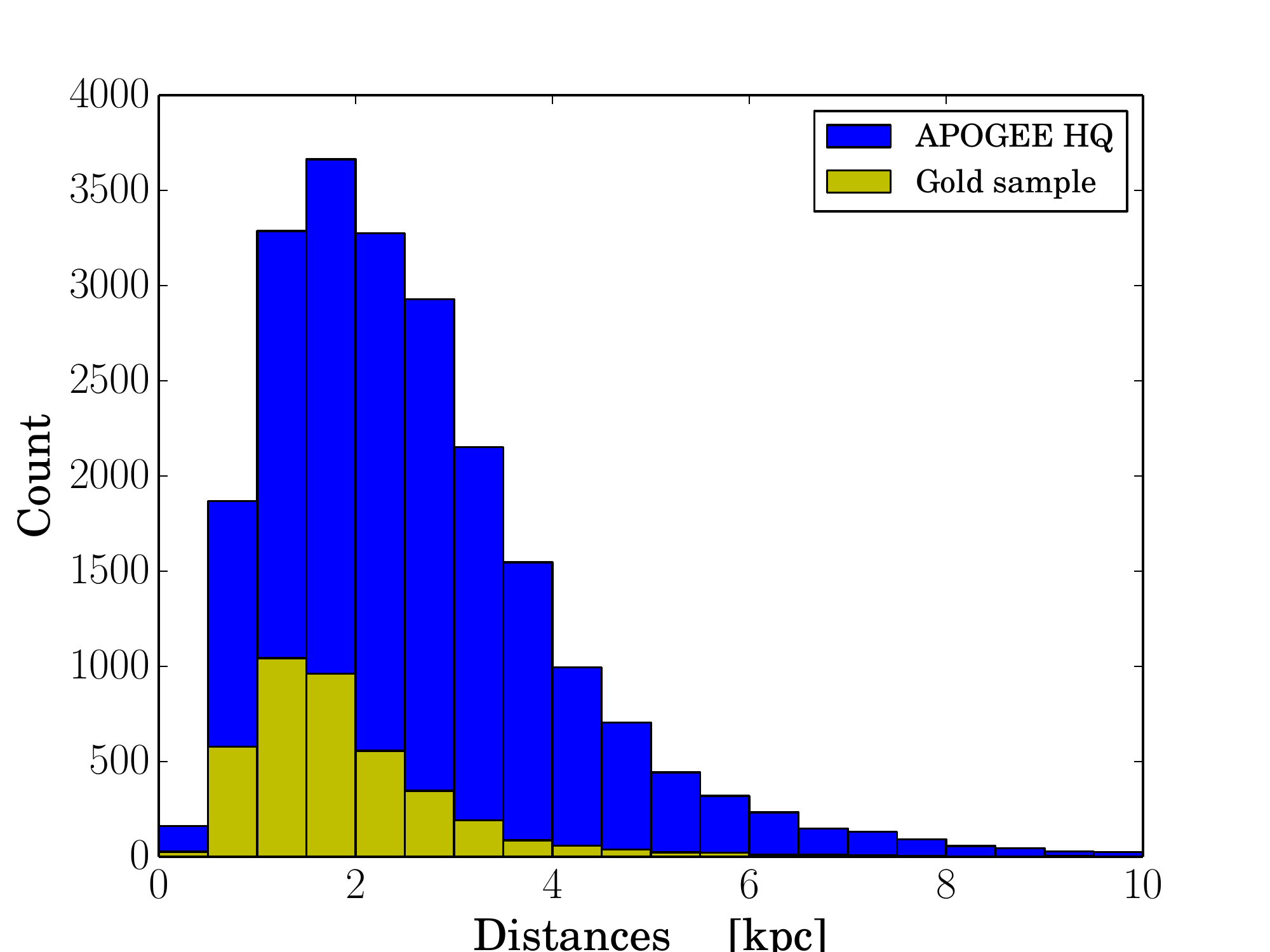}
 \includegraphics[width=0.49\textwidth]{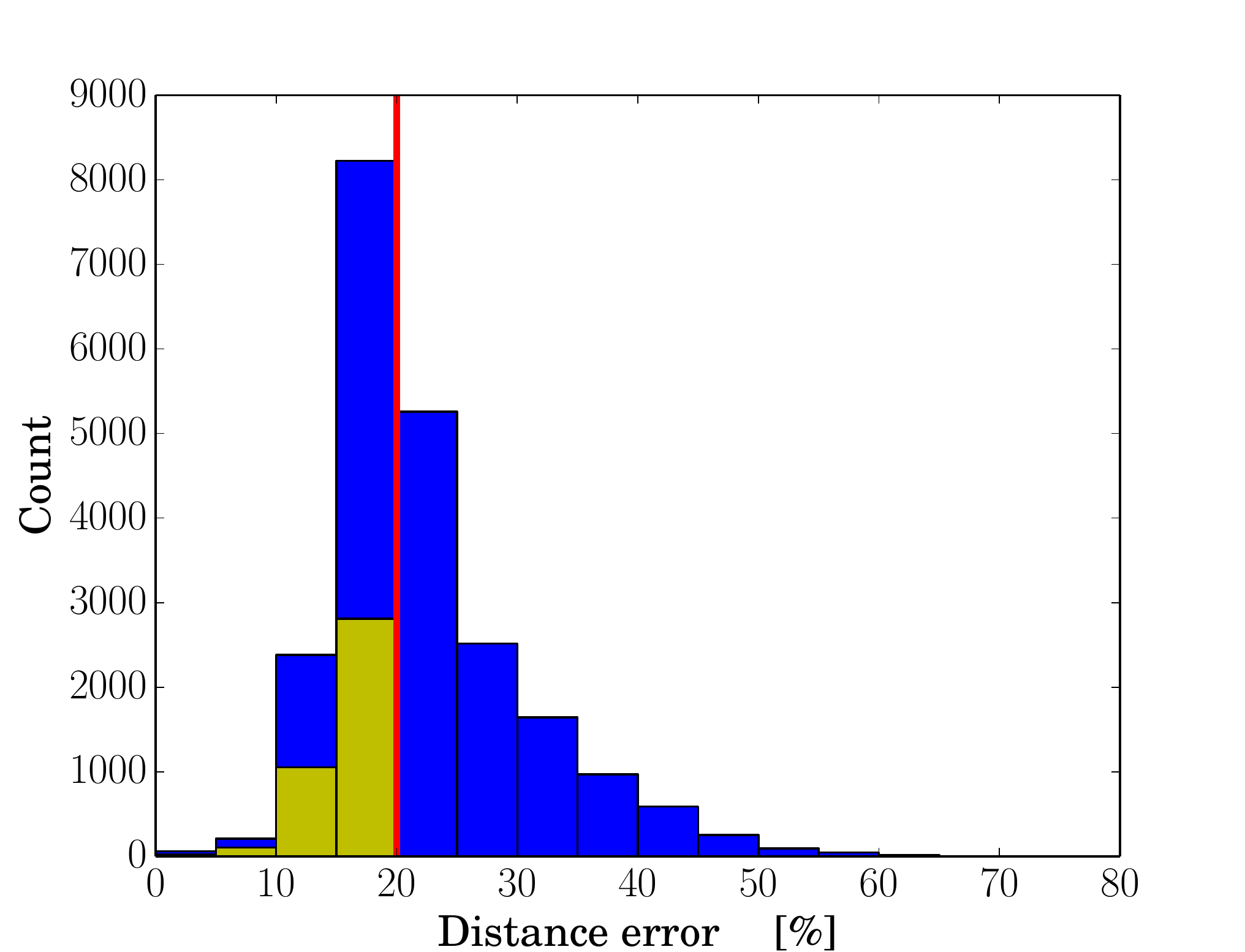}
\caption{Histogram of the distribution of spectrophotometric distances and their errors for the HQ (blue) and the Gold sample. Note that, in addition to the cut in relative distance error, indicated by the red line in the right panel, the Gold sample also satisfies a quality criterion for proper motions (see Section \ref{pm}).}
\label{disthisto}
\end{figure*}

We have computed distances for $\sim$ 21.000 stars in the HQ sample. In the left panel of Fig. \ref{disthisto}, we show the distance distribution for the APOGEE HQ sample, and the Gold sample defined in Section \ref{sample}. The Gold sample, as indicated in the right panel of Fig. \ref{disthisto}, satisfies $\sigma(d)/d<0.2$, along with a criterion on proper motion error (see Section \ref{pm}). The Gold sample consequently samples a smaller volume of the Galaxy, and the selection function for this subsample is not straightforward to calculate. 

\begin{figure*}[!h]\centering
	\includegraphics[width=0.49\textwidth]{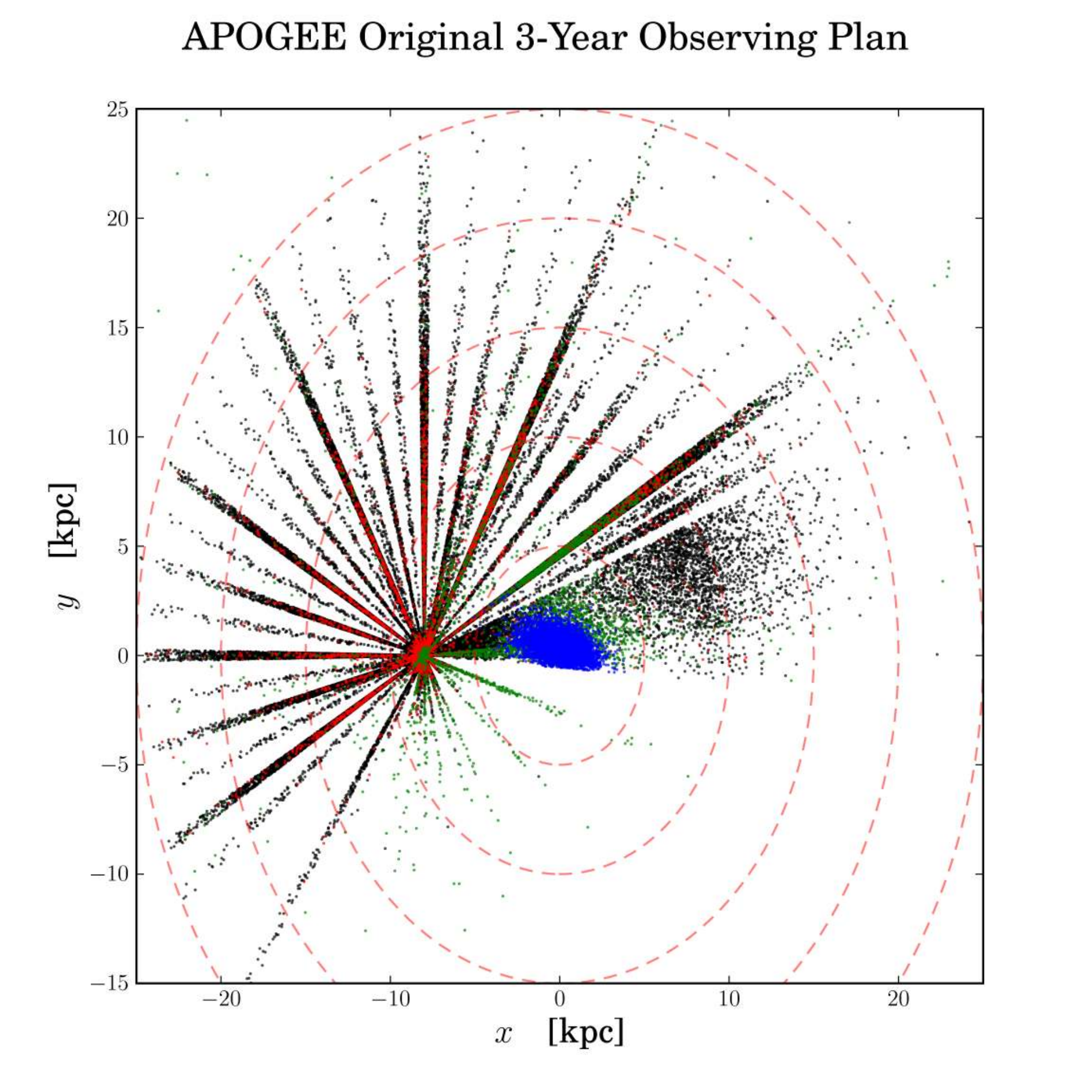}
	\includegraphics[width=0.49\textwidth]{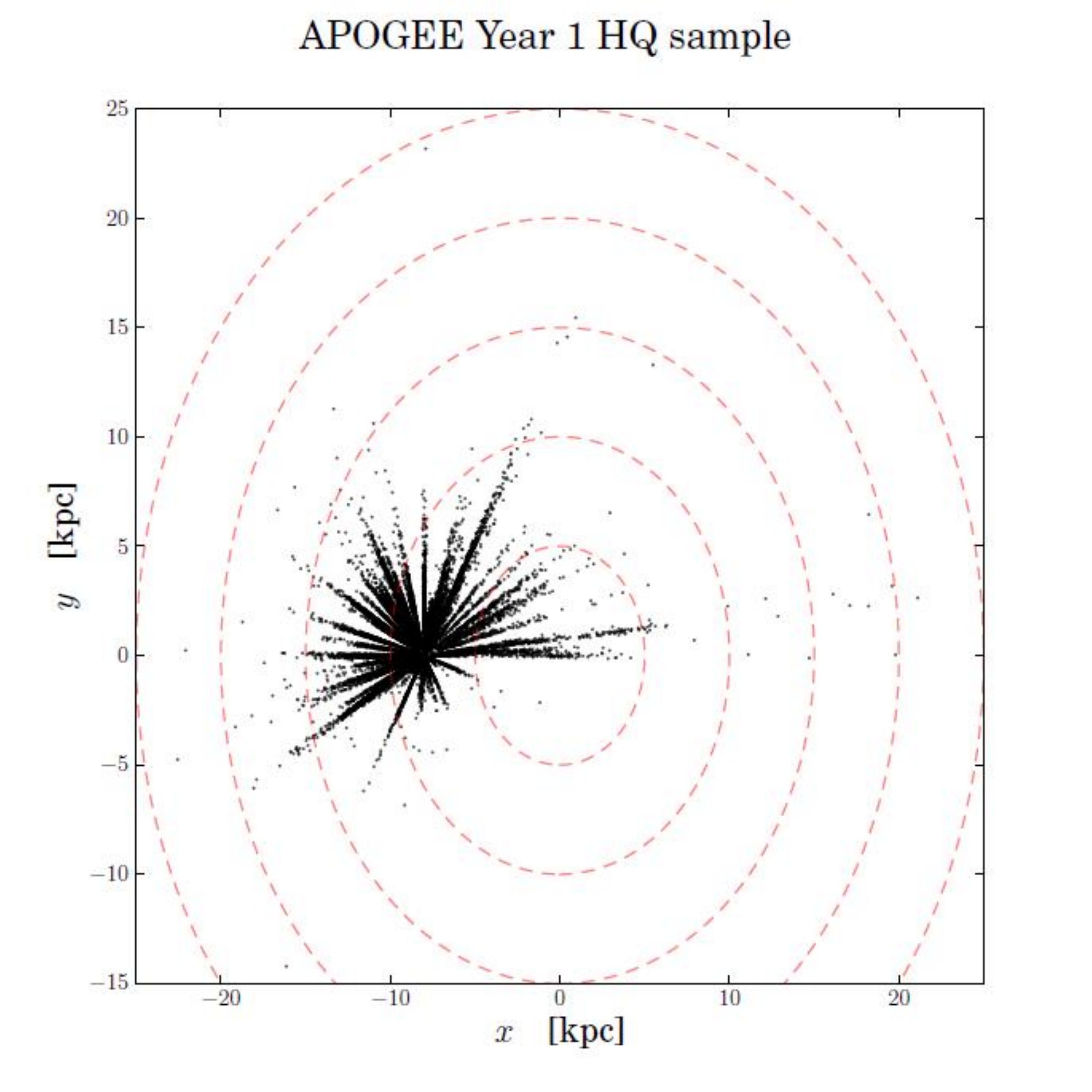}
	\includegraphics[width=0.49\textwidth]{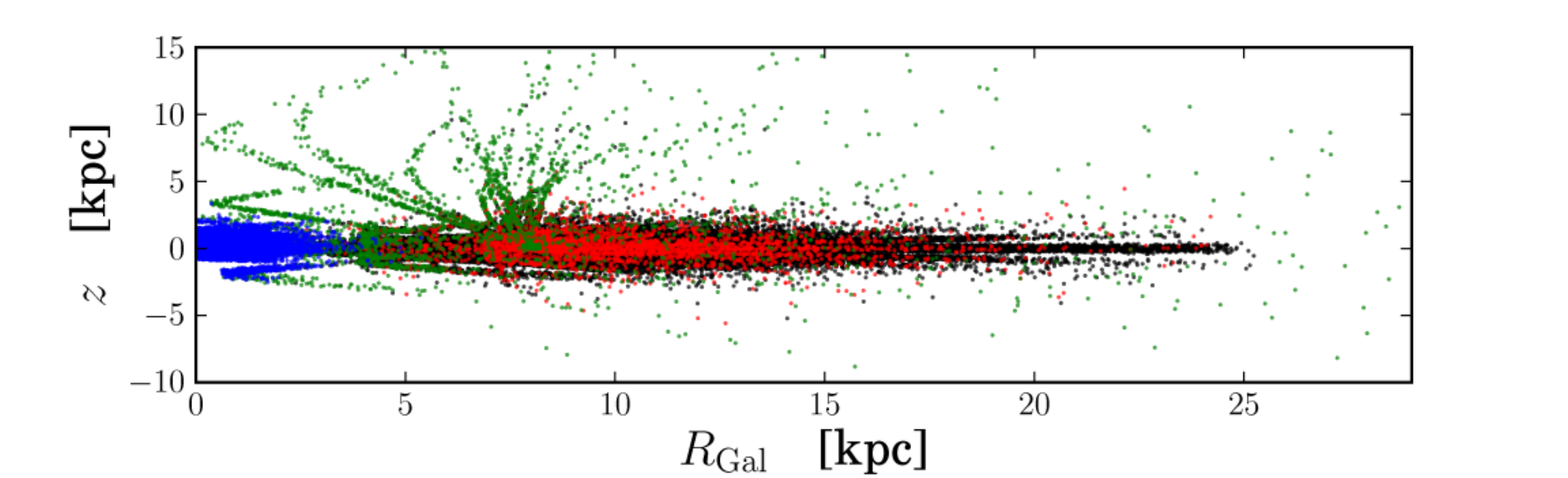}
	\includegraphics[width=0.49\textwidth]{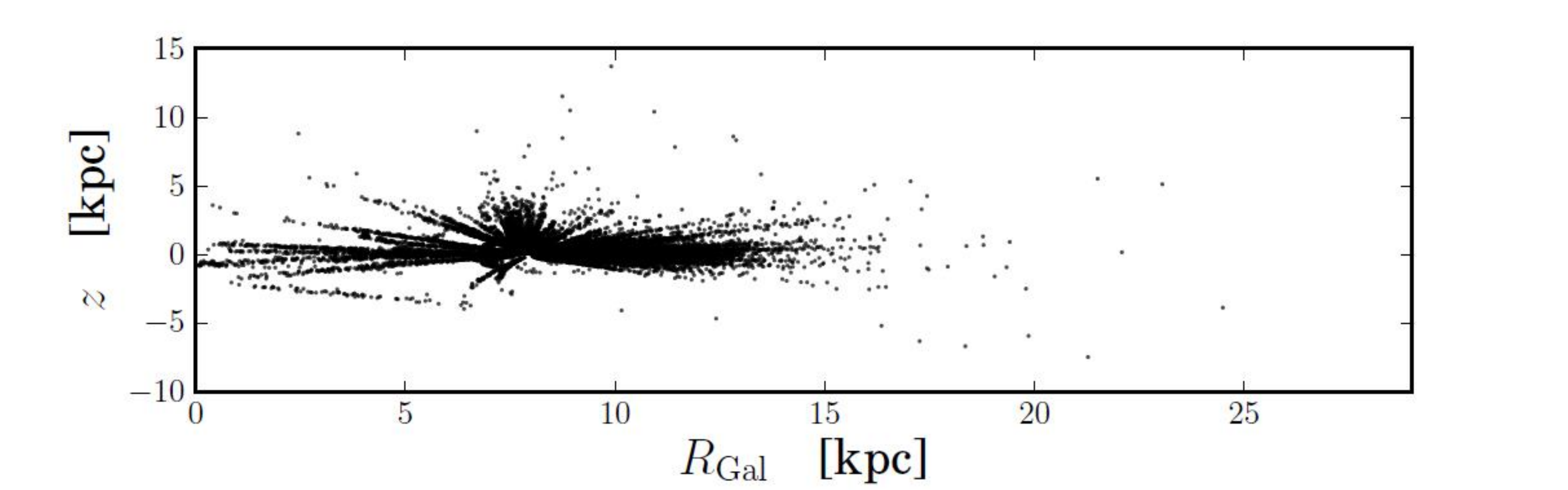}
	\caption{Left: TRILEGAL \frqq Strawman\flqq~simulation of the APOGEE 3-year survey sample in Galactocentric coordinates. Different colours correspond to different Galactic populations: blue -- bulge, green -- halo, black -- thin disc, red -- thick disc. Right: Distribution of the Year-1 APOGEE HQ sample in the same coordinates.}
	\label{tridistribution}
\end{figure*}

In Fig. \ref{tridistribution} we compare the volume covered by our Year-1 HQ sample (using our spectrophotometric distances) with the expectations for the 3-year survey data. Through multiple observations in many lines of sight, APOGEE will eventually cover a considerably larger part of the Galaxy than presented in this work.

\subsubsection{Distance validation}
To validate our code, we have compared our results with a number of completely independent distance measurements determined via asteroseismology, astrometric parallaxes and cluster isochrones. A detailed and quantitative comparison is presented in \citet{Santiago2013}. The comparison shows that the method also works reasonably well in an absolute sense. Despite a significant scatter, there is a clear one-to-one correlation with parallax and, modulo small systematic dependencies on the cluster age, with isochrone distances to open and globular clusters. The rms difference is $\lesssim 20\%$, as also expected from our error estimates. 

Additionally, our spectrophotometric distances compare favourably with the distances obtained from CoRoT data for 120 stars in the anticenter field \frqq LRa01\flqq that have been observed by APOGEE. Despite the substantial ($\sim20\%$) scatter for stars with distances $>3$ kpc and a small ($\lesssim15\%$) systematic shift in the absolute scale, a remarkable concordance of both methods is found. 

\subsection{Proper Motions}\label{pm}

Proper motions were added to the APOGEE data from an existing astrometric catalogue via crossmatching. There are two recent catalogues with sufficient sky coverage -- PPMXL \citep{Roeser2010} and UCAC-4 \citep{Zacharias2012, Zacharias2013}. The PPMXL catalogue, however, is partly based on images obtained with Schmidt photographic plates, and thus suffers from distortions in some regions of the plate and other systematic errors that are difficult to correct (e.g., \citealt{Roeser2010}). As UCAC-4 (based only on imaging with CCD cameras) also supersedes PPMXL in the achieved precision, and the number of stars in common with APOGEE for both catalogues is roughly the same (around 80\%), it was decided to use only UCAC-4 proper motions in the subsequent analyses to maintain a homogeneous catalogue. 

\begin{figure}[!h]\centering
 \includegraphics[width=0.49\textwidth]{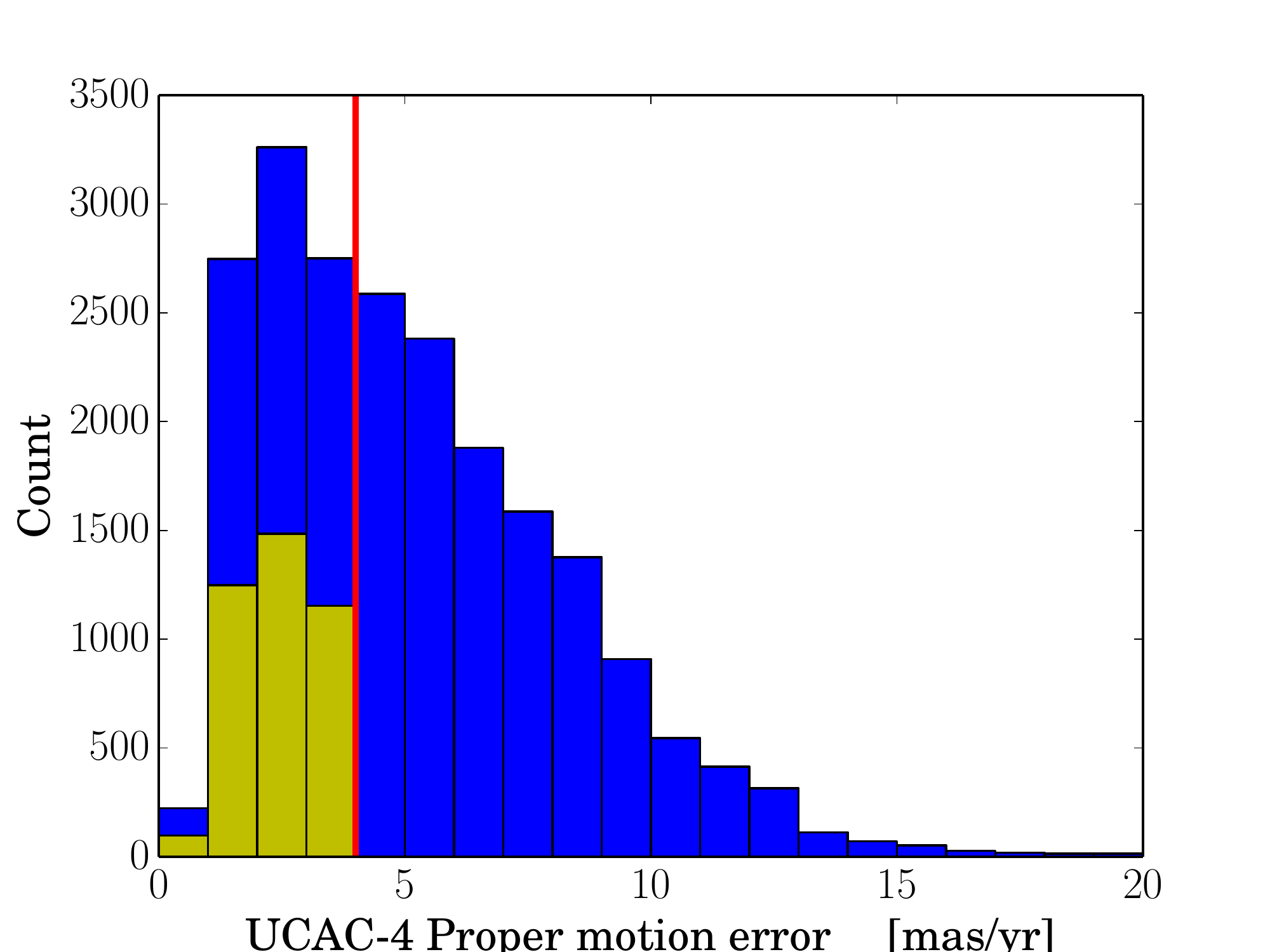}
\caption{Histogram of uncertainties in the absolute error in the UCAC-4 proper motions for the HQ sample with reliable proper motions. The quality cut for the Gold sample is indicated by the vertical red line.}
\label{pmplot}
\end{figure}

For our APOGEE stars, the following steps were taken:
\begin{enumerate}
\item We performed a multicone crossmatch with a fixed radius $r=5''$ of APOGEE's apStar302 survey data targeting file (47.622 stars) with the UCAC-4 catalogue using the VizieR crossmatch service \citep{Ochsenbein1998, Landais2012} and TOPCAT\footnote{The {\bf T}ool for {\bf OP}erating {\bf C}atalogues {\bf A}nd {\bf T}ables \citep{Taylor2005}.} to identify the nearest object. Because APOGEE targets are required to have distances to their nearest 2MASS neighbours $<6''$, this criterion is expected to result in a small number of mismatches. A match was found for 42.514 objects (89\%).
\item We used 2MASS $J, H, K_s$ magnitudes to cross-check identity: $\Delta(J),\Delta(H)$ or $\Delta(K_s)>0.01$ mag could mean confusion with a nearby 2MASS object, or careless targeting. A total of 170 such targets were found in the catalogue, and eliminated.
\item The coordinate separations between the two catalogues have also been checked: stars with separations $d>2''$ are suspicious of having problematical proper motions and have to be inspected visually using the original images. No such stars were found, however.
\item Based on the UCAC-4 input catalogue flags (from the AC2000, AGK2 Bonn, AGK2 Hamburg, Zone astrographic, 
					Black Birch, Lick Astrographic, NPM Lick, SPM Lick catalogues: $A,b,h,Z,B,L,N,S$ flags $<2$; 37.004 objects), and the UCAC-4 Hipparcos flag identifying known double stars from the Hipparcos \citep{Perryman1997, vanLeeuwen2007} and Tycho-2 \citep{Hog2000} catalogues ($H \ne {2,4,5}$; 42.362 objects), a combined \frqq UCAC-4 reliability flag\flqq~was assigned (PMflag $=1$, if the star suffices all the criteria, PMflag $=0$, if not). This flag determines 6.913 of the 42.514 matched objects as problematical -- meaning that for 75 \% of the survey data we have reliable proper motions. The percentage for the HQ sample is even higher (79\%), because the applied S/N cut effectively removes fainter targets, which are less likely to have (reliable) UCAC-4 proper motion measurements.
\end{enumerate}
Figure \ref{pmplot} shows the typical uncertainties of UCAC-4 data for our samples. Our Gold sample, as indicated in this Figure and Fig. \ref{disthisto}, includes only stars with absolute proper motion errors below 4 mas/yr and distance errors below 20\%.

\subsection{Orbital parameters}
\label{orbits}

It has been known for decades \citep[e.g.,][]{Eggen1962,Scheffler1982} that different stellar populations may be characterised by their orbital properties. From the full phase-space information ($\alpha, \delta,d,\mu_{\mathrm{\alpha}},\mu_{\mathrm{\delta}},v_{\mathrm{los}}$), the stellar orbits for our samples were calculated using the Python module {\it galpy}\footnote{http://github.com/jobovy/galpy}, developed and maintained by J. Bovy (IAS Princeton).

We have assumed a standard Milky Way type potential, consisting of an NFW-type dark matter halo \citep*{Navarro1997}, a Miyamoto-Nagai disc \citep{Miyamoto1975} and a Hernquist stellar bulge \citep{Hernquist1990}, in such a way that a flat rotation curve is achieved for the model Galaxy, and that the correct value for the circular velocity at the solar position($R_0=8.0$ kpc) is recovered \citep[$v^\odot_{\mathrm{circ}}=220$ km/s, see e.g.][]{Bovy2012a}. The Solar motion with respect to the local standard of rest have been adopted from \citet{Hogg2005}: $(U,V,W)_{\odot}=(10.1,4.0,6.7)\,\mathrm{km/s}$.
The stellar motions are integrated with the {\it scipy}\footnote{http://www.scipy.org/} routine \texttt{odeint} over at least 2.5 Gyr and 6 revolutions around the Galaxy.

Various tests have shown that the small deviations in the form of the potential do not lead to significant changes in the properties of the computed orbits, and the time step size for the integration has been chosen sufficiently small that stable and smooth orbits are recovered, but not too small to pose an issue for the required computing time resources.\\

From the integrated Galactic orbits, characterizing quantities such as orbital eccentricity $e$, median and mean Galactocentric radii $R_{\mathrm{med}},R_{\mathrm{mean}}$, apo- and pericenter $R_{\mathrm{apo}}, R_{\mathrm{peri}}$, maximum vertical amplitude $z_{\mathrm{max}}$, rotational velocity $v_{\phi}$ as well as the energy $E$, angular momentum $L_z$ and actions. We currently limit our analysis to the widely used parameter set ($e, R_{\mathrm{med}},z_{\mathrm{max}}$).

\subsubsection{Uncertainties}

The most likely orbital parameters and their uncertainties are estimated using a simple Monte Carlo procedure \citep[similar to, e.g.,][]{Gratton2003, Boeche2013} in the following manner. For each star, 100 orbits are computed under variation of the initial conditions (distance modulus, proper motions and radial velocity) according to their estimated errors, where the errors were assumed to follow a Gaussian distribution\footnote{The error distribution for distance (in contrast to the distance modulus) is {\it not} Gaussian!}. From the 100 realisations, the median value of each orbital parameter and its 1$\sigma$ quantiles are used to estimate the most likely value and its uncertainties.

\begin{figure*}[ht]\centering
 \includegraphics[width=7cm]{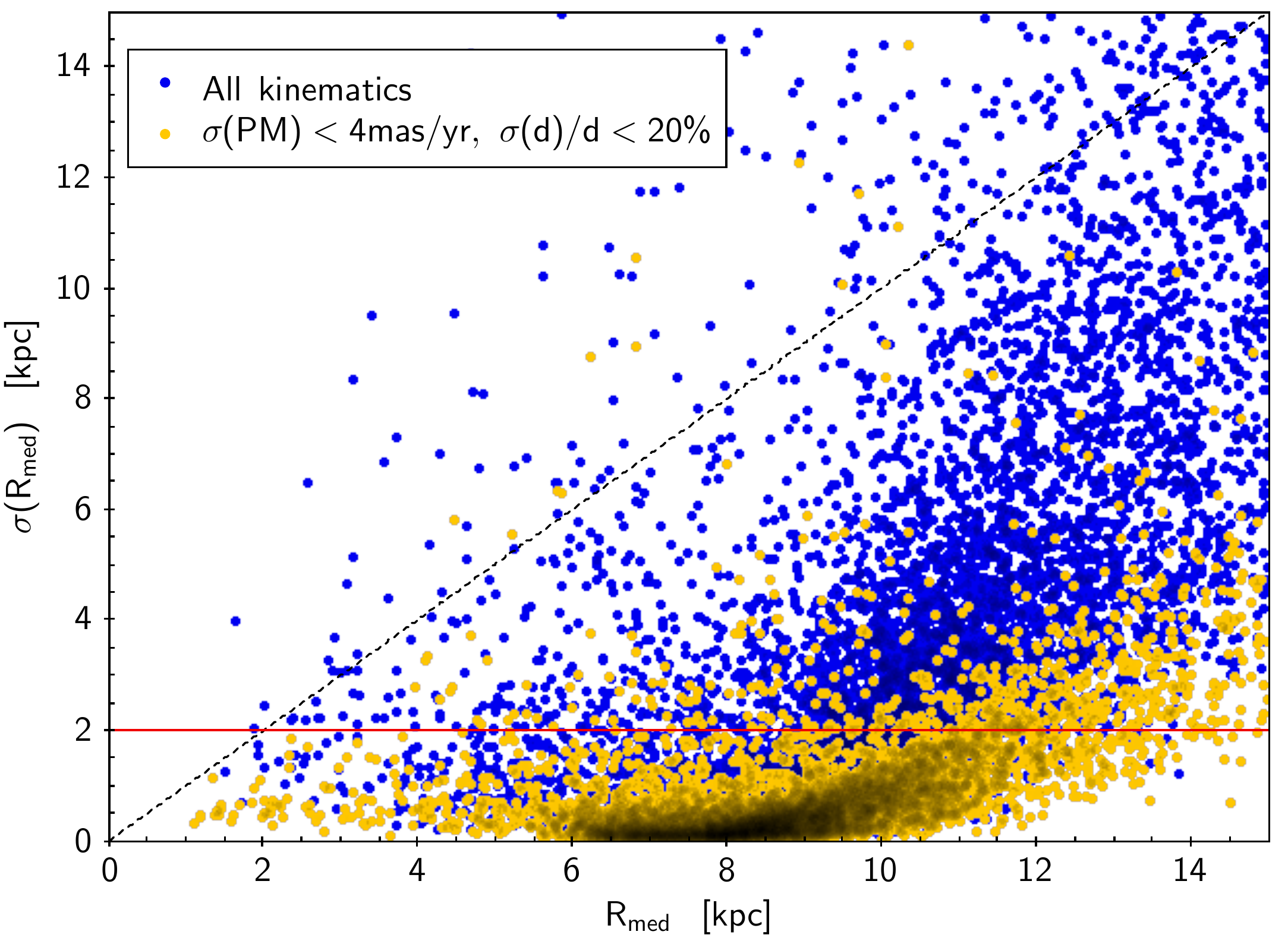}
 \includegraphics[width=8cm]{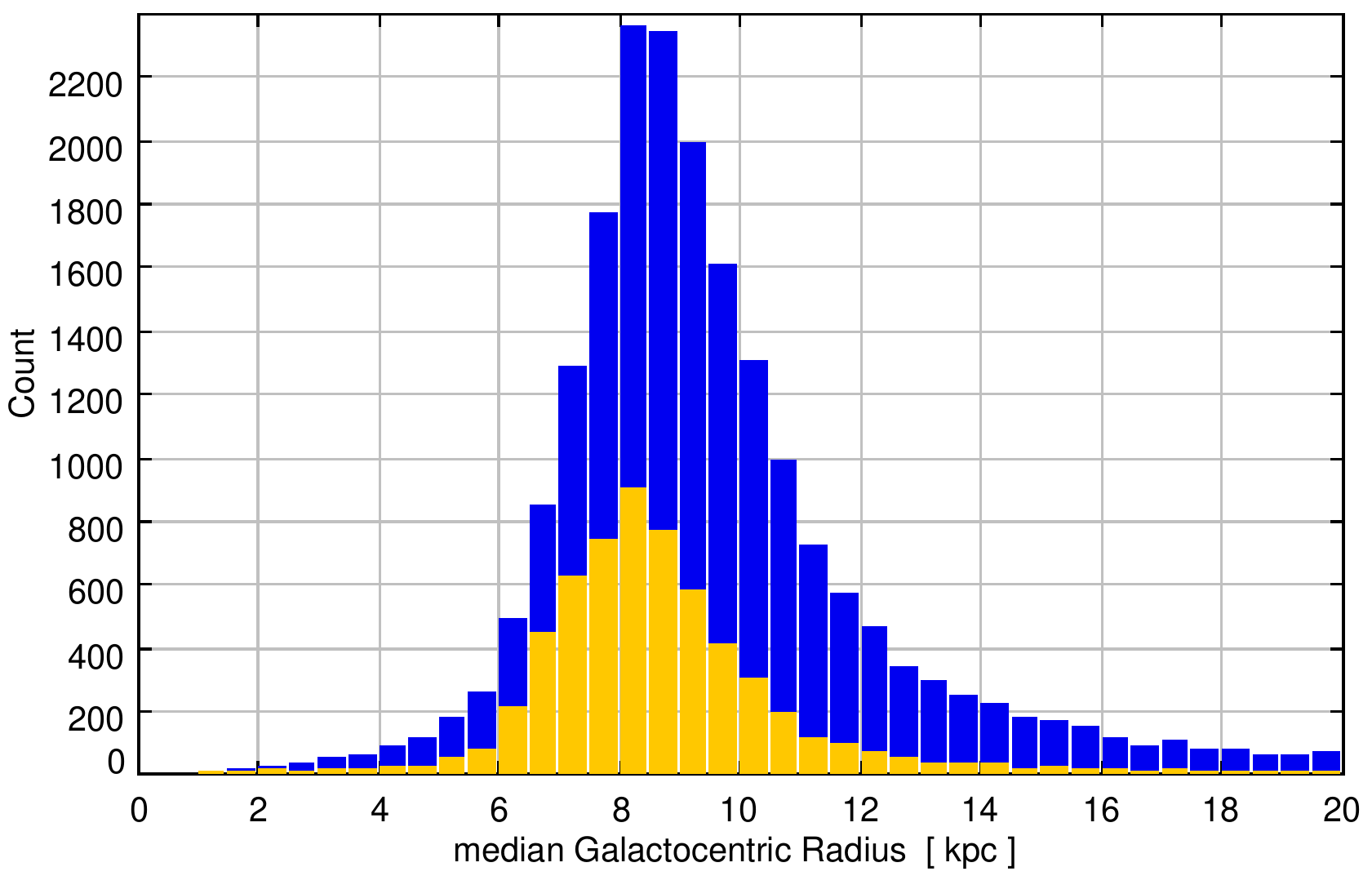}
 \includegraphics[width=7cm]{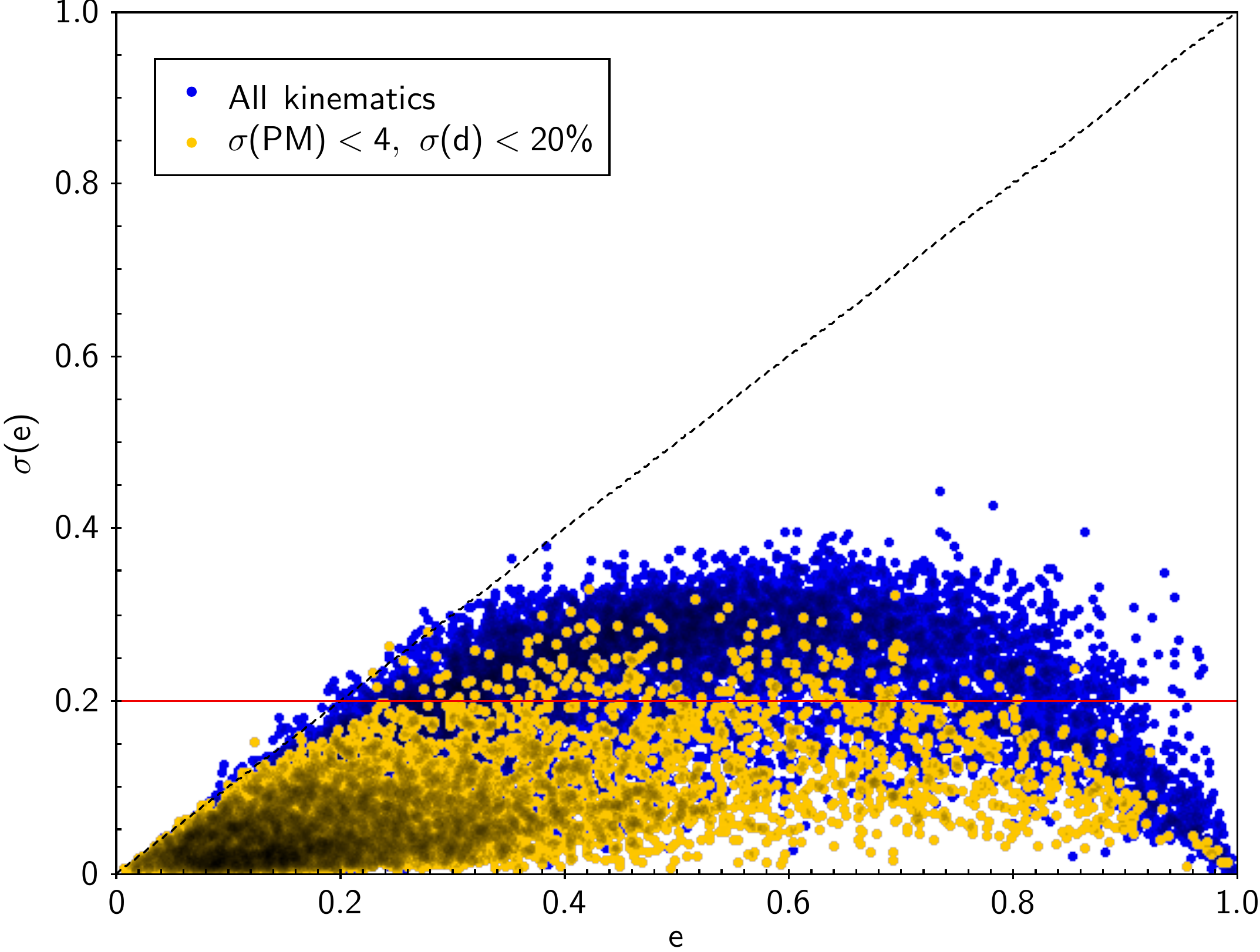}
 \includegraphics[width=8cm]{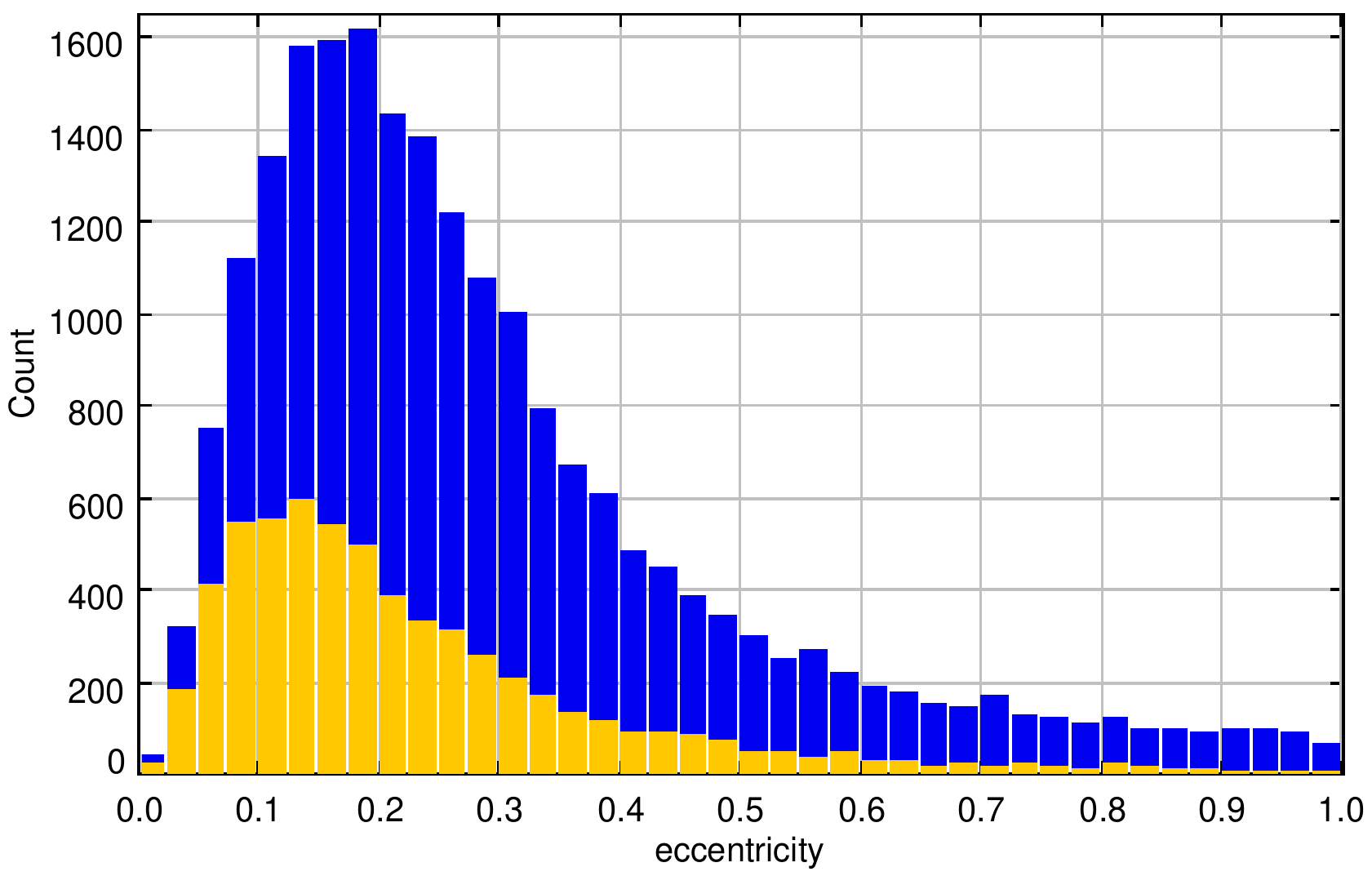}
 \includegraphics[width=7cm]{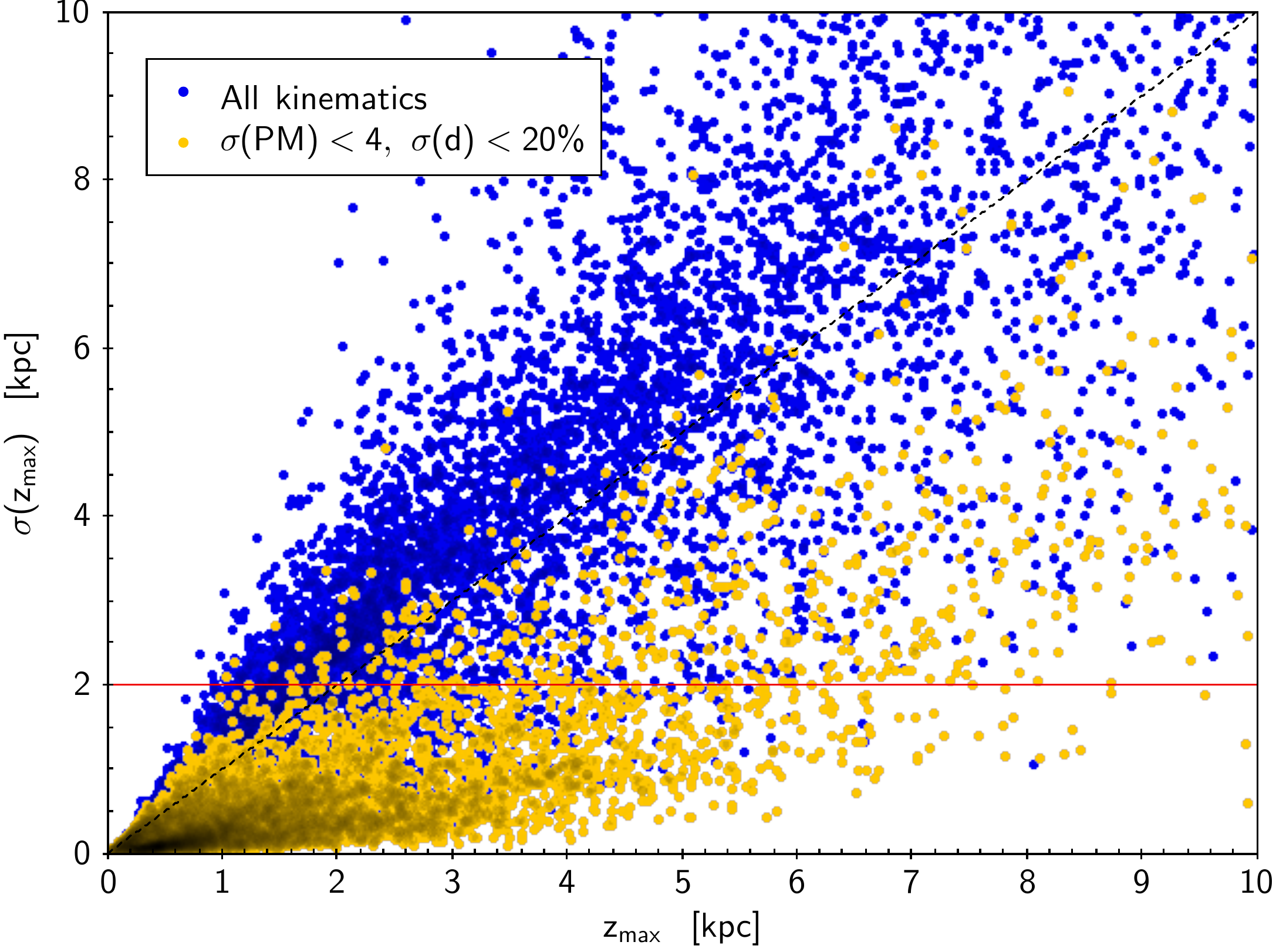}
 \includegraphics[width=8cm]{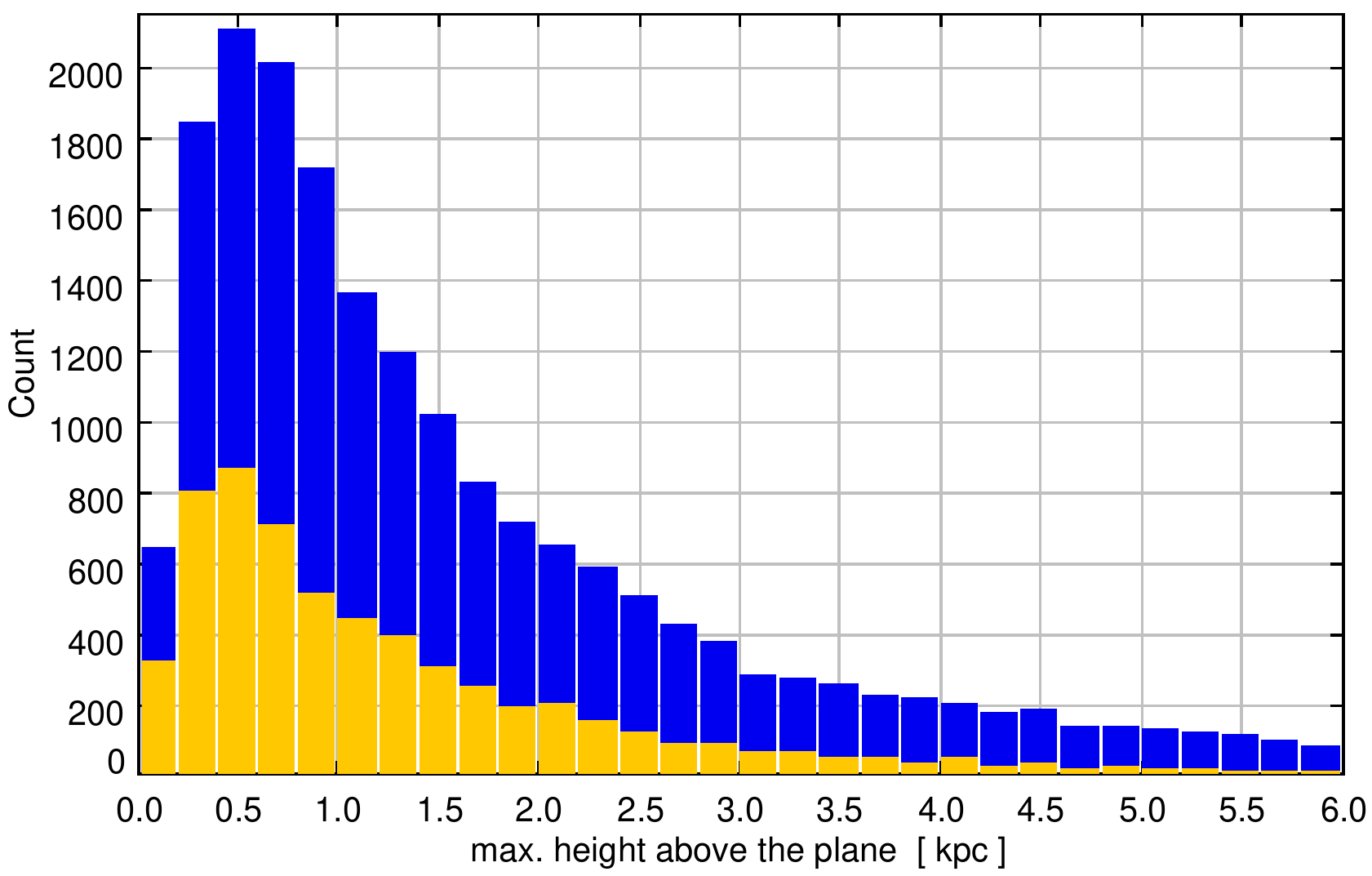}
\caption{Left column: Calculated Monte Carlo uncertainties for the three commonly used orbital parameters median Galactocentric radius $R_{\mathrm{med}}$, eccentricity $e$ and maximum height above the plane $z_{\mathrm{max}}$ (from top to bottom), for both the HQ$^k$ and the Gold samples, as a function of the corresponding median value. Right column: Histograms of the corresponding median orbital parameters, showing the estimated dynamical properties of our samples.}
\label{orbiterrors}
\end{figure*}

The left column of Fig. \ref{orbiterrors} shows the calculated uncertainties for the main parameters Galactocentric radius $R_{\mathrm{med}}$, eccentricity $e$ and maximum height above the plane $z_{\mathrm{max}}$. These plots provide the justification for the introduction of the Gold sample. Whereas the error distributions for the whole HQ$^k$ sample are unsatisfactory (often the orbital parameter uncertainties are far too large to allow for any meaningful interpretation, even in a statistical sense), the additional distance and proper motion quality cuts applied for the Gold sample result in considerably more reliable orbital data for this subset. 

Based on tests like these, the final decisions on the definition of the Gold sample were made, essentially as a trade-off between sample size and high-precision parameters. The decision to cut in the observational parameters $\sigma(\mu)$ and $\sigma(d)$, rather than the actual orbital parameter errors, is motivated by the idea to keep the selection function as simple as possible. In the near future, we are planning to simulate the selection of this sample, which also requires a careful modeling of these observational uncertainties.


\section{Results}\label{results}

We now have the full 6-dimensional phase-space coordinates of the stars in our HQ sample for which proper motions were available (the HQ$^k$ sample), and particularly reliable orbital parameters and distances for a sub-sample of it (the Gold sample). With this information we can perform a first chemodynamical analysis of APOGEE's first-year data. 

Our sample is unique with respect to previous samples used in the literature. Indeed, before APOGEE (and GES), high-resolution spectroscopic surveys of the Galactic disc have been limited to very small Galactic volumes -- 25 pc in the case of Fuhrmann's Solar neighbourhood survey \citep{Fuhrmann1998, Fuhrmann2002, Fuhrmann2004, Fuhrmann2008, Fuhrmann2011}, $\sim100$ pc in the case of \citet{Bensby2003} and \citet{Adibekyan2011}, a small number of pencil beams in the case of \citet{Kordopatis2011a} and \citet{Bensby2011}. Although low- and medium-resolution data from SEGUE, RAVE and ARGOS \citep{Ness2012} have significantly extended the volume covered by spectroscopic stellar surveys, key observables of chemical evolution such as radial metallicity gradients in the disc are still confined to Heliocentric distances of $\sim 2-$3 kpc,\footnote{Perhaps with the exception of \citet{Cheng2012}, who cover a large range of the outer Galactic disc with SEGUE main-sequence turn-off stars. Samples of HII regions, open clusters, cepheids and young stellar objects still cover a larger volume (e.g., \citealt{Cescutti2007}), but in contrast to red giants and long-lived dwarfs, these tracers do not cover the Galaxy uniformly in age. Another possibility is to use planetary nebulae as tracers of chemical evolution \citep{Maciel1994, Maciel1994a}, although their ages and even their abundances are still subject to considerable uncertainties \citep{Stasinska2010}.} and often affected by non-trivial selection biases (e.g., \citealt{Bovy2012b, Schlesinger2012}). Instead, the sample studied here extends over larger volumes, and can be used to complement previous works. Biases are certainly still present, and we will carefully discuss results that might suffer from these biases, although in the case of APOGEE we expect them to be small (a detailed study of the possible biases will be the topic of our next paper).

Here we focus on the results obtained with a local subsample of our main HQ sample (to discuss the Solar vicinity) and then extend our results to a larger portion of the disc (as explained in Section~\ref{sample}).

\subsection{The Solar Vicinity}

\subsubsection{What is a \frqq local sample\flqq?}

\begin{figure}[!h]\centering
	\includegraphics[width=0.49\textwidth]{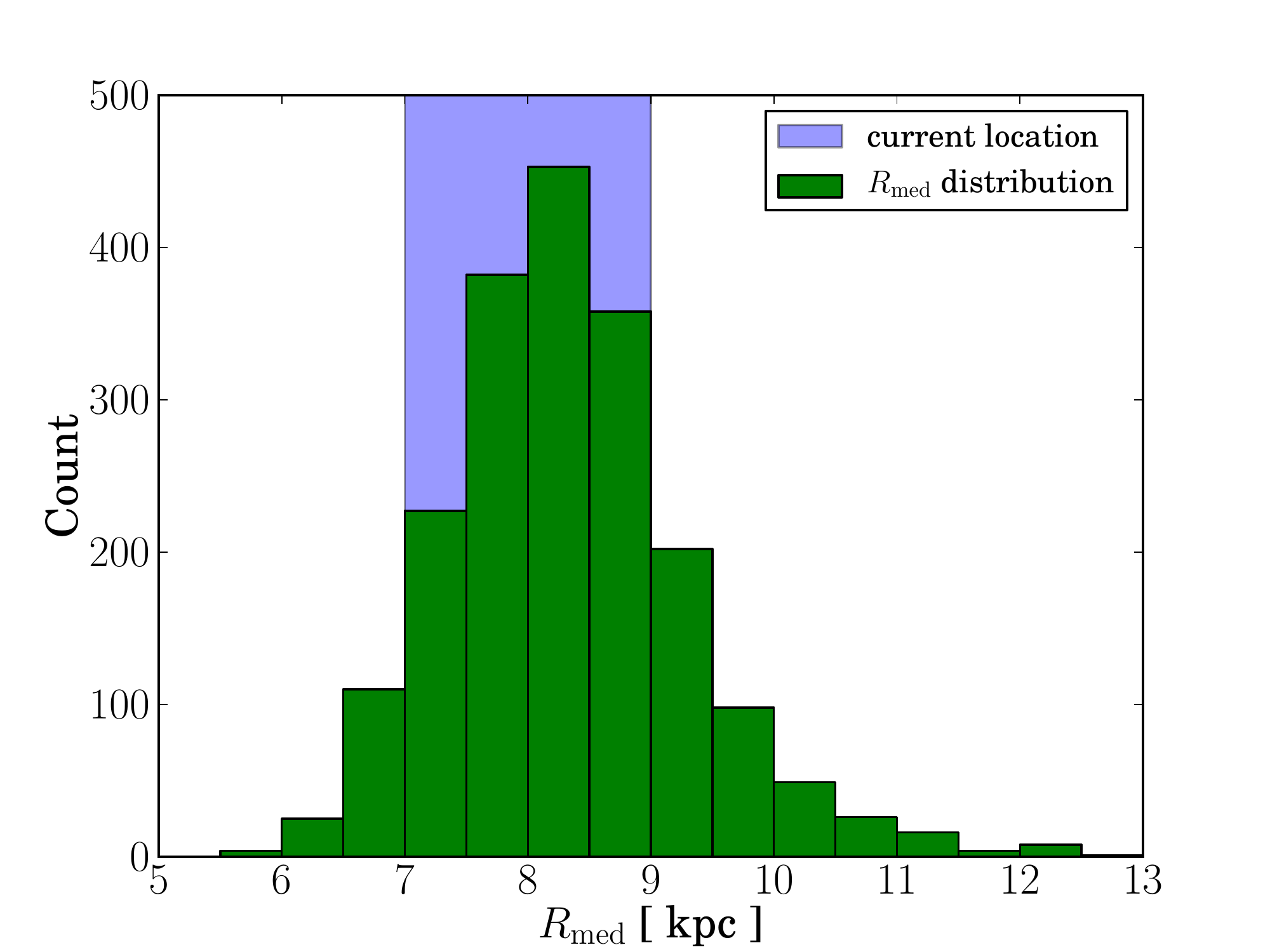}
	\caption{Illustration of the \frqq blurring\flqq~effect: A sizeable fraction of stars observed to be located less than 1 kpc from the Sun's current position (with $7<R_{gal}<9$ kpc; blue-shaded region) move on eccentric inner or outer disc orbits, and are only passing through the Solar neighbourhood.}
	\label{blurring}
\end{figure}

To separate kinematically hot ``visitor stars'' from inner and outer Galactic regions that are passing through the (extended) Solar neighbourhood on highly eccentric orbits, we can make use of the computed orbital parameters. Fig. \ref{blurring} shows a histogram of the median Galactocentric radii of APOGEE HQ$^k$ giants currently located within a 1 kpc sphere around the Sun ($d<1$ kpc).
The Figure illustrates that both stars with guiding radii in the inner as well as the outer disc contribute to the \emph{local}~field population as they are passing by on eccentric orbits. 

Radial migration is radically different from this effect, because it cannot be recognised from the present kinematics of a star if it has migrated from its birthplace. A migrated star on a cool disc orbit can only be distinguished from a locally born star by using chemistry (e.g., \citealt{Freeman2002}), but only if the chemical imprints of their birth places differ by measurable amounts (which are, however, expected to be small). In particular, extreme migrators will then appear in the wings of the \emph{cleaned local}~metallicity distribution, defined as stars with median orbital radii $R_{\mathrm{med}}$ (or similarly, mean Galactocentric radii or angular momenta) close to the Solar value. We will therefore often use $R_{\mathrm{med}}$ instead of the current Galactocentric radius $R$.

\subsubsection{The Metallicity Distribution Function}

The metallicity distribution function (MDF) of the extended Solar neighbourhood is one of the most important and widely used observables to constrain chemical evolution models. 

\begin{figure}[!ht]\centering
	\includegraphics[width=0.48\textwidth]{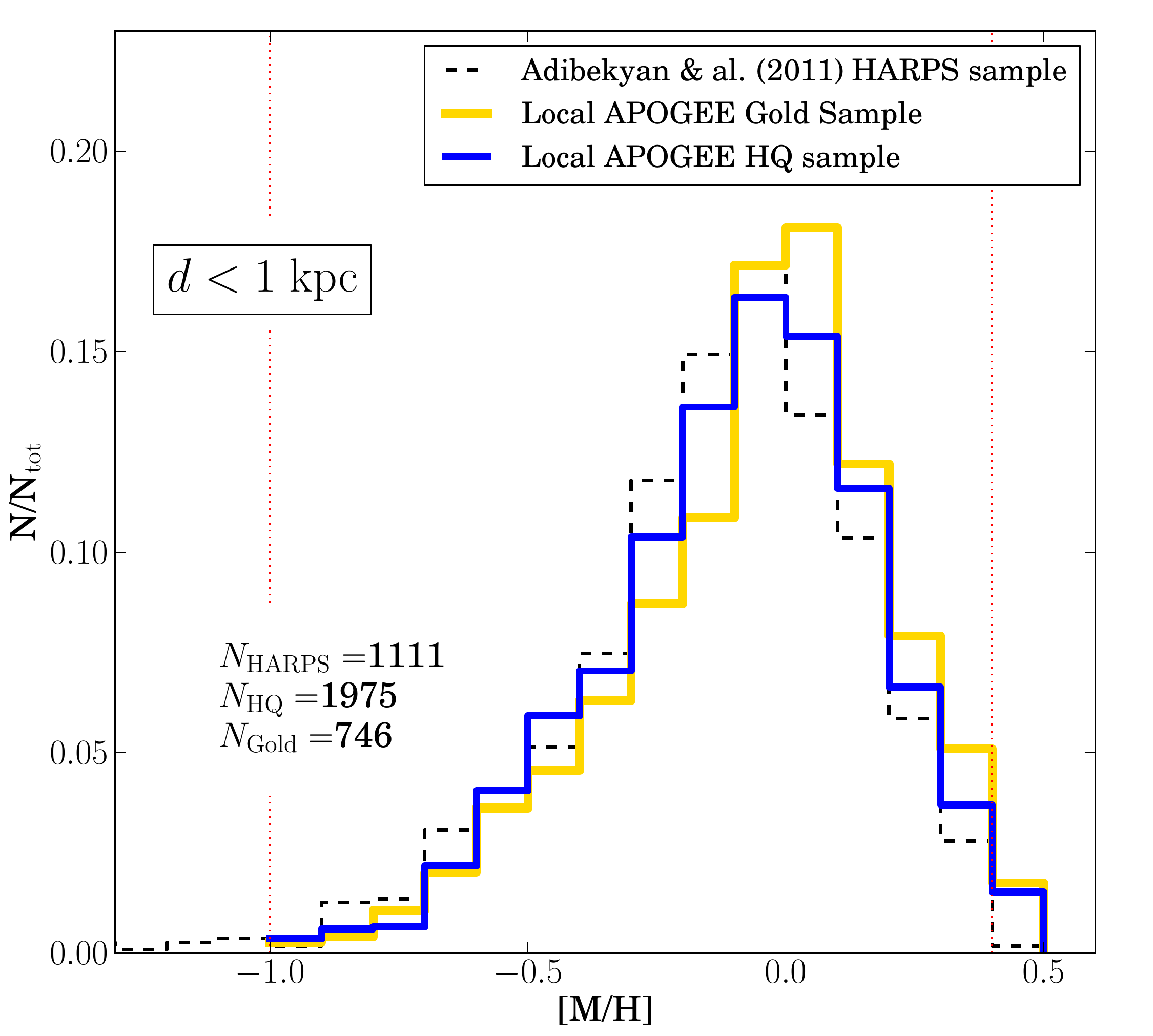}
	\caption{The \frqq local\flqq~metallicity distribution for the HARPS FGK dwarf sample of \citet{Adibekyan2011} and the APOGEE HQ and Gold red giant samples (blue and gold histograms). The red dotted vertical line at [Fe/H]$=-1.0$ indicates our adopted metallicity limit for the HQ sample, while the line at $+0.4$ indicates a possible upper reliabilty limit for ASPCAP metallicities. 
}
	\label{localmetals}
\end{figure}

In Fig. \ref{localmetals}, we compare the local MDF of the high-resolution HARPS FGK dwarf sample of \citet{Adibekyan2011} with the \frqq local\flqq~APOGEE HQ and Gold samples. The overall concordance is quite remarkable: both the HQ and the HARPS sample exhibit a peak at metallicity slightly below the Solar value, and their low-metallicity tails agree well within statistical uncertainties. However, a slight discrepancy is found in the percentage of super-Solar metallicity stars. The MDF for the Gold and the HQ sample differ somewhat in this regime, owing to the fact that the additional selection criteria for the Gold sample introduce some subtle biases. Careful modelling of the selection criteria is expected to resolve these discrepancies.

Here, the reader should be reminded that APOGEE's \emph{local}~ HQ sample still extends to 1000 pc (and has almost no stars with d $<$ 250 pc, see Fig.~\ref{disthisto}), whereas the HARPS sample is confined to $\sim 60$ pc, so that the similarity of the MDFs may not be straightforward to explain. 

\subsubsection{The chemical plane}\label{localchemplane}

Stellar chemical-abundance ratio diagrams can be rich in information about the chemical evolution of a galaxy, as they encode the star-formation and chemical-enrichment history of the ISM at the time of a star's birth. Particularly widely used is the [$\alpha$/Fe] vs. [Fe/H] diagram, because iron and the $\alpha$-elements are produced and returned to the ISM on different timescales.\footnote{For example, the $\alpha$-element oxygen is mainly produced by type II SNe, i.e., in short-lived massive stars, whereas type Ia SNe produce predominantly more iron \citep{Matteucci1990}.} Comparing these two abundance ratios for a statistically significant sample constrains the formation history of different Galactic components, the shape of the IMF, stellar yields, the efficiency of dynamical mixing and other parameters (see, e.g., \citealt{Pagel2009, Matteucci2012a}). 

The usefulness of abundance-ratio diagrams for Galactic Archaeology purposes has been recently challenged by the fact that stellar radial migration can mix stars born at different Galactocentric radii \citep{Sellwood2002,Rovskar2008,Schonrich2009}. The quantification of the effects of radial stellar migration and its causes is thus of crucial importance (see \citealt{Minchev2013} for a discussion). It is also known that pure chemical evolution models fail to explain the existence of local \emph{super-metal-rich (SMR) stars}\footnote{Stars whose atmospheric metal abundance is significantly higher than the local interstellar medium, first found by \citet{Grenon1972}.} (see, e.g., \citealt{Chiappini2009} and references therein), and that dynamical mixing mechanisms may affect stellar orbits by heating and/or radial migration.\footnote{Or, in the terminology of \citet{Schonrich2009}: \frqq blurring\flqq~and \frqq churning\flqq.} Whereas (radial) heating mainly changes the eccentricity of a star and does not significantly alter its guiding radius, radial migration shifts the angular momentum and thus the guiding radius of a stellar orbit, while it may remain on a circular orbit. In fact, radial migration has been shown to preferentially affect stars on kinematically cool orbits \citep{Minchev2012}. Heating can be caused by, e.g., scattering off of giant molecular clouds \citep{Spitzer1951, Mihalas1981}, by interaction with the bar and spiral arms \citep{Minchev2010, Minchev2006}, or by merging satellites \citep{Quinn1993, Villalobos2008}. Similarly, several scenarios have been proposed to trigger radial migration, although their relative importance is still under discussion. 

The consensus view is that even in the presence of radial migration the chemical diagrams are still extremely useful, and sometimes abundance ratios can be less prone to migration effects than absolute abundances, as shown in \citet{Minchev2013}. In the following we discuss the abundance plots obtained with our samples, as this is the first time we can study the chemical plane close to the disc, in a region extending far beyond the Solar vicinity, and with large statistics.\\

{\it Comparison with other local high-resolution samples}\\

\begin{figure*}[ht]\centering
	\includegraphics[width=0.47\textwidth]{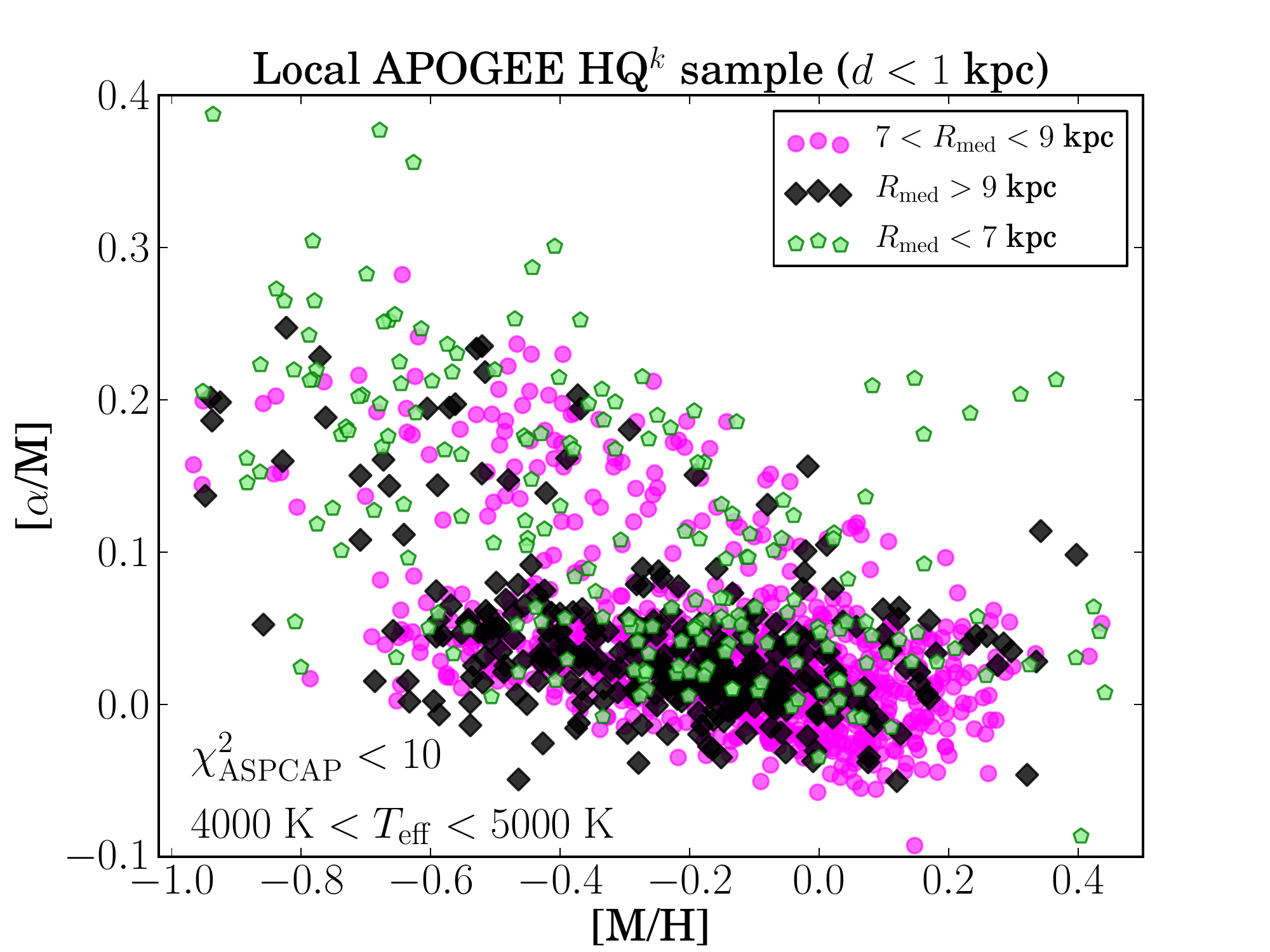}
	\includegraphics[width=0.49\textwidth]{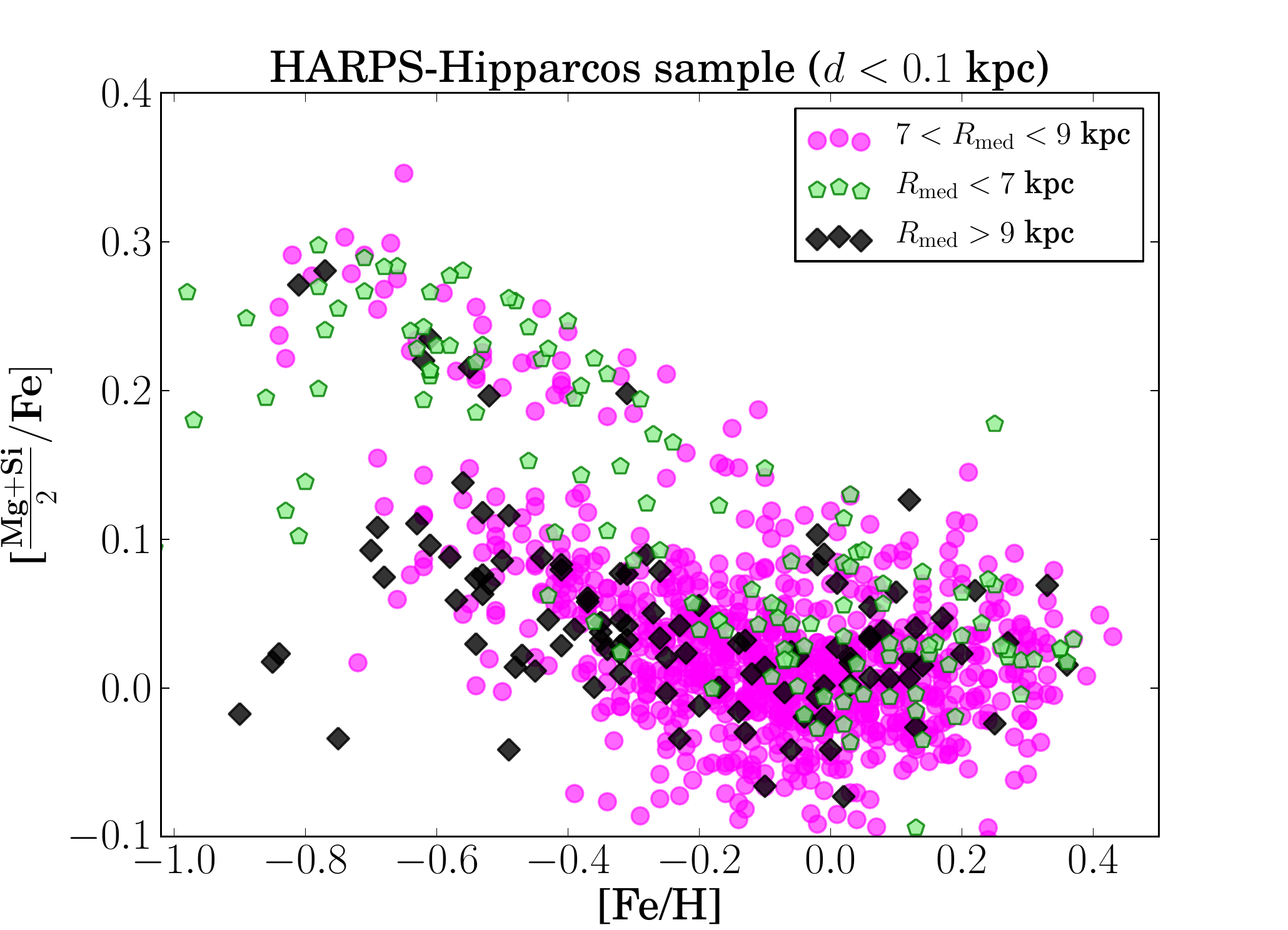}
	\caption{The chemical plane using \frqq local\flqq~APOGEE HQ$^k$ stars ($d<1$ kpc, left) compared to the {\it very} local high-resolution HARPS FGK dwarf sample of \citet{Adibekyan2011}. Orbits for the HARPS sample were computed in the same manner as for the APOGEE stars. Although the sampled volumes are quite different in size, the general resemblance of both plots is reassuring. Both plots exhibit the intriguing \frqq gap\flqq~between high- and low-$\alpha$ population, and in both samples it is not straightfroward to explain by selection effects.}
	\label{verylocalchemplane}
\end{figure*}

Local high-resolution studies have found a significant gap in the [$\alpha$/Fe] vs. [Fe/H] chemical-abundance plane, whose origin is still under discussion. The high-resolution volume-complete FOCES sample obtained by K. Fuhrmann (e.g., \citealt{Fuhrmann2011}) seems to imply that this gap corresponds to a star formation hiatus as advocated by the Two-Infall model \citep{Chiappini1997}. Similar analyses carried out recently by \citet{Haywood2013} and \citet{Adibekyan2013}, using the HARPS sample of \citet{Adibekyan2011}, lead to the same conclusion, identifying the two regimes in [$\alpha$/Fe] as chemical signatures of the different formation epochs of thin and thick disc. The recent study by \citet{Bensby2013a}, analyzing high-resolution spectra of more than 700 solar-neighbourhood dwarf stars, also points into this direction. The authors find that the different abundance trends for thin and thick disc, and hence the gap, are subject to less scatter when discarding more uncertain chemical abundance data. APOGEE appears to confirm the reality of the gap, displaying a similar gap in the [$\alpha$/M] vs. [M/H] diagram (see Fig. \ref{verylocalchemplane}).  

Figure \ref{verylocalchemplane} displays the APOGEE chemical abundance plane ([$\alpha$/M] vs. [M/H]) for stars with $d<1$ kpc, and compares this picture with the high-resolution ($R\sim40,000$) high-$S/N$ HARPS sample of \citet{Adibekyan2011}, using their individual abundances for Mg, Si and Fe.\footnote{Although APOGEE in principle tracks all $\alpha$-elements, it is expected to be most sensitive to atomic lines like Mg~I and Si~I in the temperature regime corresponding to the lower giant branch, and thus to smaller distances.} The similarity of the plots may serve as an initial validation of the ASPCAP pipeline for [M/H] and [$\alpha$/M].
In both the APOGEE and the HARPS sample there is no a-priori reason to expect the observed gap to be caused by selection biases, because unlike in SEGUE, RAVE or the high-resolution studies of \citet{Bensby2003} and \citet{Ramirez2013}, the thick disc was not targeted preferentially by these surveys. However, we cannot ultimately confirm nor dismiss this statement until the selection function for APOGEE is properly accounted for (as will be shown in a forthcoming paper).

In Fig.~\ref{verylocalchemplane} (left panel) the APOGEE stars are labelled according to three groups of $R_{\mathrm{med}}$ (again showing that the \emph{local} sample contains stars on eccentric orbits whose most probable birth radii, apart from radial migration, are outside/inside the Solar circle~$7<R_{\mathrm{med}}<9$ kpc). The high [$\alpha$/M] cloud is more populated by stars coming from the inner regions (see discussion on this particular point in Section~\ref{out}). On the other hand, the low [$\alpha$/M] cloud extends down to [M/H]$\sim -$0.8, independently of the studied $R_{\mathrm{med}}$ bin, in an almost flat manner. This behaviour is different from what is seen in the \emph{thin-disc-like} stars from HARPS where the low [$\alpha$/Fe] cloud shows an increase of [$\alpha$/Fe] towards low metallicities. This difference, most probably, arises from the different biases present in the HARPS and APOGEE sample used here (as both samples have used different colour and temperature cuts). Another contributing factor is that the HARPS data were analysed using an equivalent-width pipeline (ARES; \citealt{Sousa2007}), whereas ASPCAP uses a cross-correlation technique.\\

{\it The kinematical properties of a chemically-divided disc}\\

\begin{figure}[!h]\centering
	\includegraphics[width=0.49\textwidth]{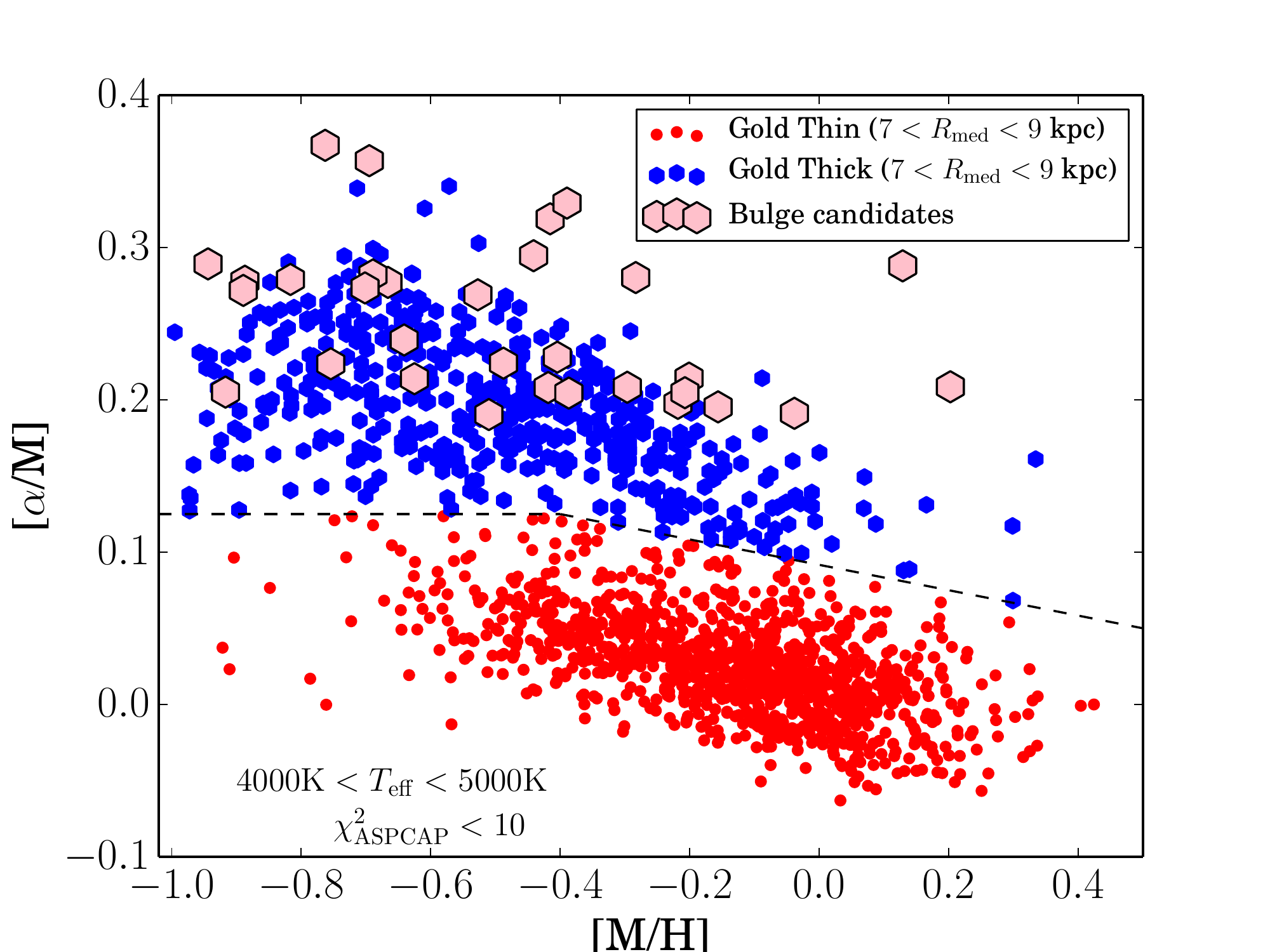}
	\caption{The APOGEE chemical plane at the \frqq Solar circle\flqq~($7<R_{\mathrm{med}}<9$ kpc) for the Gold sample. To avoid spurious [$\alpha$/M] data, we only show stars satisfying $\chi^2<10$ and $4000<T_{\mathrm{eff}}<5000$ K. A possible (purely chemical) definition of thin and thick disc, consistent with, e.g., \citet{Lee2011}, is indicated by the division into the red and blue points and the dashed line. For comparison, we also plot kinematically selected candidate bulge stars (pink hexagons).}
	\label{thinthickbulge}
\end{figure}

\begin{figure*}[ht!]\centering
	\includegraphics[width=0.48\textwidth]{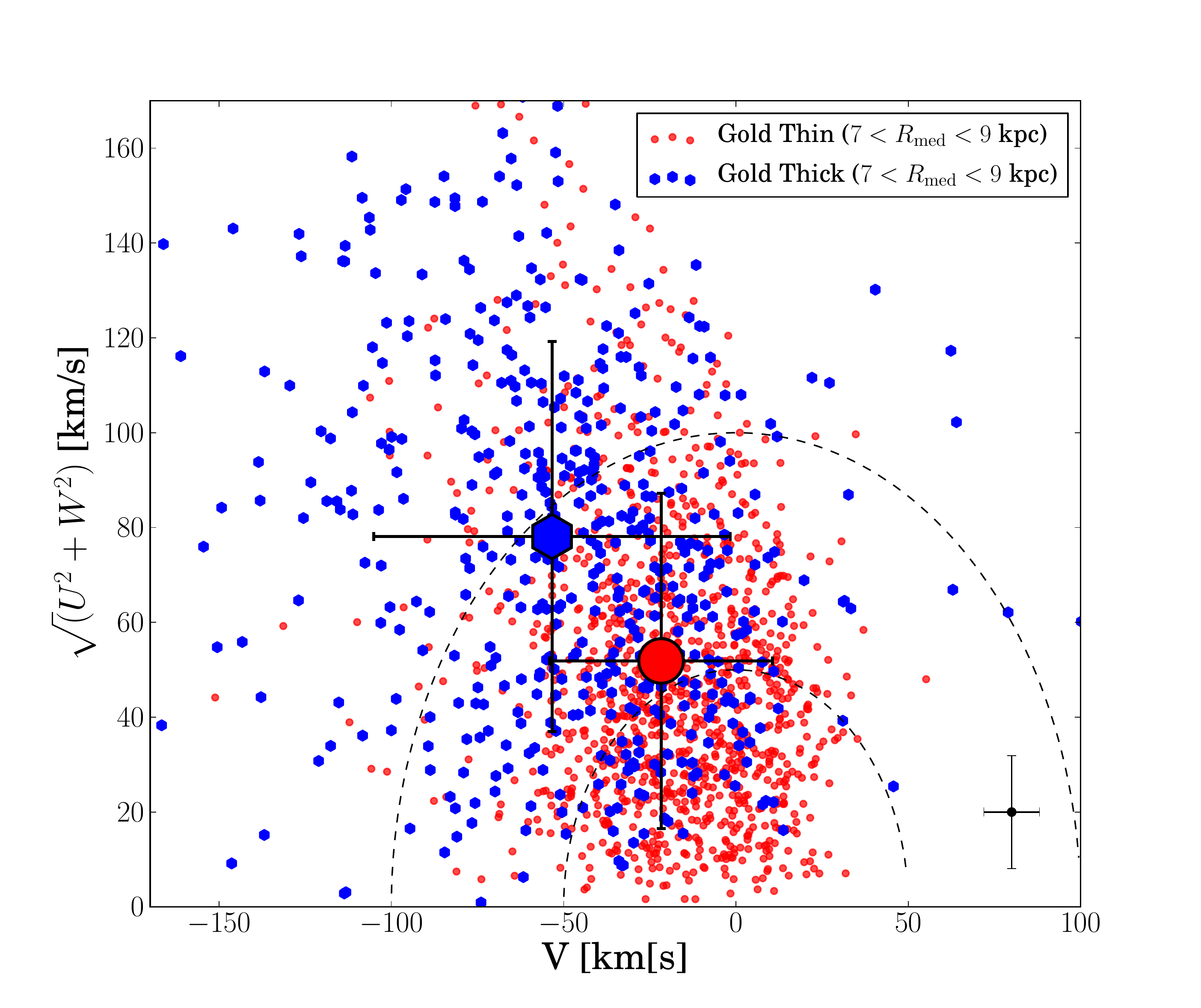}
	\includegraphics[width=0.48\textwidth]{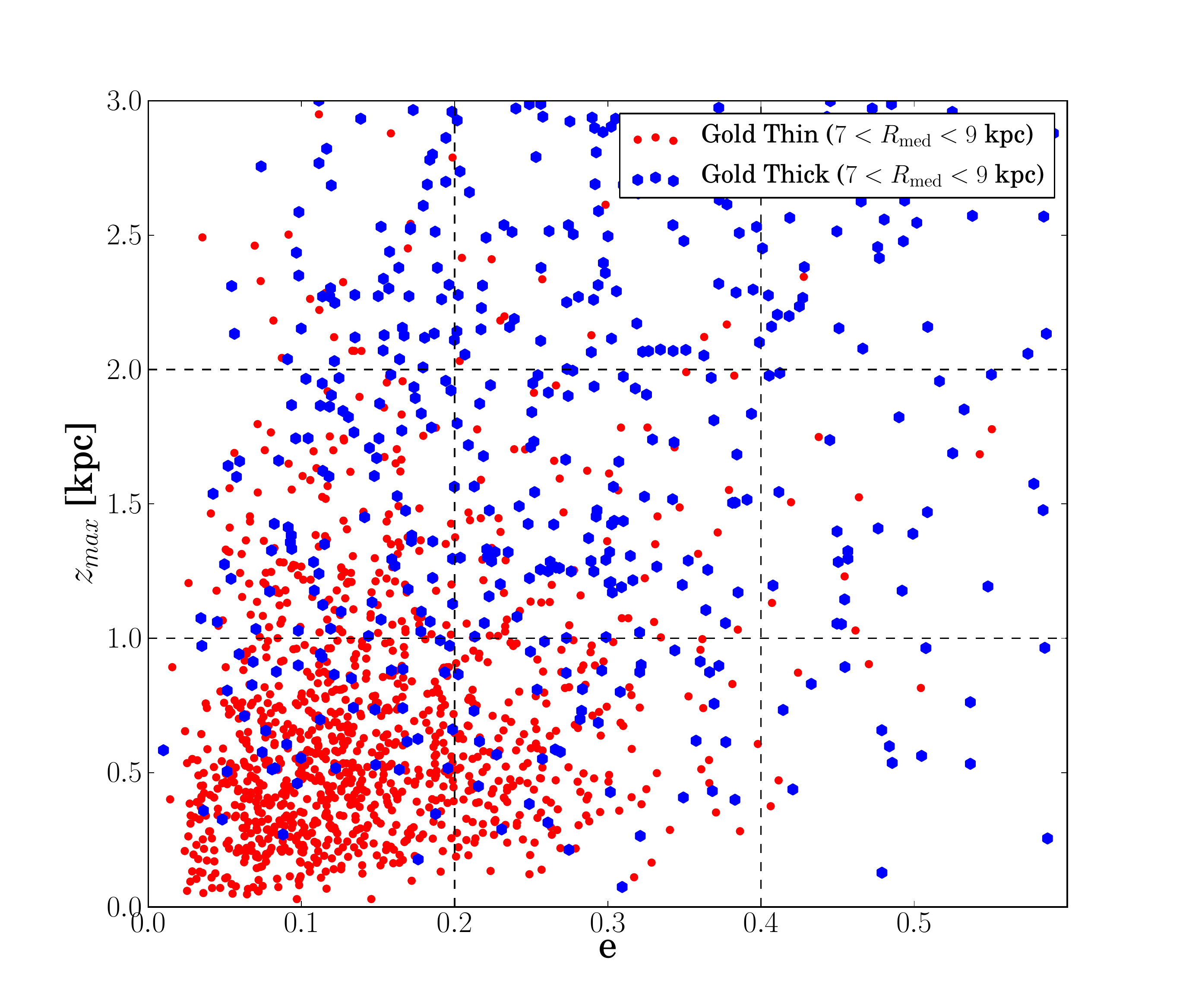}
	\caption{
Kinematical properties of the local chemically defined \frqq thick\flqq~and \frqq thin\flqq~discs (cf. Figure \ref{thinthickbulge}). Middle: The Toomre diagram ($\sqrt{U^2+W^2}$ vs. $V$, see \citealt{Feltzing2003}) of Galactocentric velocities (with respect to the Local Standard of Rest). The dashed curves indicate constant space motion. We can confirm the rotational lag of the thick disc found by numerous other studies, but find the value of this lag to be very much dependent on the exact separation between thick and thin disc in the chemical plane ($\Delta V\approx20-70$ km/s). Right: The $e-z_{\mathrm{max}}$ plane used by \citet{Boeche2013} to separate stellar populations into orbital families, indicated by the dashed lines. We will use this kinematical division in future analyses to compare with their findings.}
	\label{thinthickbulge_kinematics}
\end{figure*}

It is tempting to interpret the two \frqq clouds\flqq~in the [$\alpha$/M] vs. [M/H] diagram as two distinct stellar populations (i.e., {\it chemical} thin and thick discs\footnote{Another possibility is to separate populations on the basis of kinematics (e.g., \citealt{Bensby2003})}). Here, we will briefly explore this approach, and divide the chemical plane in a similar way to \citet{Lee2011} and \citet{Adibekyan2011}, as illustrated in Fig. \ref{thinthickbulge}. For the moment, we focus only on stars whose median Galactocentric radius (as determined by the orbit integration routine) is near the \frqq Solar circle\flqq~($7<R_{\mathrm{med}}<9$ kpc). It is now also interesting to see where the two populations defined above are located in orbital-parameter space: In Fig. \ref{thinthickbulge_kinematics}, we show how our chemically-divided local sample distributes kinematically (see caption for details). \\

A few characteristics can be noticed immediately from Figures 8--10:

\begin{itemize}
\item The local sample spans a wide range in metallicities, from below [M/H]$=-1$ to above +0.3.\footnote{Although our sample is currently restricted to [M/H]$>-1.0$.}
\item When dividing the sample according to the [$\alpha$/Fe] cut shown in Fig.~\ref{thinthickbulge}, we find that the peak of the metallicity distribution of the chemical \frqq thin disc\flqq~is at [M/H]$\sim-0.1$, and that of the thick disc is at [M/H] $\sim-0.5$, in concordance with the Geneva-Copenhagen survey and high-resolution spectroscopy literature(e.g., \citealt{Nordstrom2004, Holmberg2007, Rocha-Pinto1996, Kotoneva2002}).
\item The thin disc's spread in [$\alpha$/M] for a given metallicity is comparable to the quoted observational scatter ($\sim0.08$ dex). This result implies that, provided the gap is \frqq real\flqq, random uncertainties can in principle account for the [$\alpha$/M] scatter in the thin disc. While this result at first sight leaves little room for radial migration, \citet{Minchev2013} have shown that the presence of strong radial migration does not necessarily imply a large scatter in the abundance ratios.
\item The $[\alpha$/M] ratio in the thick disc increases as the metallicity decreases, reaching a plateau of [$\alpha$/M]$\sim+0.2$ at [M/H]$\sim-0.6$. Also, the scatter in $[\alpha$/M] increases with decreasing [M/H].
\end{itemize}

In a forthcoming paper, we will study \frqq orbital families\flqq~(groups of stars with similar orbital properties, see, e.g., right panel of Figure \ref{thinthickbulge_kinematics}) to be able compare with the RAVE red giant sample of \citet{Boeche2013}. Similar to their results, we find orbital parameter distributions like the Toomre diagram \citep{Feltzing2003} of chemically-defined thin and thick disc to change considerably with slight variations of the cut in the [$\alpha$/M] vs. [M/H] plane (see caption of Fig. \ref{thinthickbulge_kinematics}). We therefore plan to study \frqq mono-abundance populations\flqq ~\citep{Bovy2012b} in the near future, to investigate if, instead of a rigid dichotomy in the kinematics, a smooth transition from thick to thin disc exists, and to compare these findings with results from RAVE and SEGUE.

\subsection{Outside the Solar vicinity}
\label{out}

\subsubsection{The locus of bulge stars selected only by kinematics/position}

Although APOGEE's first-year data contain a rather small number of HQ$^k$ stars in the Galactic bulge, we also show where purely kinematically-selected HQ$^k$ bulge star candidates (i.e., stars with $R_{\mathrm{med}}<4$ kpc, $z_{\mathrm{max}}<3$ kpc) fall in Figure \ref{thinthickbulge}. The bulge candidates (which could also be members of the inner disc) seem to display yet a different chemical-abundance pattern from the thick disc. From our small sample, we tentatively suggest that they are generally more $\alpha$-enhanced than the local thick disc at a fixed metallicity, and that the so-called \frqq knee\flqq~in the chemical-abundance plane, corresponding to the metallicity value of the ISM at the time of the bulk contribution of SNe type Ia, might be located at a higher metallicity. 
These preliminary results, while in agreement with earlier studies by, e.g., \citet{Zoccali2006}, \citet{Fulbright2007} and \citet{Lecureur2007}, are somewhat different from the more recent homogeneous abundance analyses of \citet{Melendez2008}, \citet{Alves-Brito2010}\footnote{Indeed, \citet{Alves-Brito2010} re-analysed the same equivalent widths of \citet{Fulbright2007} and found Solar $\alpha$-element abundances instead of elevated [$\alpha$/Fe].} and \citet{Gonzalez2011} who find a similar abundance pattern for bulge and thick disc giants for [Fe/H]$<-0.2$, and need to be confirmed or dismissed with future APOGEE data for more stars. Similar to our findings, the recent study of microlensed bulge dwarfs by \citet{Bensby2013} suggests that the bulge stars are slightly more $\alpha$-enhanced than the local thick disc. If true, these observations would imply either a) a different IMF for the bulge and the thick disc (e.g., \citealt{Ballero2007}), and/or b) a different origin for the bulge and the local thick disc, where the bulge formed in a shorter timescale than the thick disc.

\subsubsection{The chemical plane at three different radial bins}

It was first shown by the high-resolution observations of \citet{Edvardsson1993} that disc stars at different Galactocentric guiding radii differ also in their chemical abundance patterns. With APOGEE, we are now able to systematically scan the Galaxy to large distances, eventually creating a chemo-dynamical map. In this section we present a few useful examples. 

\begin{figure}[!h]\centering
	\includegraphics[width=0.5\textwidth]{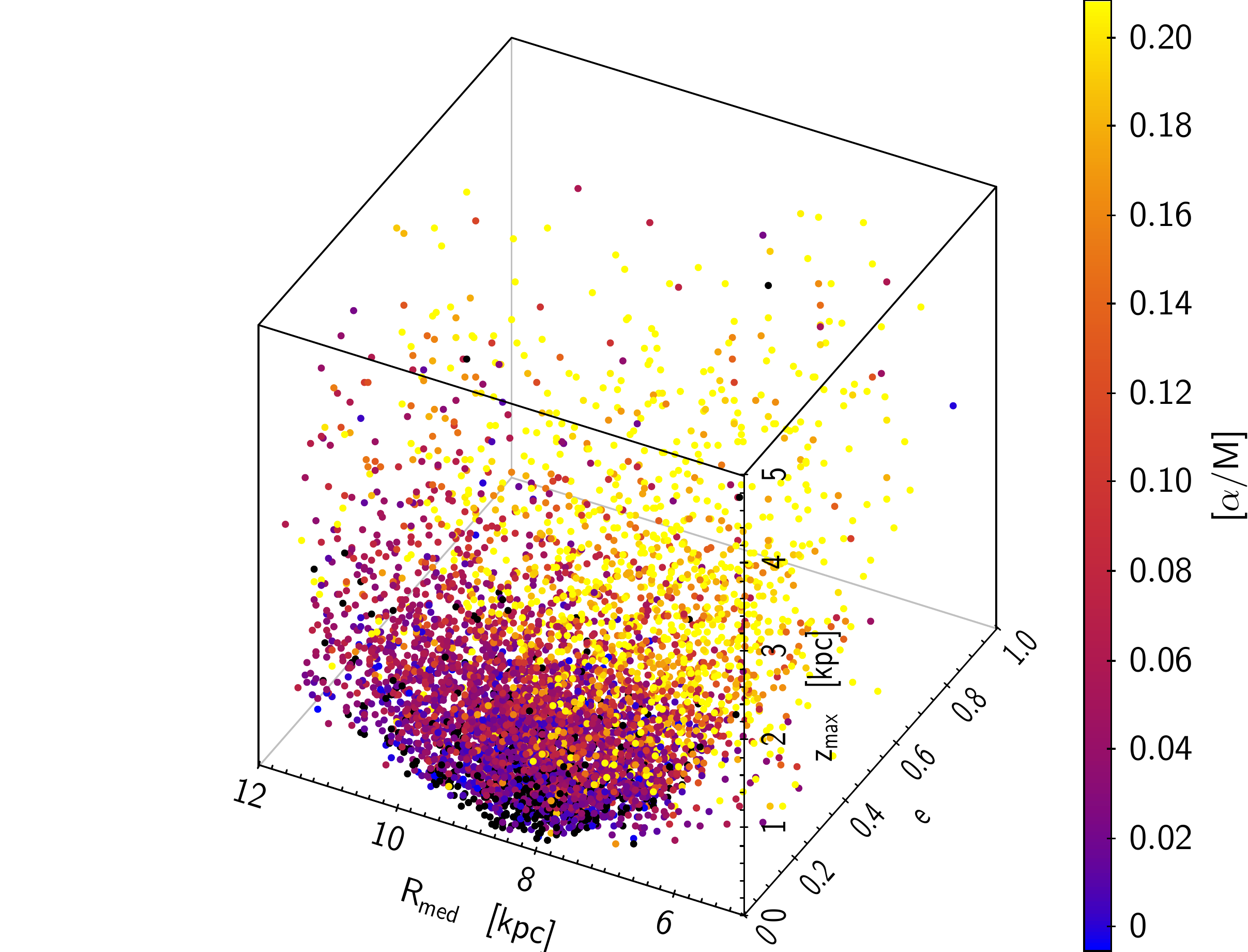}
	\caption{Distribution of the Gold sample in orbital-parameter space ($e, R_{\mathrm{med}}, z_{\mathrm{max}}$), colour-coded by $\alpha$-element abundance. As expected, $\alpha$-enhanced stars are on vertically hotter and more eccentric orbits. Also, as previously suggested by \citet{Bensby2011}, the density of $\alpha$-enhanced stars (the \frqq chemical thick disc\flqq) rapidly decreases with Galactocentric orbital radius. This latter result does not appear to depend critically on selection biases.}
	\label{erzcube}
\end{figure}

\begin{figure*}[!h]\centering
	\includegraphics[width=0.4\textwidth]{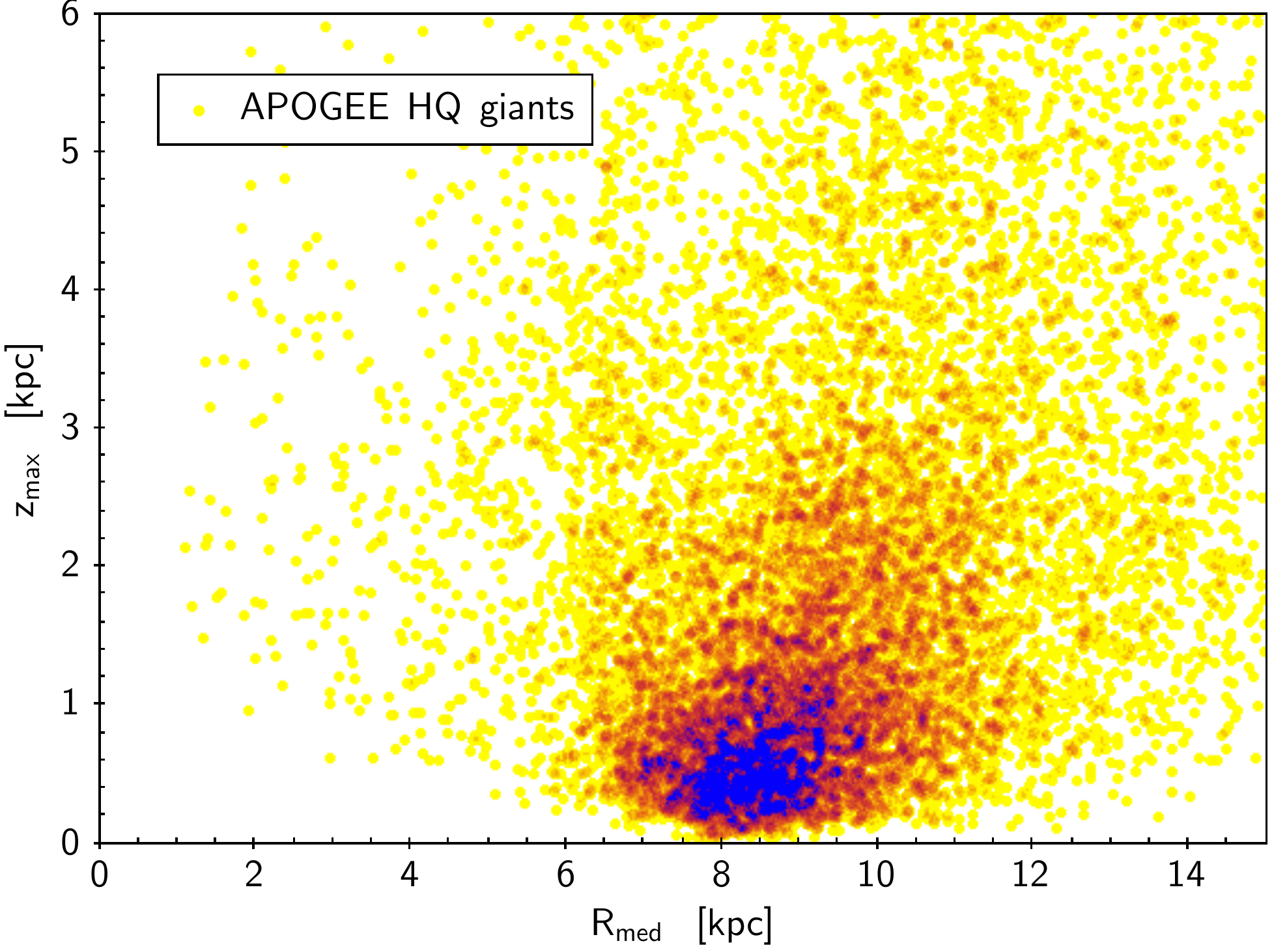}
	\includegraphics[width=0.4\textwidth]{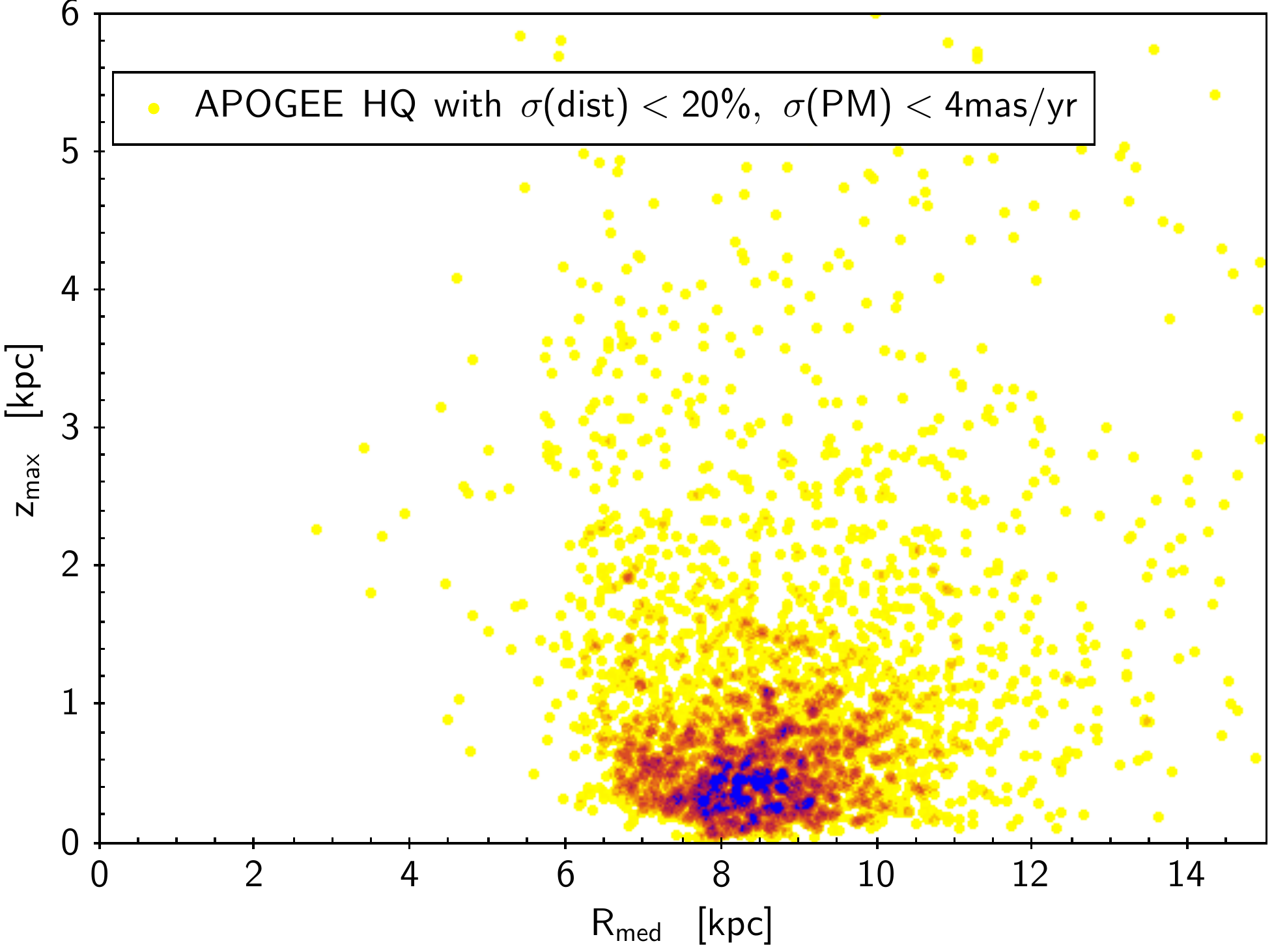}
	\caption{Density distribution of the HQ (left) and the Gold sample (right) in the $R_{\mathrm{med}}- z_{\mathrm{max}}$ plane (light colours denote low density). There is a striking deficiency of Gold-sample stars with inner-Galaxy kinematics ($R_{\mathrm{med}}<6$ kpc).}
	\label{rzdistributions}
\end{figure*}

Figs. \ref{erzcube} and \ref{rzdistributions} show the distribution of our samples in orbital-parameter space ($e, R_{\mathrm{med}}, z_{\mathrm{max}}$). In particular, Fig. \ref{erzcube} nicely displays how stellar kinematics correlate with chemical properties. In the following, we will use projections of this cube to extract and highlight some of these relationships, focussing mainly on the $R_{\mathrm{med}}- z_{\mathrm{max}}$ and the $e- z_{\mathrm{max}}$ planes.

One major drawback of the current Gold sample constructed from Year-1 APOGEE data is its lack of stars in the inner parts of the Galaxy (Fig. \ref{rzdistributions}).\footnote{This is expected to improve slightly when Year-2 data are added, and especially with the additional APOGEE dark-time observations of the inner Galaxy in spring 2014.} We will therefore often use the HQ sample to accomplish a statistically robust sample, separating stars into wide $R_{\mathrm{med}}$ bins. At this point, the reader is reminded that the uncertainties in the orbital parameters can be quite sizeable (see Fig. \ref{orbiterrors}), and that orbital parameters of the HQ sample should generally be used in wide bins, and only for statistical purposes.

\begin{figure*}[!h]\centering
	\includegraphics[width=0.8\textwidth]{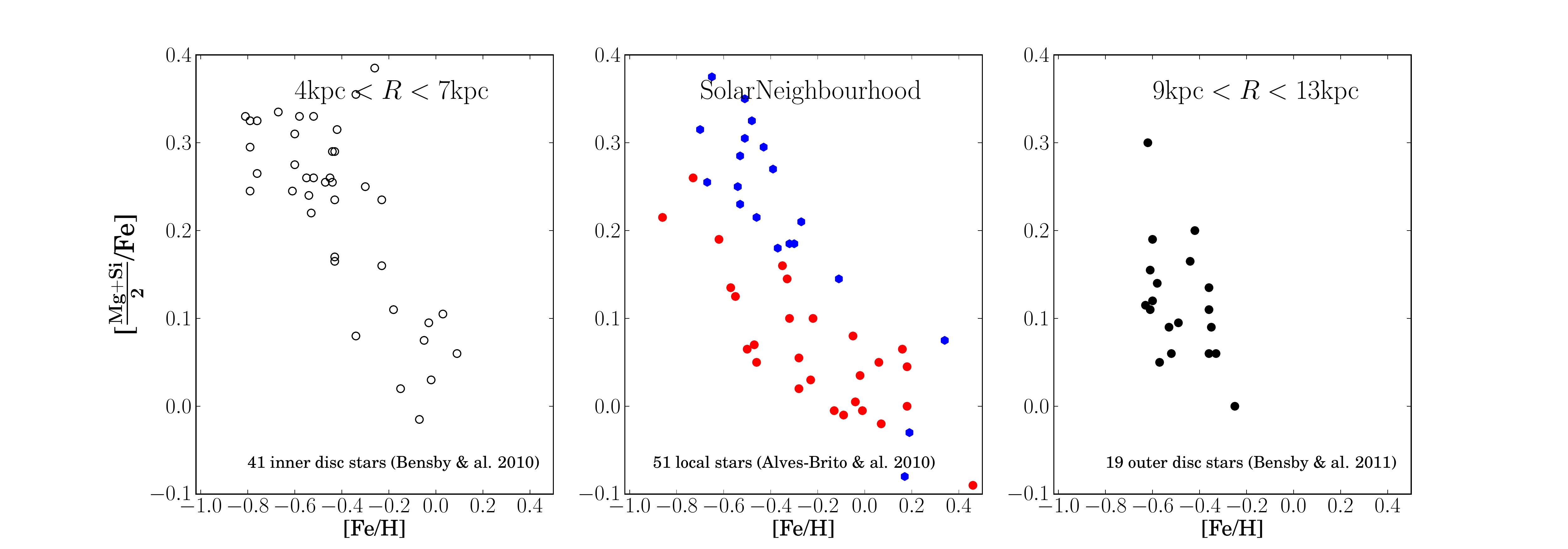}\\
	\includegraphics[width=0.8\textwidth]{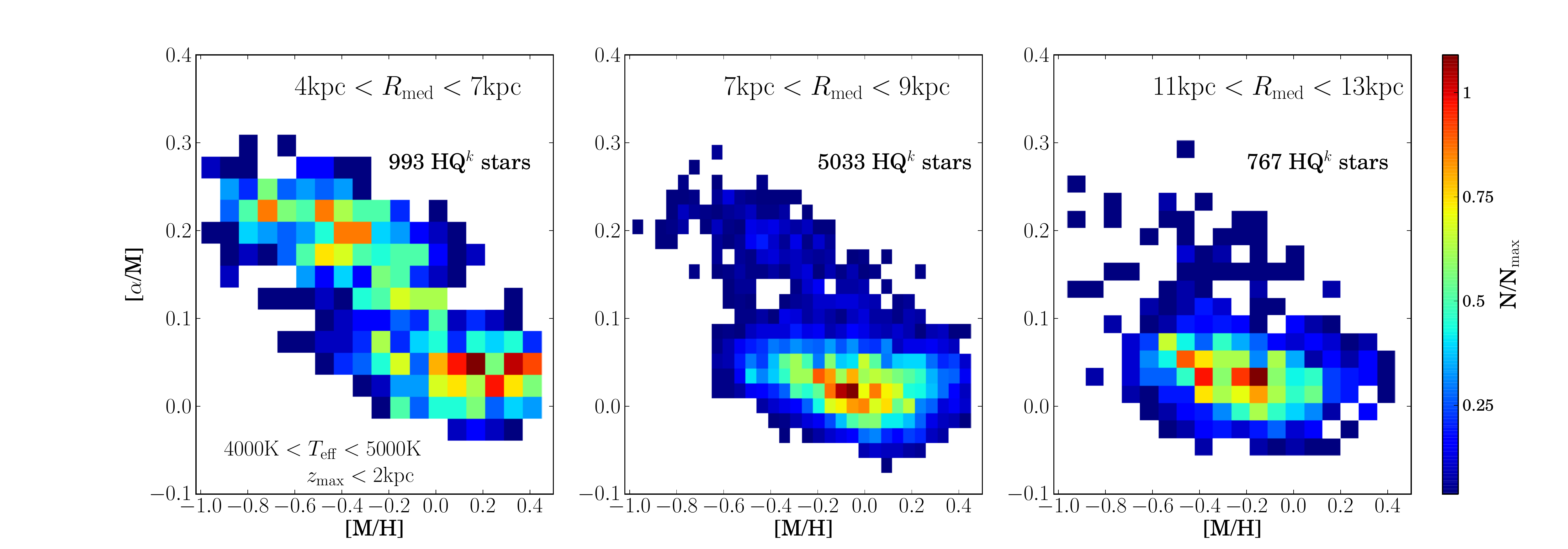}\\
	\includegraphics[width=0.8\textwidth]{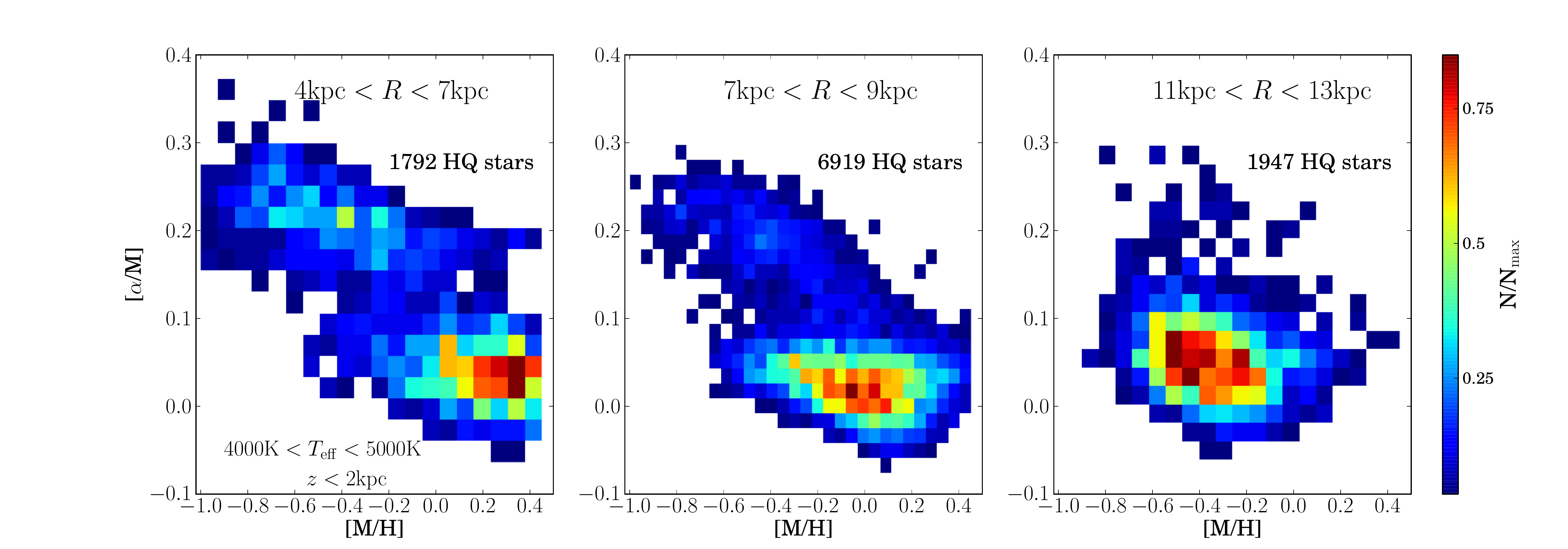}
	\includegraphics[width=0.8\textwidth]{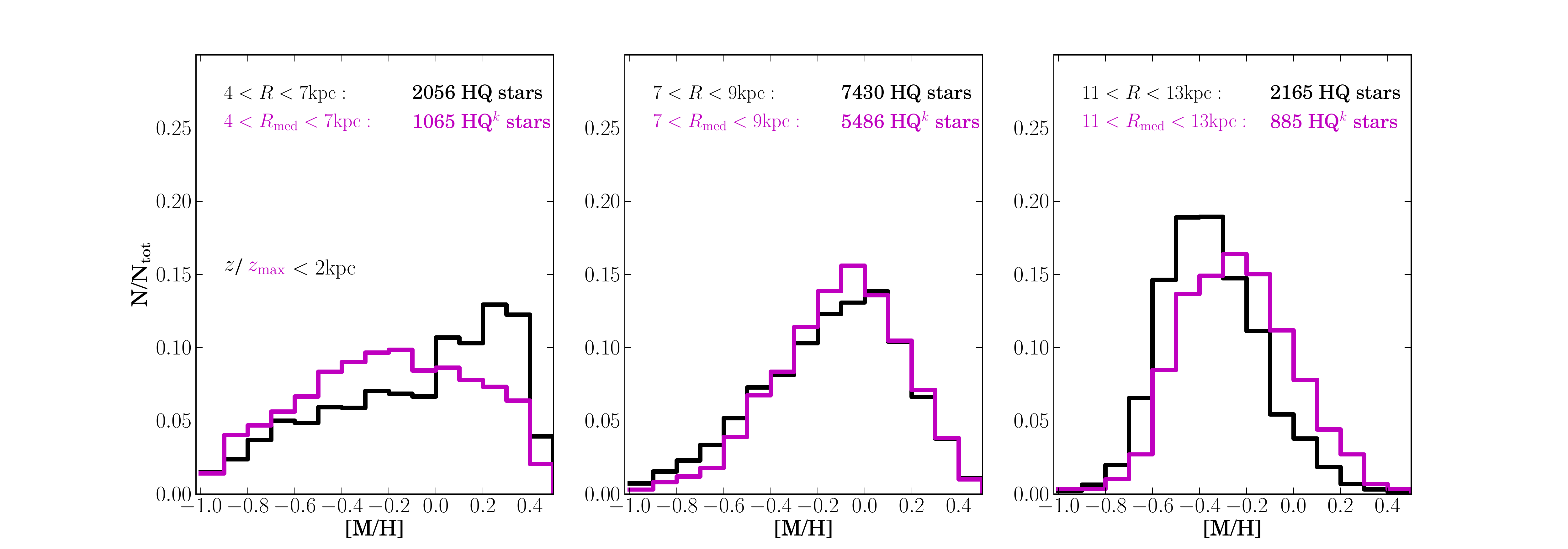}\\
	\caption{Chemical abundances of red giant stars in the Galactic disc in three bins of Galactocentric radius.
	Top row: [$\alpha$/Fe] vs. [Fe/H] diagrams for the high-resolution samples of \citet{Bensby2010a, Alves-Brito2010} and \citet{Bensby2011}. The authors collected high-resolution spectra and performed a manual spectroscopic analysis for their sample. 
	Second and third row: Density plot of the chemical abundance plane in the same radiall bins for the APOGEE HQ$^k$ and HQ samples, with respect to the orbital parameters ($R_{\mathrm{med}}, z_{\mathrm{max}}<2$ kpc) and the real-space coordinates ($R, z<2$ kpc), respectively. As before, in this plot we restrict these samples to a smaller temperature range ($4000~\mathrm{K}<T_{\mathrm{eff}}<5000~\mathrm{K}$), for which ASPCAP currently gives the most reliable values for the [$\alpha$/Fe] abundance ratio. We confirm the result of \citep{Bensby2011} that the radial scale length of the thick disc is much shorter than that of the thin disc: In the $11<R_{\mathrm{med}}<13$ kpc bin, almost no stars with thick disc abundance pattern are present.
	Bottom row: MDFs for the three radial bins, again with respect to orbital (magenta) and real-space (black) coordinates (here we are using the full temperature range of the HQ sample defined in Table~\ref{selectionsummary}).}
	\label{chemplane_rbins}
\end{figure*}

To highlight APOGEE's potential in chemical mapping, we compare the APOGEE [$\alpha$/Fe] vs. [Fe/H] abundance plane in different bins of $R_{\mathrm{med}}$ with the recent high-resolution study of disc field red giants by \citet{Bensby2011} reproduced here in the upper row of Fig. \ref{chemplane_rbins}. 
Several characteristics can be noted immediately:

\begin{itemize}
\item By comparing the compilation of \citet[][first row in Fif. \ref{chemplane_rbins}]{Bensby2011} with what is obtained with our first-year APOGEE data (second and third rows), we see a general agreement of the abundance trends. However, the Bensby et al. data extend to larger [$\alpha$/Fe] ratios than our APOGEE sample (by no more than $\sim$ 0.1~dex in the inner and solar neighborhood subsamples). The main differences between the Bensby et al. sample and ours are caused by different abundance analysis techniques and the narrower $J-K_s$ colour range considered by Bensby et al. in order to estimate reliable photometric distances.
\item In the plots shown in the second row, our sample was divided into wide bins in $R_{\mathrm{med}}$, in order to minimize the contamination by stars moving on very eccentric orbits, whose most probable guiding radii lie outside the defined bins (``blurring"). This allows us to conclude that the local thin disc extends from quite low ([M/H] $\sim-0.7$) to super-solar metallicities ([M/H] $\sim+0.4$) which may be currently, but not definitively explained by radial migration. Also in the outer disc, we find a sizeable number of super-metal-rich (SMR) stars ([Fe/H] $>$ 0.2) which probably originate from an inner Galactic region. Notice that these stars are not observed in the corresponding Bensby et al. sample shown in the first row, most probably because of low statistics. For comparison, the corresponding diagrams where the ``blurring" contamination has not been taken into account are shown in the third row. 
\item The proportion of thin disc to thick disc increases with Galactocentric orbital radius. In the left panels (corresponding to the inner disc), the large fraction of high-$\alpha$ stars as well as the significant difference between the abundance distributions when using orbital parameters ($R_{\mathrm{med}}, z_{\mathrm{max}}$) instead of real-space coordinates ($R, z$) may in part be explained by a selection bias in the inner-disc sample, as we preferentially detect stars passing through the Solar neighbourhood on eccentric orbits -- and these tend to be older, $\alpha$-enriched stars from the inner disc. This bias should be small in the other two panels, suggesting that the scale length of the thick disc is shorter than that of the thin disc \citep{Bensby2011, Bovy2012d, Cheng2012a}. 
\item The metallicity distributions in the different radial bins are shown in the last row of Fig. \ref{chemplane_rbins}. Again, a clear difference is seen between the distributions when defining the bins with respect to orbital median radius or real space coordinates. For instance, a clear contamination from stars with different guiding radii is seen on the left panel where the large contribution from high-metallicity stars disappears once $R_{\mathrm{med}}$ is used instead of $R$.
\item As predicted by pure chemical-evolution models for the thin disc (e.g., \citealt{Chiappini2001}), the metallicity distribution is broader in the inner disc than towards the outer parts. This happens because of the shorter infall timescales assumed for the inner regions which produces a larger number of metal-poor stars (also known as the G-dwarf problem). In the outer parts, where the star formation is less strong (and the infall timescales are longer), the resulting metallicity distribution is narrower. The predicted change in the metallicity distributions peak are small in the galactocentric distance range considered here. The data shown in the last row of the figure, when using $R_{\mathrm{med}}$, does not show a strong peak variation and shows that 
the MDF is broader in the inner regions when compared to the outer ones. This is also in good agreement with the recent predictions of the chemodynamical model of \citet{Minchev2013,Minchev2014} (but see below).
\item Another crucial constraint on chemodynamical models is the percentage of SMR stars at the different radial bins. Unfortunately, the biases involved in our sample could be playing an important role when determining this observable (as they will certainly influence the final shape of the MDFs shown in this row). Although we must currently refrain from quantitative interpretations of the MDF before taking into account all the selection effects involved in our samples, we find that the fraction of SMR stars increases with decreasing Galactocentric distance. Indeed, it is not clear how ASPCAP contributes with further biases in the high-metallicity regime (e.g., some of the SMR stars could have been cut out by our colour, temperature and $\chi^2$ selections; further ASPCAP difficulties at metallicities beyond $\sim +$0.4 are currently not fully understood). One could then imagine the number of SMR stars seen in the present figure to represent lower limits on the fraction of SMR stars in the respective Galactic regions.
\end{itemize}

\subsubsection{Disc abundance gradients and variations of the MDF with height above the plane}\label{gradients}

Chemical gradients are among the main observables constraining chemical-evolution models, determining the relative enrichment history of different Galactocentric annuli, the amount of gas infall \citep{Chiappini2001}, radial mixing \citep{Schonrich2009}, etc. To date, however, the main tracers used to determine the chemical gradients of the Galaxy are young objects, and often suffer from low number statistics (see, e.g., \citealt{Stasinska2012}). Red giant stars span a wide range of ages and are therefore a better tool to reconstruct star-formation histories \citep{Miglio2013a}. \\

{\it The metallicity gradient and the MDF at different distances from the Galactic plane} \\

\begin{figure*}[!h]\centering
	\includegraphics[width=0.44\textwidth]{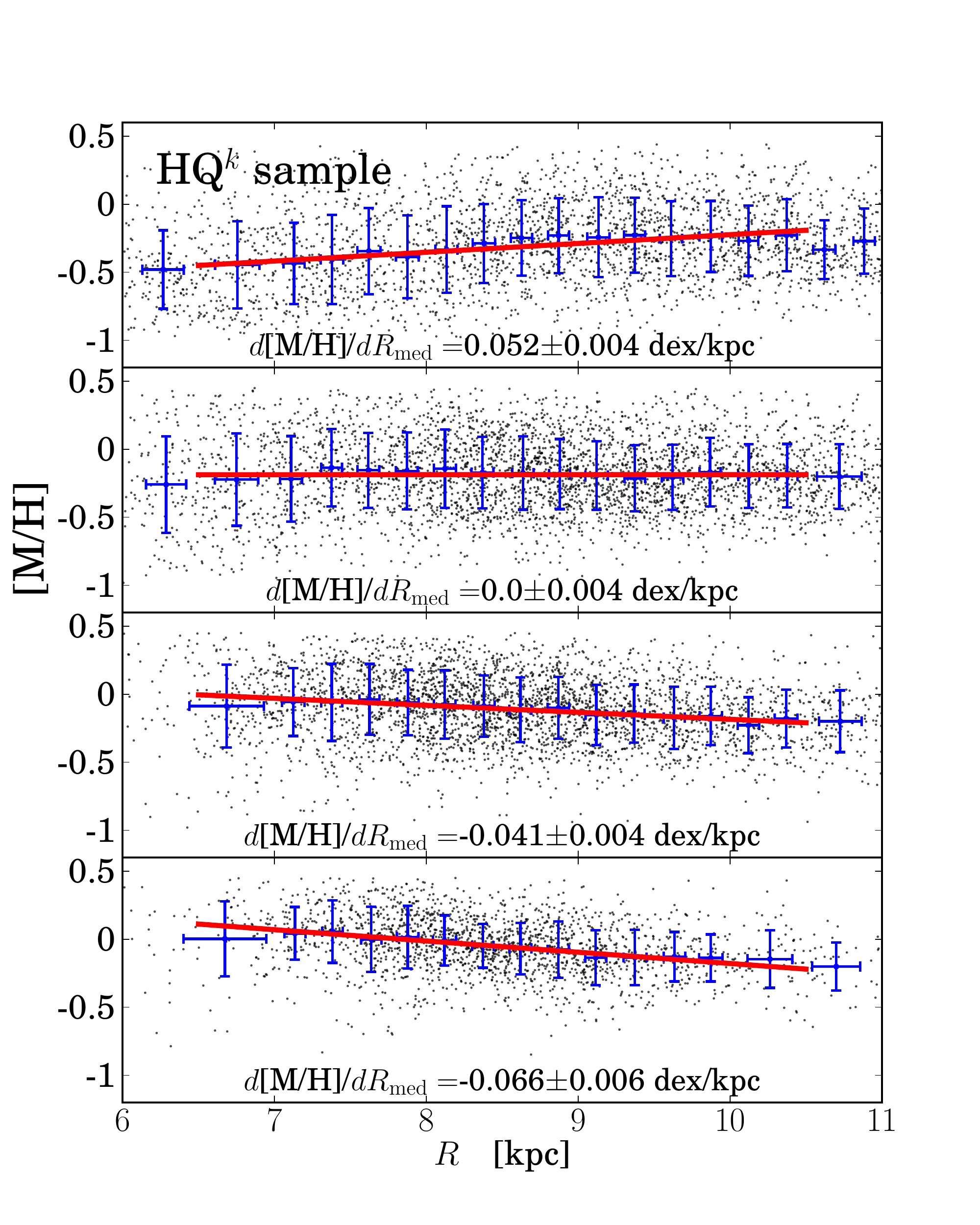}
	\includegraphics[width=0.44\textwidth]{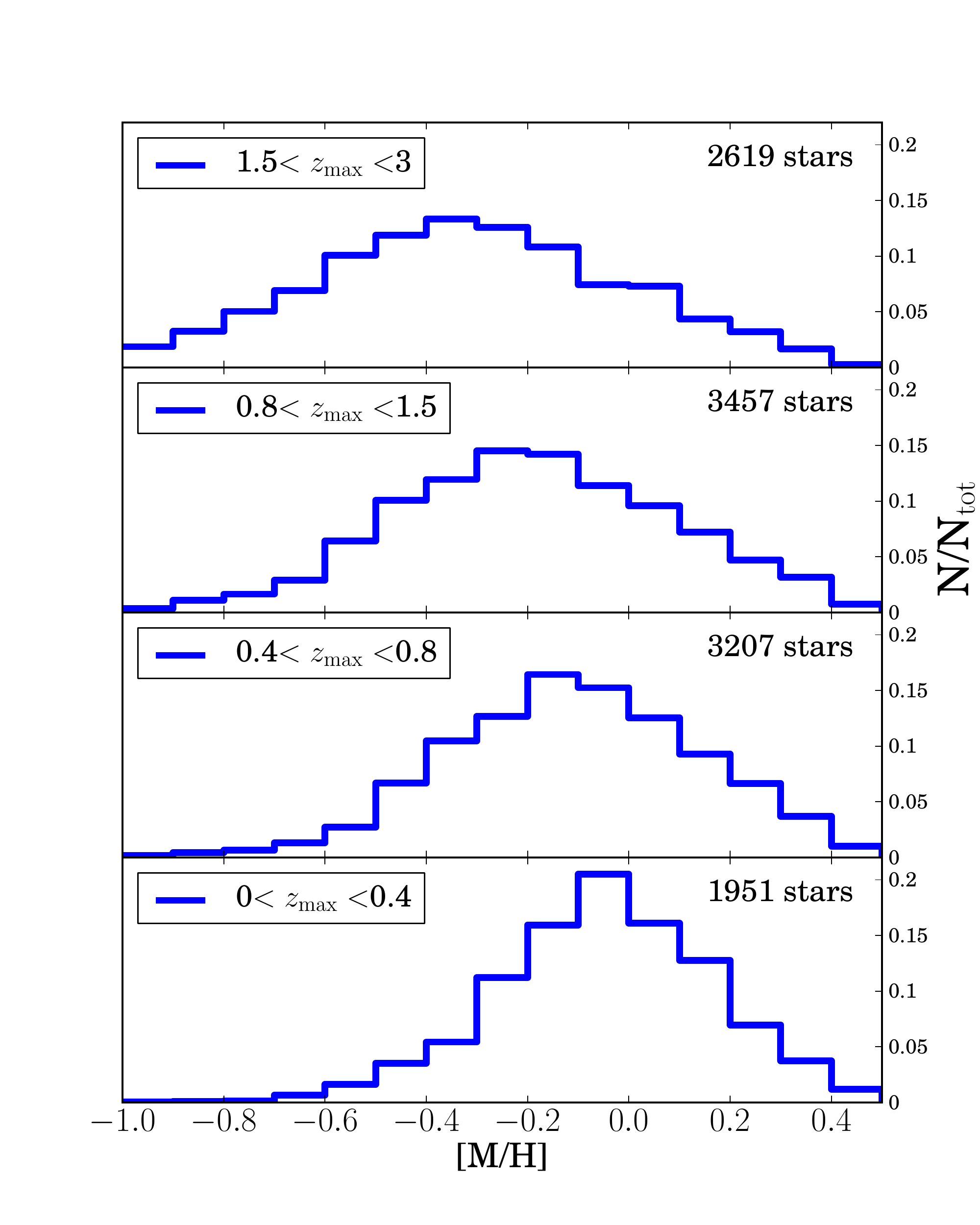}
	\vspace{-1em}
	\includegraphics[width=0.44\textwidth]{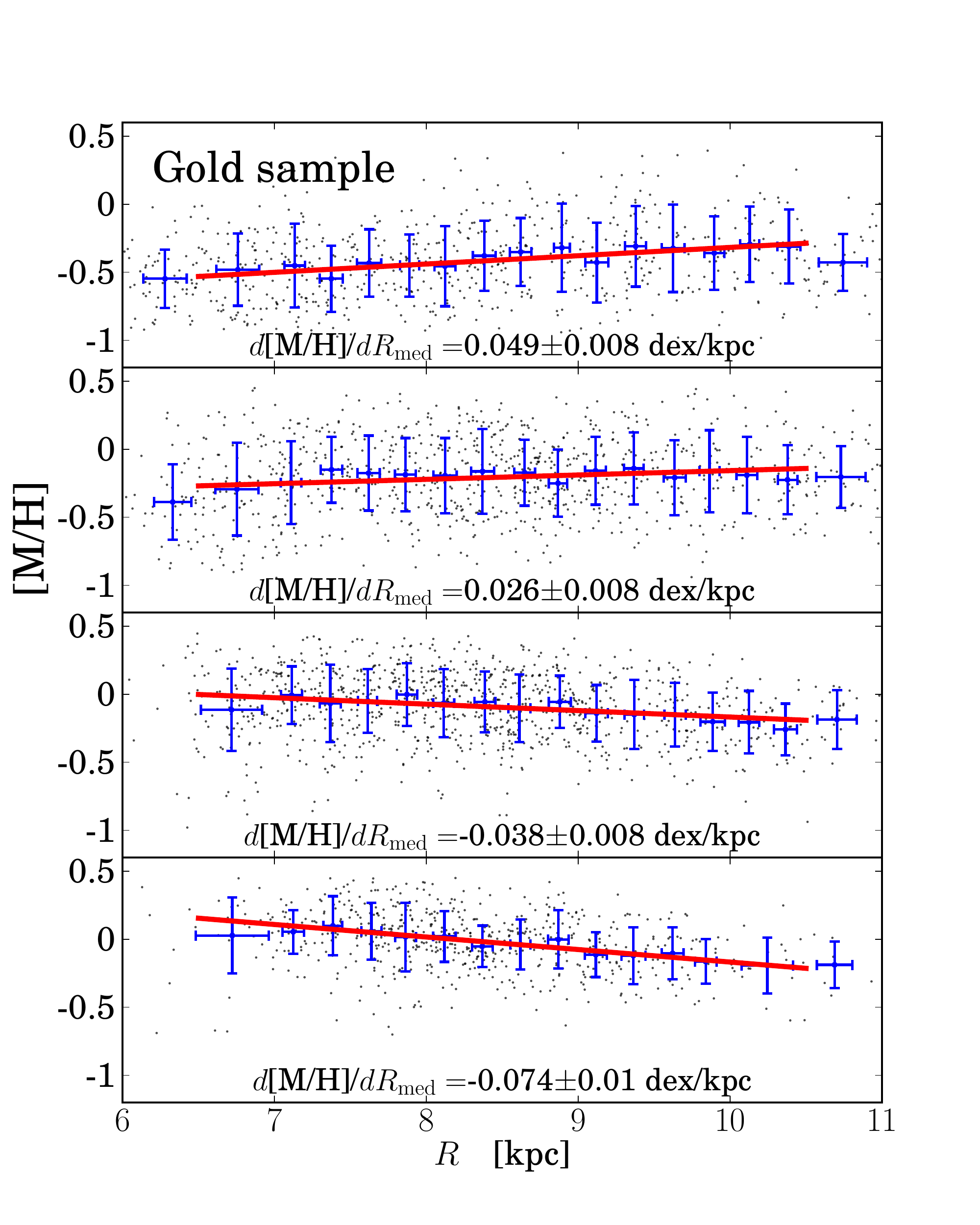}
	\includegraphics[width=0.44\textwidth]{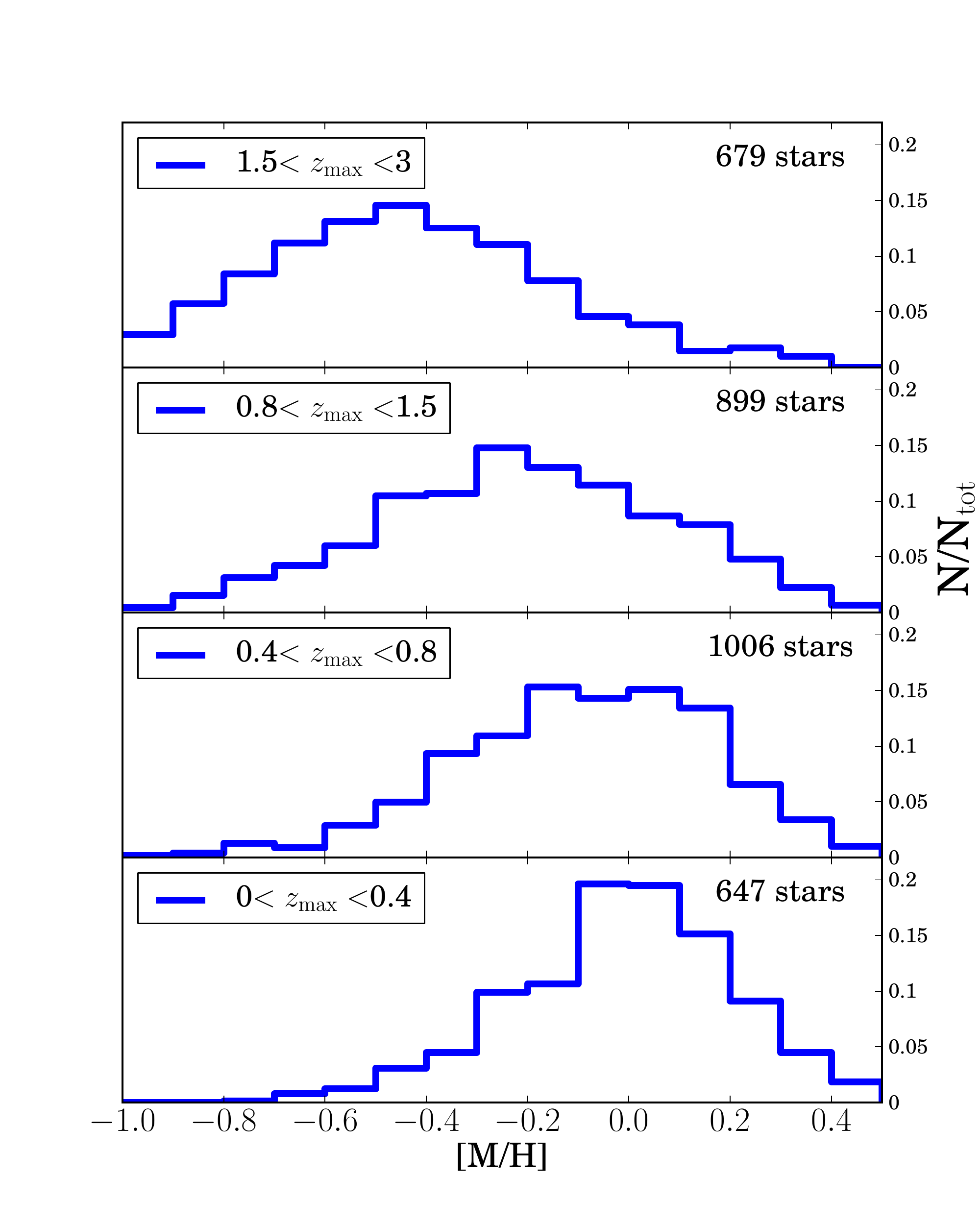}
	\caption{Top: Radial metallicity gradients (using the median orbital radii $R_{\mathrm{med}}$) and metallicity distribution functions as a function of $z_{\mathrm{max}}$ for the HQ sample. The gradients were computed using a simple least-squares optimisation, errors were estimated via bootstrapping. Note that we still do not account for any selection biases. Bottom: The same for the Gold sample.}
	\label{b1}
\end{figure*}

In Fig. \ref{b1}, we show results for the radial metallicity gradient and the MDF as a function of maximum height above the plane, for both the HQ and the Gold samples (for a complementary work, extending to more inner Galactocentric distances -- but without kinematics -- see \citealt{Hayden2013}). 

In the recent paper by \citet{Boeche2013a}, the authors compare the gradients obtained from a RAVE dwarf sample with those of the Geneva-Copenhagen survey (similar to our approach, the authors provide their results with respect to the orbital parameter space ($R_{\mathrm{g}}, z_{\mathrm{max}}$)\footnote{$R_{\mathrm{g}}\approx R_{\mathrm{med}}$ is the orbital \frqq guiding radius\flqq, a quantity directly related to the angular momentum of a star (\citealt{Boeche2013}).}, but only for three bins of $z_{\mathrm{max}}$). For comparison, their results are summarised in Table \ref{Fegradtable}, along with our measured values. The agreement between the APOGEE and RAVE samples used here is remarkable. Despite the use of different tracer populations, different surveys with vastly different selections, different distance estimates and a different orbit integration codes assuming different MW potentials, the tendencies for the gradients found for dwarfs and giants agree. 

As reported in previous works, our results show that the kinematically coolest stellar population ($z_{\mathrm{max}}<0.4$ kpc) exhibits the steepest (negative) radial gradient ($\frac{d\mathrm{[Fe/H]}}{dR_\mathrm{g}}=-0.066\pm0.006$ dex/kpc); as we move to higher $z_{\mathrm{max}}$, the gradient flattens (\citealt{Carrell2012, Cheng2012, Boeche2013a}). Furthermore, thanks to the fact that our sample extends well above the plane (compared with previous works), we can confirm that the gradient changes its sign ($\frac{d\mathrm{[Fe/H]}}{dR_\mathrm{g}}\simeq+0.05$ dex/kpc) for $1.5<z_{\mathrm{max}}<3$ kpc. 
The latter result as well as the overall trend of the metallicity gradient with height above the plane, is seen in both the Gold and the HQ$^k$ sample, suggesting that the measured gradients do not critically depend on potential selection biases.\footnote{Indeed, \citet{Boeche2013a} show that different cuts in $R_g$ result in only small differences of their abundance gradients.} The measured gradients for the Gold and the HQ$^k$ sample differ significantly only in one $z_{\mathrm{max}}$-bin. We suggest this to be caused the additional kinematical selection of the Gold sample, along with contamination of the high-$z_{\mathrm{max}}$ panels of Fig. \ref{b1} by thin-disc stars with poorly-determined orbital parameters (see discussion below).

While the general consistency of the radial abundance trends of RAVE and APOGEE may suggest that the measured value of the abundance gradients at low Galactic latitudes is a rather robust observable, the agreement of both surveys with GCS results is only of qualitative nature. The metallicity gradient values at different distances from the Galactic plane measured by \citet{Boeche2013a} for the GCS sample typically differ from the corresponding APOGEE and RAVE values by $+0.03$ dex/kpc (see \citealt{Boeche2013a} for a discussion). 

From these considerations, we suggest that the inversion of the [M/H] gradient above $z\sim z_{\mathrm{max}}\approx1.5$ kpc could be:
\begin{itemize}
\item A consequence of the smaller scale length of the thick disc with respect to the thin disc. In this case, the more metal poor stars of the thick disc would be concentrated towards smaller Galactocentric distances, creating the impression of a positive gradient \citep{Boeche2013a}, or 
\item Due to yet another selection effect related to the inhomogeneous coverage of the Galactic disc(s) by finite-sightline observations \citep{Bovy2012d}, which is present in all the currently available large-scale Galactic survey data (APOGEE, RAVE and SEGUE).\footnote{Although RAVE as a hemisphere survey should be less affected by this type of bias.} Initial simulations for a SEGUE sample with the stellar population synthesis model TRILEGAL have shown that selection effects may well produce a significant gradient that is not present in the underlying simulation (Brauer et al. 2014, in prep.).
\end{itemize}

The observed flattening of the gradient with height above the plane does not depend on the choice of $z_{\mathrm{max}}$ instead of the stars' \frqq current\flqq~height $z$ above the Galactic plane (for the corresponding figure, using the current R and z positions, see Fig. \ref{b2}). On the other hand, the exact values of the gradients do very much depend on the set of (orbital-) space coordinates used. See Appendix A for a discussion.\\

Although we do not exclude the possibility that the gradient inversion may be a ``real'' characteristic of the Galactic disc at intermediate Galactocentric distances ($6\lesssim R\lesssim11$ kpc), which could in this case be related to the flaring of young stellar populations in the outer disc (as previously seen in dynamical simulations, e.g., \citealt{Minchev2012}), we caution the reader about the physical reality of of this feature. \\

\begin{table*}[t]
\caption[]{Radial [Fe/H]\tablefootmark{a} gradients with respect to the orbital guiding radius\tablefootmark{b} in the range $6<R_{g}<11$ kpc, for four ranges of $z_\mathrm{max}$.}
\label{Fegradtable}
\vskip 0.3cm
\centering
\begin{tabular}{l|ccccc}
\hline\hline
\noalign{\smallskip}
$\frac{d\mathrm{[Fe/H]}}{dR_\mathrm{g}}$ [dex/kpc]       &  APOGEE HQ$^k$  & APOGEE Gold  & GCS dwarfs \tablefootmark{c} & RAVE dwarfs\tablefootmark{a} \\
\noalign{\smallskip}
\hline
\noalign{\smallskip}
0.0$\leq z_\mathrm{max}~\mathrm{[kpc]}<$0.4           &$-0.066\pm0.006$ & $-0.074\pm0.010$& $-0.043\pm0.004$ &$-0.065\pm0.003$\\

0.4$\leq z_\mathrm{max}~\mathrm{[kpc]}<$0.8           &$-0.041\pm0.004$ &$-0.038\pm0.008$& $-0.008\pm0.011$ &$-0.059\pm0.005$\\

0.8$\leq z_\mathrm{max}~\mathrm{[kpc]}<$1.5& $+0.000\pm0.004$&$+0.026\pm0.008$ & $+0.056\pm0.019$ & $+0.006\pm0.015$\\

1.5$\leq z_\mathrm{max}~\mathrm{[kpc]}<$3.0& $+0.052\pm0.004$&$+0.049\pm0.008$& -- & --\\
\noalign{\smallskip}
\hline
\end{tabular}
\tablefoot{\tablefoottext{a}{For the APOGEE data: [M/H]$_{\mathrm{calib}}$}; \tablefoottext{b}{For the APOGEE w.r.t. the median orbital Galactocentric radius $R_{\mathrm{med}}$}. The 1$\sigma$--uncertainties are computed using a bootstrap method. ; \tablefoottext{c}{Values from \citet{Boeche2013a}.}}
\end{table*}

\begin{table*}[t]
\caption[]{Radial [$\alpha$/Fe]\tablefootmark{a} gradients with respect to the orbital guiding radius\tablefootmark{b} in the range $6<R_{g}<11$ kpc, for four ranges of $z_\mathrm{max}$ }
\label{alphagradtable}
\vskip 0.3cm
\centering
\begin{tabular}{l|ccccc}
\hline\hline
\noalign{\smallskip}
$\frac{d[\alpha/\mathrm{Fe}]}{dR_\mathrm{g}}$ [dex/kpc]       &  APOGEE HQ$^k$  & APOGEE Gold  & GCS dwarfs\tablefootmark{c} & RAVE dwarfs\tablefootmark{a}\\
\noalign{\smallskip}
\hline
\noalign{\smallskip}
0.0$\leq z_\mathrm{max}~\mathrm{[kpc]}<$0.4           &$-0.005\pm0.001$ & $-0.005\pm0.002$& $+0.010\pm0.002$ &$-0.004\pm0.001$\\

0.4$\leq z_\mathrm{max}~\mathrm{[kpc]}<$0.8           &$-0.009\pm0.001$ &$-0.007\pm0.002$& $-0.006\pm0.005$ &$-0.005\pm0.002$\\

0.8$\leq z_\mathrm{max}~\mathrm{[kpc]}<$1.5& $-0.019\pm0.001$&$-0.022\pm0.002$ & $-0.023\pm0.007$ & $-0.020\pm0.005$\\

1.5$\leq z_\mathrm{max}~\mathrm{[kpc]}<$3.0& $-0.031\pm0.001$&$-0.023\pm0.002$& -- & --\\
\noalign{\smallskip}
\hline
\end{tabular}
\tablefoot{\tablefoottext{a}{For the APOGEE data: [$\alpha$/M]}; \tablefoottext{b}{For the APOGEE data: the median orbital Galactocentric radius $R_{\mathrm{med}}$. The 1$\sigma$--uncertainties are computed using a bootstrap method}; \tablefoottext{c}{Values from \citet{Boeche2013a}.}}

\end{table*}

{\it The [$\alpha$/Fe] gradient and distribution function at different distances from the plane} \\

\begin{figure*}[!h]\centering
	\includegraphics[width=0.44\textwidth]{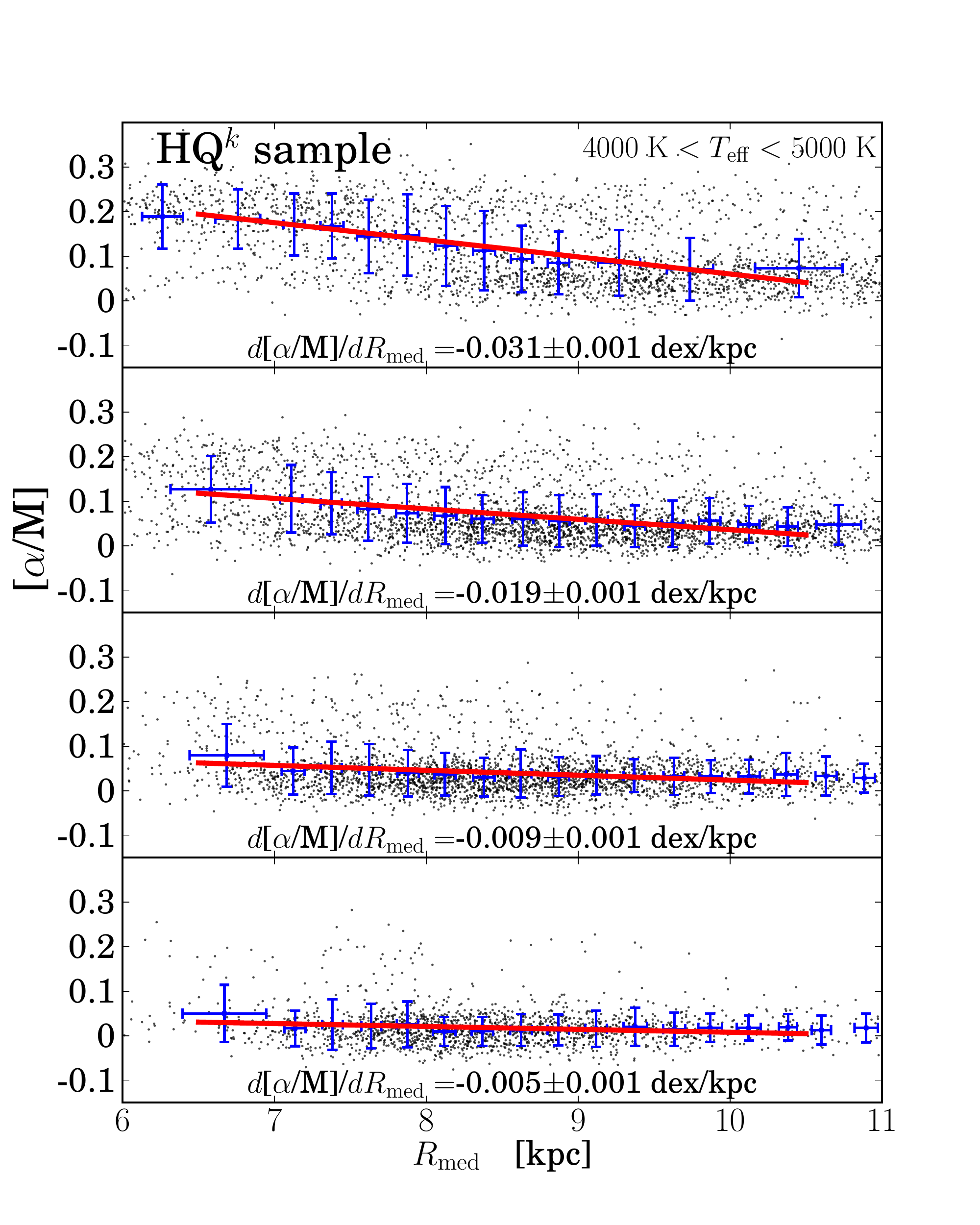}
	\includegraphics[width=0.44\textwidth]{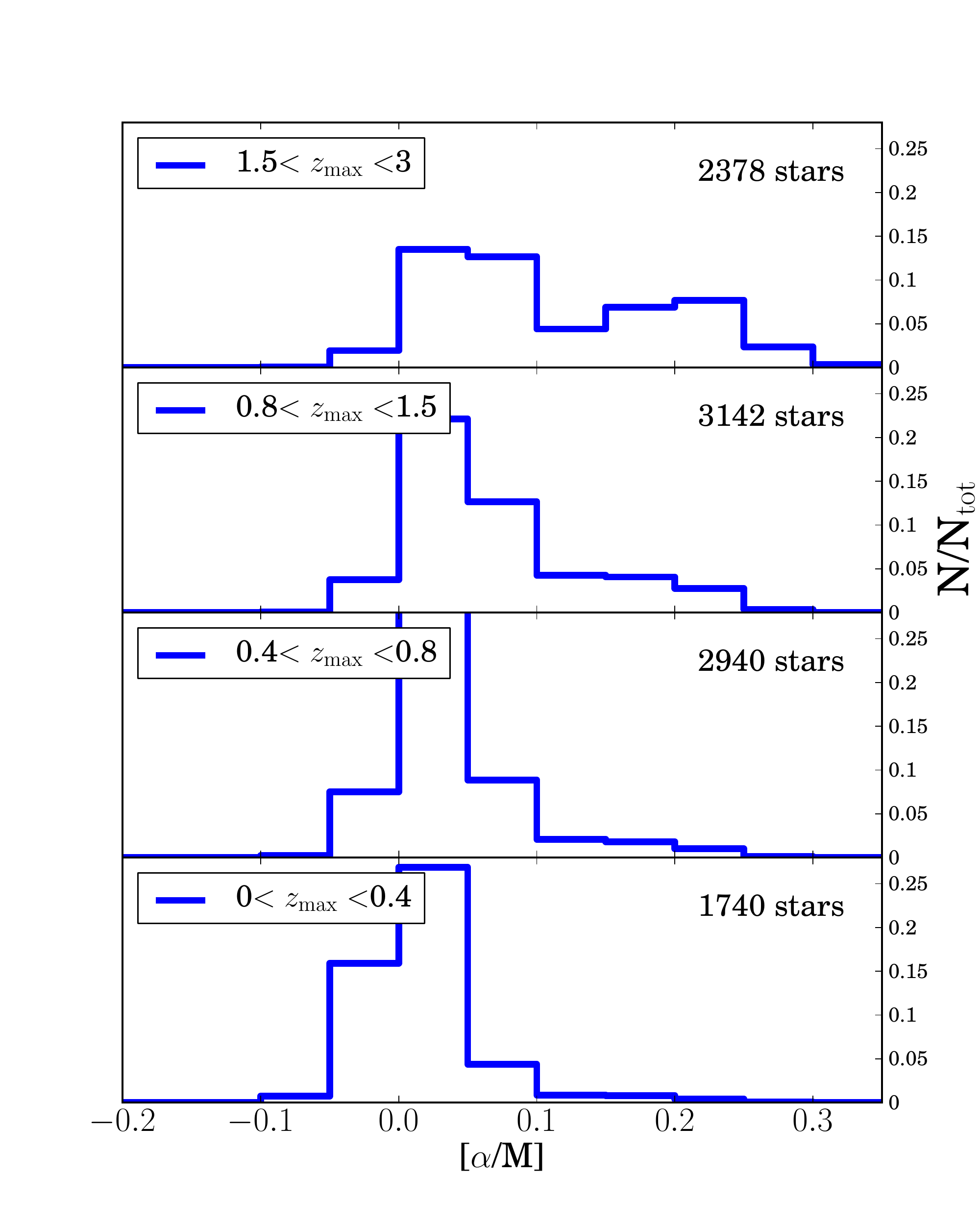}
	\vspace{-1em}
	\includegraphics[width=0.44\textwidth]{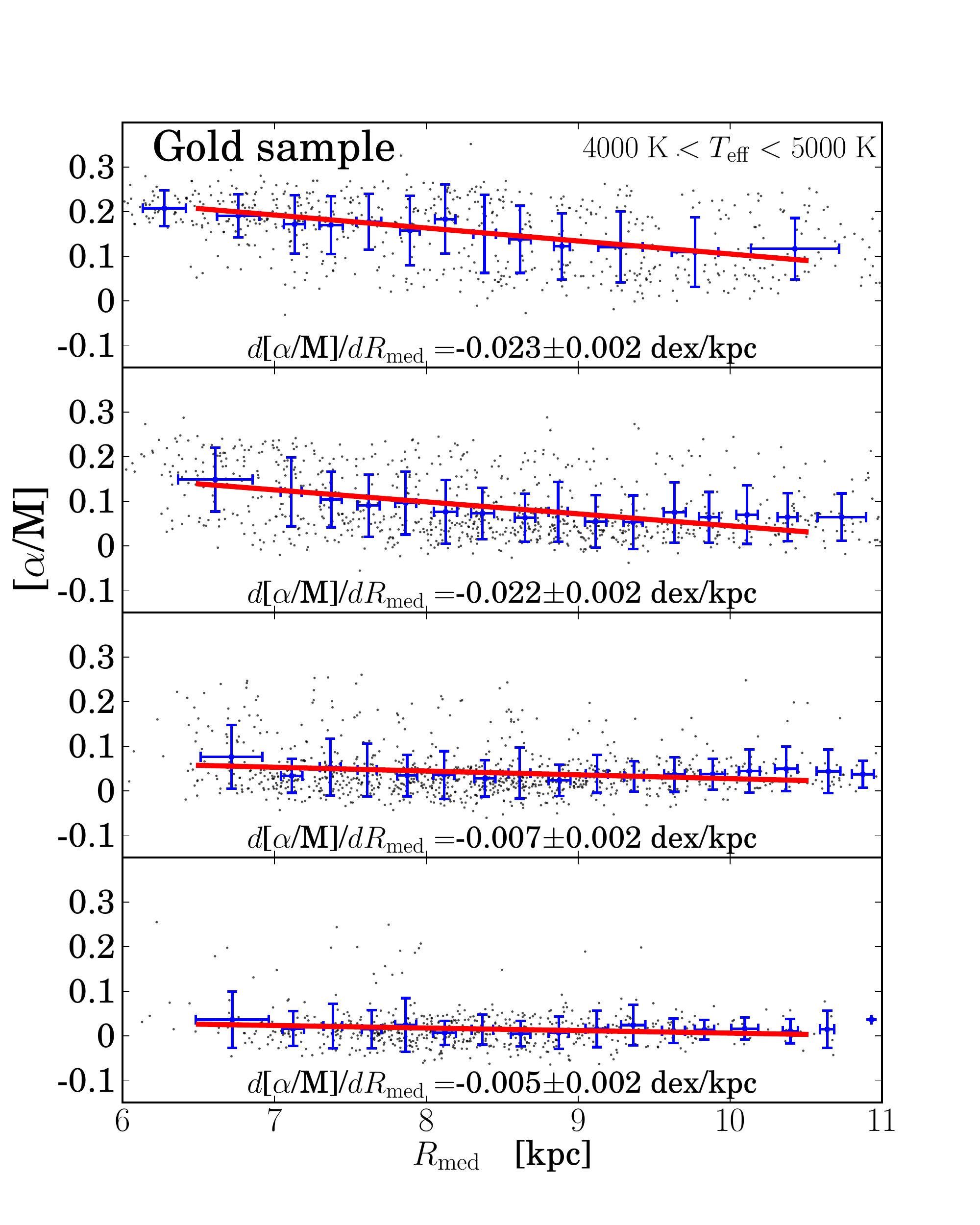}
	\includegraphics[width=0.44\textwidth]{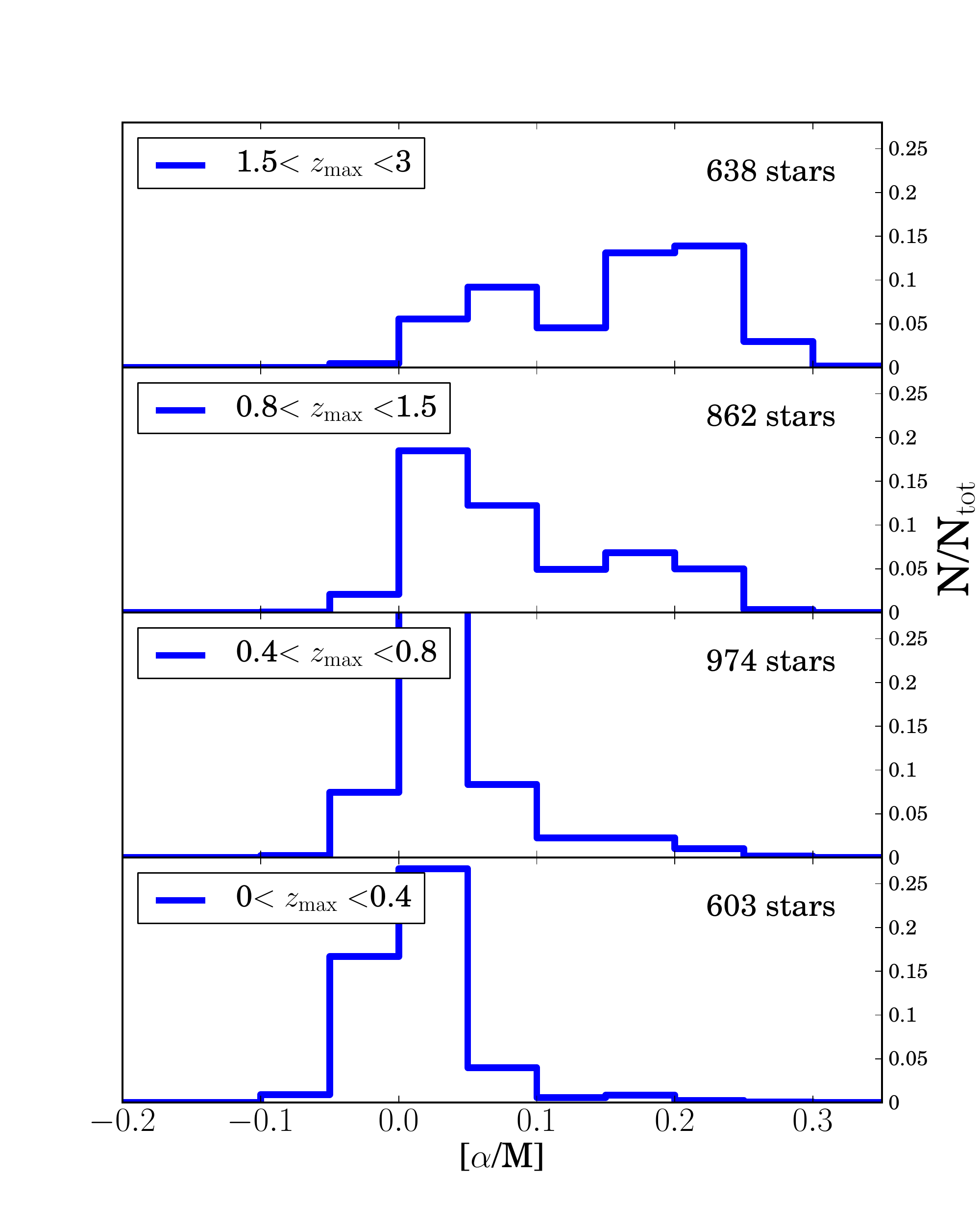}
	\caption{Same as Figure \ref{b1}, but for [$\alpha$/M].}
	\label{b3}
\end{figure*}

Fig. \ref{b3} presents the gradients and distributions in the [$\alpha/$M] abundance ratio for the APOGEE HQ$^k$ and the Gold sample, in the same fashion as Fig. \ref{b1}. The radial trend for small Galactic heights is slightly negative but almost flat, and that the negative trend increases with $z_{\mathrm{max}}$. Again, our measured gradients are fully consistent with the results of \citet{Boeche2013a} for the RAVE dwarf sample; the values agree within 1$\sigma$-uncertainties. The general trend of the steepening gradient is also found in the GCS data\footnote{However, the photometric [$\alpha$/Fe] estimates for the Geneva-Copenhagen survey used by \citet{Boeche2013a} are from \citet{Casagrande2011}, and should only be treated as proxies for [$\alpha$/Fe].}. As before, the corresponding figure using the current $z$ and $R$ values is given by Fig. \ref{b2}.\\

For the two highest bins in $z$ or $z_{\mathrm{max}}$, there are quite sizeable differences in the MDFs as well as in the [$\alpha$/M] distributions for the HQ$^{(k)}$ samples.
While for the ($z_{\mathrm{max}},R_{\mathrm{med}}$) plots shown in Fig. \ref{b3}, the low-$\alpha$ population dominates up to large distances from the plane, this is not the case for the corresponding diagram in the ($R, z$) plane (see Fig. \ref{b2}, bottom). Again, this is true both for the HQ$^k$ and Gold samples.
Given the considerable errors in the orbital parameter $z_{\mathrm{max}}$ for a sizeable fraction of our sample (especially for the HQ$^k$ sample at high distances from the Galactic plane), we suggest that this result may be due to the contamination of the upper panels by thin disc stars with poorly-determined orbits. This effect also has an impact on the exact value of the gradient at these Galactic heights. By enlarging our sample, we expect to explore this issue in more detail.

In particular, the [$\alpha$/M] distribution at high $z$ (see upper panels in the lower right plot of Fig. \ref{b2}) set rather tight limits on the effect of flaring of the thin disc, at least in our Galactocentric radial range. At high distances from the plane, we see essentially no low-[$\alpha$/M] stars. Because this figure is not subject to large uncertainties in the orbital parameters, we are close to seeing the real proportion of high-to-low [$\alpha$/Fe] stars here.\\

\section{Conclusions}
\label{conclusions}
In this first paper of a series of APOGEE papers, we have begun to explore the chemo-kinematical properties of the Milky Way disc using data from the first year of SDSS-III/APOGEE. We have compared our findings with results from local optical high-resolution samples in the literature as well as lower resolution surveys such as GCS and RAVE. In this section, we briefly summarize the main results of our work.

First, APOGEE appears to deliver reliable chemical abundances for [M/H] and [$\alpha$/M], and confirms many results previously obtained with smaller high-resolution spectroscopic samples. Together with the Gaia-ESO survey, APOGEE extends the Galactic volume covered by high-resolution spectroscopy from the inner disc and bulge to the outskirts of the disc.

We obtained the metallicity distribution function (MDF) of stars within 1~kpc from the Sun (d $<$ 1~kpc). This MDF turned out to be remarkably similar to the one obtained with the high-resolution HARPS FGK dwarf sample of \citet{Adibekyan2011}, despite the different volumes covered by the two samples. In both cases the MDF peaks at a metallicity slightly below the Solar value, and show comparable tails towards lower metallicities.

We can confirm the \frqq gap\flqq~in the [$\alpha$/Fe] vs. [Fe/H] diagram reported by previous works and argue that, similar to the volume-complete sample of \citet{Fuhrmann2011}, it is unlikely to be caused by a selection effect. Using our large sample of red giants stars, we corroborate the results obtained by \citet{Bensby2011,Bovy2012d} and \citet{Cheng2012} who found evidence for a shorter scale-length of the thick disc.

Although we have only of a small number of bulge candidates in our sample, APOGEE data appear to indicate different chemical signatures for the bulge and the thick disc, 

Motivated by similar results of \citet{Boeche2013a} using dwarf stars from RAVE and the Geneva-Copenhagen survey, we measure an inversion of the radial [M/H] gradient for stars at greater Galactic heights. We interpret this partly as a signature of inside-out formation of the Galactic disc, and partly as an effect of selection biases. An overall quantitative agreement with results from RAVE is still hampered by the radically different selection functions for RAVE and APOGEE.

Performing initial tests with the population synthesis code TRILEGAL, we confirm the need for a careful modelling of the survey selection function for future analyses.

The coming papers of this series will focus on a more detailed comparison with the chemo-dynamical Galaxy simulation of \citet{Minchev2013}, and include simulations of the APOGEE HQ and Gold samples with TRILEGAL and the Besan\c{c}on model \citep{Robin2003}. We also plan to employ a newly developed selection interface (Piffl et al. 2014, in prep.) to create mock surveys from a full chemo-dynamical MW model.

\bibliographystyle{aa}
\bibliography{ChemoAPOGEE1_accepted}

\begin{acknowledgements}
It is a pleasure to thank R.-D. Scholz for helpful discussions on astrometry, and J. Bovy \& K. Schlesinger for their valuable input on the discussion. Furthermore, we thank J. Bovy for making his code publically available.\\

TCB acknowledges partial support from grant PHY 08-22648; Physics Frontier Center/ JINA, awarded by the U.S. National Science Foundation. KC acknowledges support from the National Science Foundation (AST-0907873). PMF is supported by an NSF grant AST-1311835.\\

Funding for the Brazilian Participation Group has been provided by the Ministério de Ciência e Tecnologia (MCT), Funda\c{c}\~{a}o Carlos Chagas Filho de Amparo à Pesquisa do Estado do Rio de Janeiro (FAPERJ), Conselho Nacional de Desenvolvimento Científico e Tecnológico (CNPq), and Financiadora de Estudos e Projetos (FINEP).\\

Funding for SDSS-III has been provided by the Alfred P. Sloan Foundation, the Participating Institutions, the National Science Foundation, and the U.S. Department of Energy Office of Science. The SDSS-III web site is \url{http://www.sdss3.org/}.\\

SDSS-III is managed by the Astrophysical Research Consortium for the Participating Institutions of the SDSS-III Collaboration including the University of Arizona, the Brazilian Participation Group, Brookhaven National Laboratory, Carnegie Mellon University, University of Florida, the French Participation Group, the German Participation Group, Harvard University, the Instituto de Astrofisica de Canarias, the Michigan State/Notre Dame/JINA Participation Group, Johns Hopkins University, Lawrence Berkeley National Laboratory, Max Planck Institute for Astrophysics, Max Planck Institute for Extraterrestrial Physics, New Mexico State University, New York University, Ohio State University, Pennsylvania State University, University of Portsmouth, Princeton University, the Spanish Participation Group, University of Tokyo, University of Utah, Vanderbilt University, University of Virginia, University of Washington, and Yale University.\end{acknowledgements}

\appendix
\section{Gradients with respect to $(R, z)$}

\begin{figure*}[!h]\centering
	\includegraphics[width=0.44\textwidth]{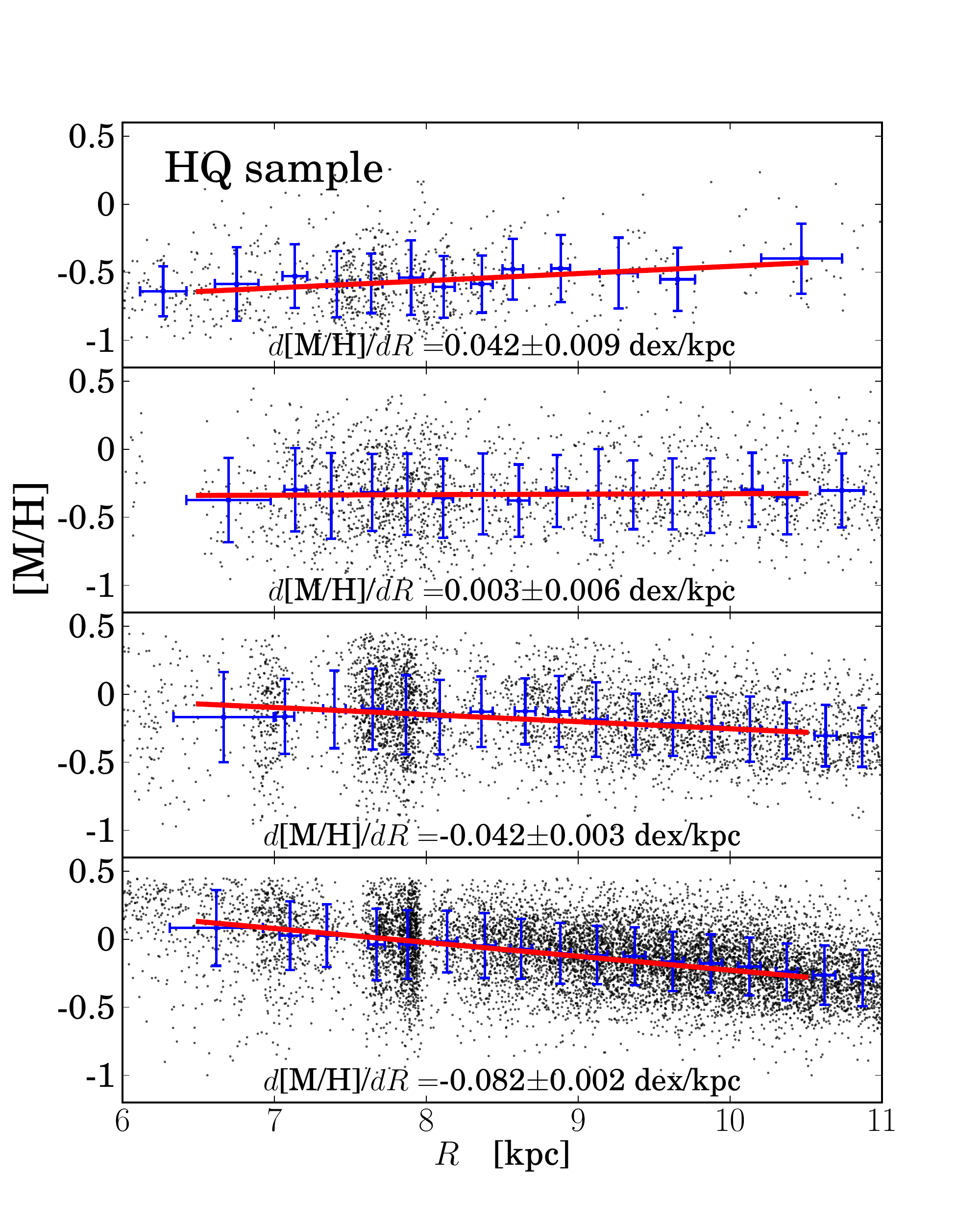}
	\includegraphics[width=0.44\textwidth]{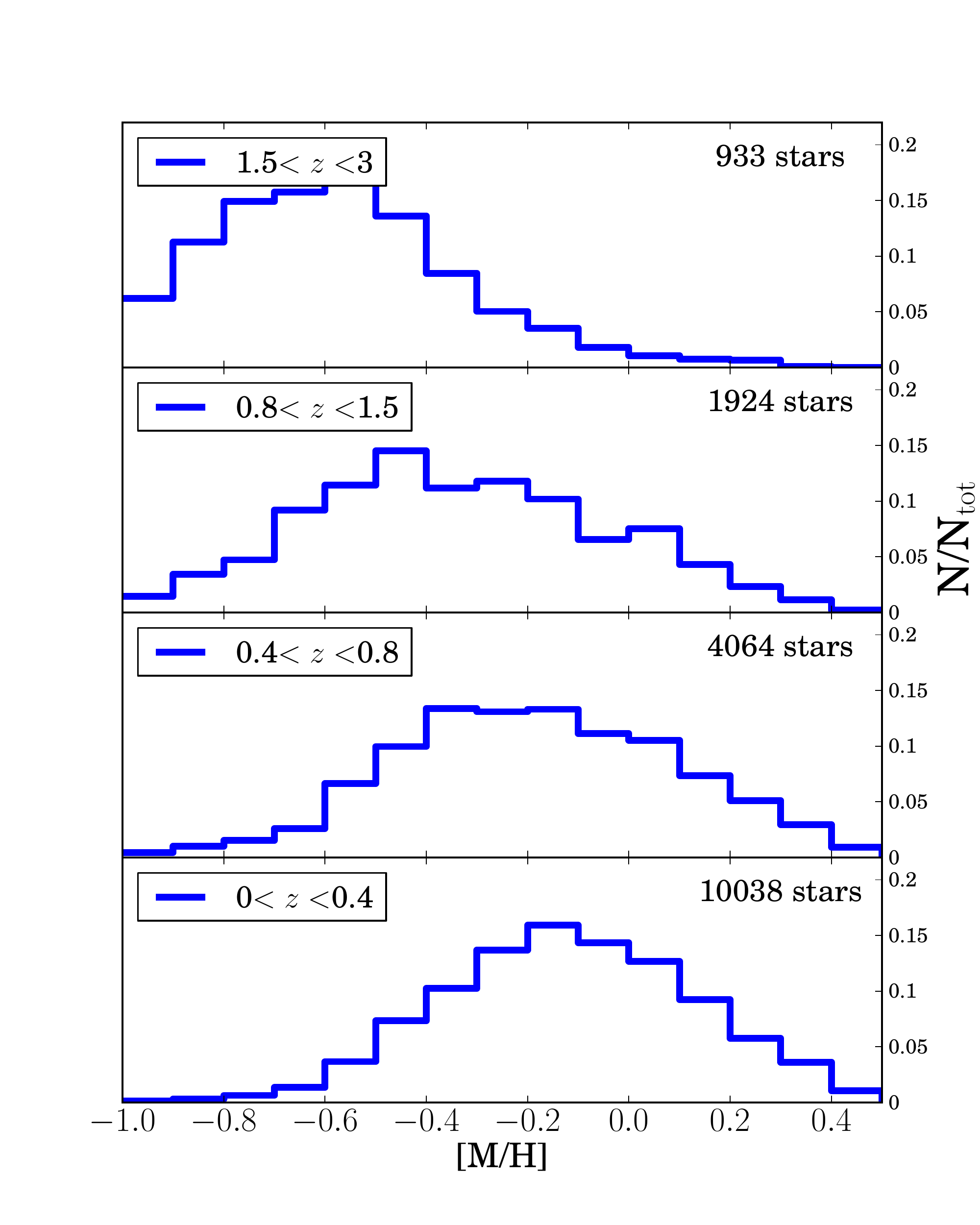}
	\vspace{-1em}
	\includegraphics[width=0.44\textwidth]{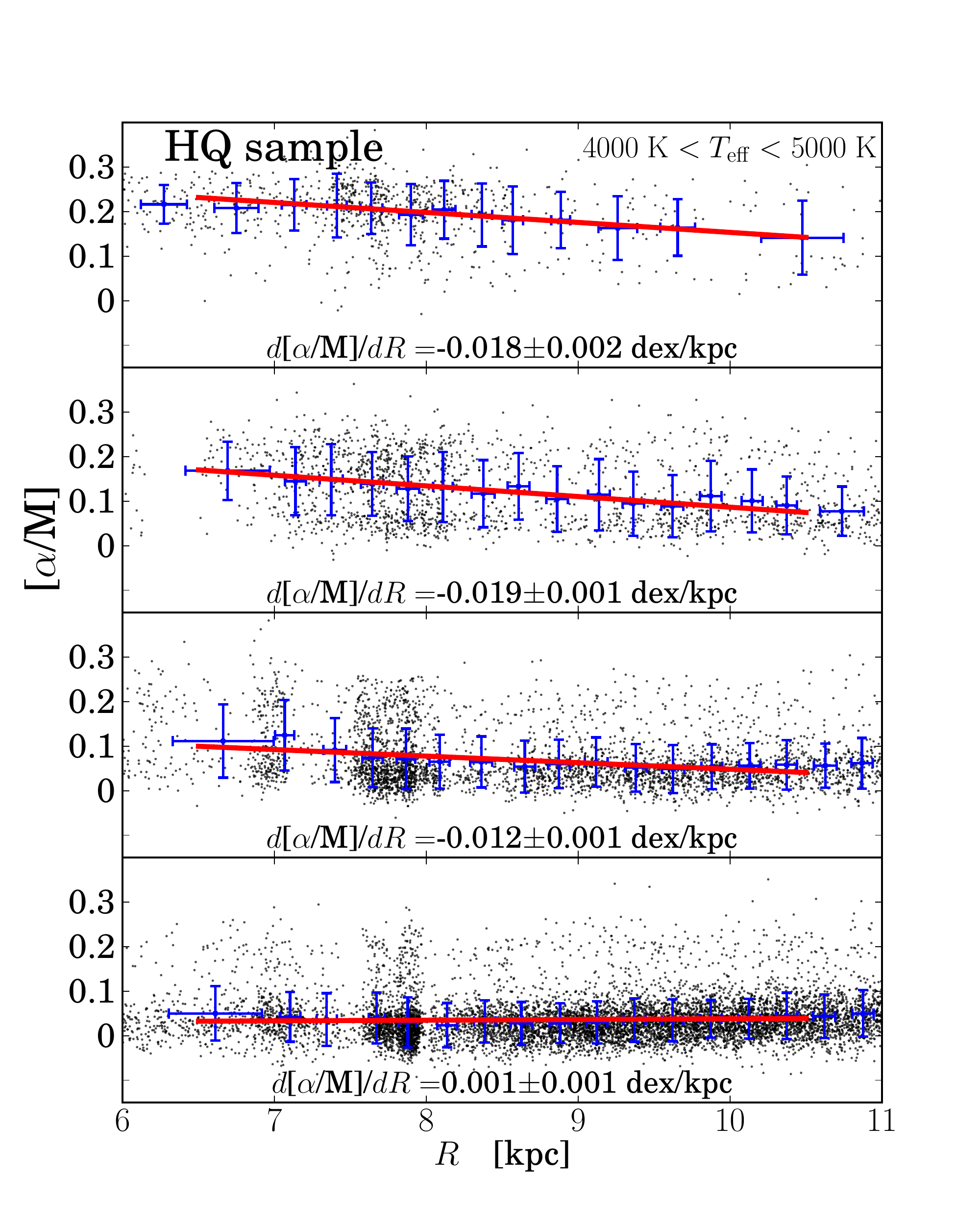}
	\includegraphics[width=0.44\textwidth]{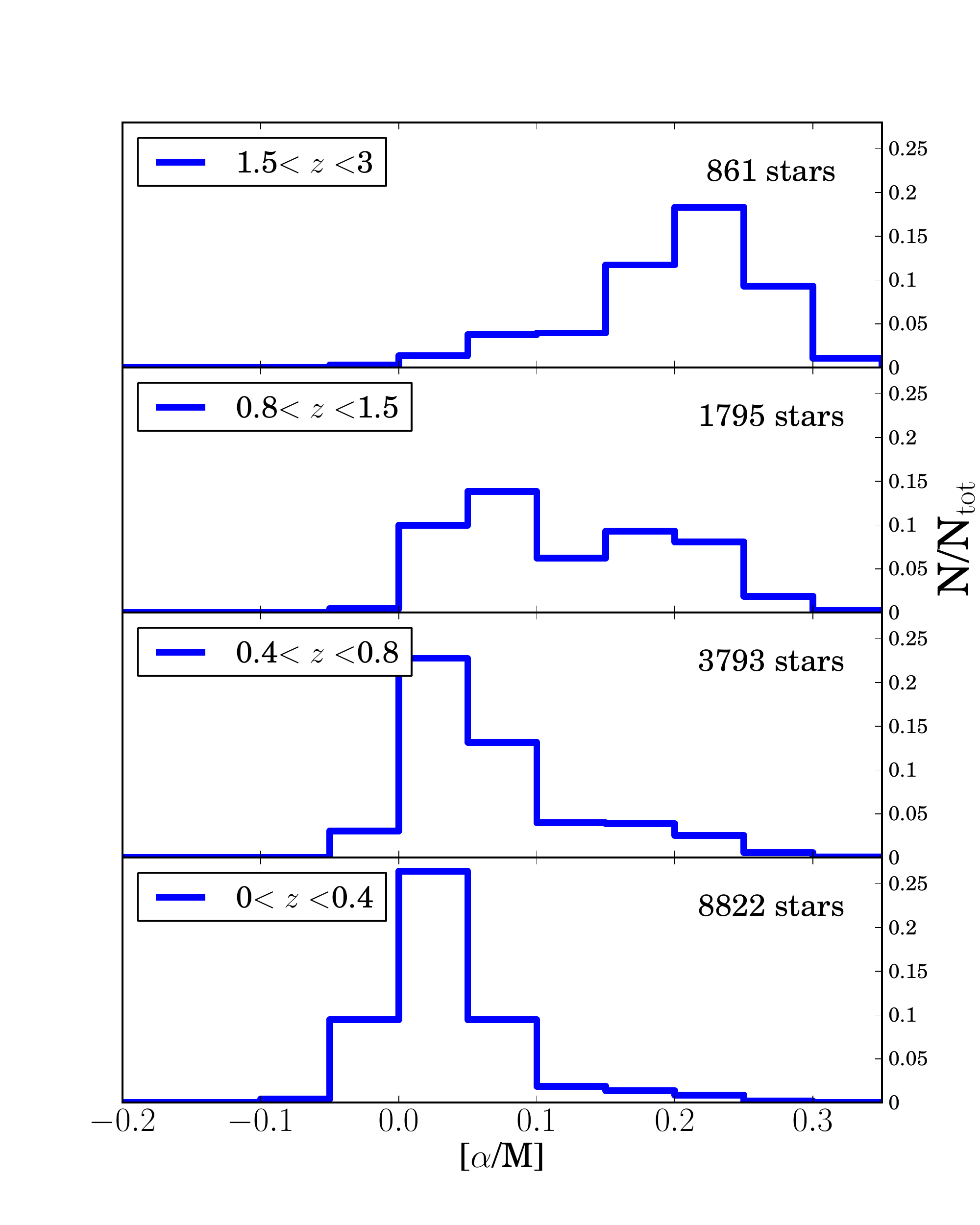}
	\caption{Top: Radial metallicity gradients (using now the {\it current} Galactocentric distance $R$) and metallicity distribution functions as a function of {\it current} Galactic height $z$ for the Gold sample. Bottom: The same for the HQ sample.
Bottom: Same as above, for the [$\alpha$/M] distribution functions.}
	\label{b2}
\end{figure*}

In order to estimate the effect of \frqq blurring\flqq~and the influence of orbital parameter uncertainties on our measured abundance gradients, we also computed the [M/H] and [$\alpha$/M] abundance gradients with respect to the current Galactocentric distance $R$, for different bins in current distance from the Galactic plane $z$. The results are shown in Fig. \ref{b2}. 

Since $\frac{d\mathrm{[Fe/H]}}{dR}$ provides a more direct observable than $\frac{d\mathrm{[M/H]}}{dR_g}$, which depends also weakly on the adopted Galactic potential and are influenced by subtle volume-based kinematic biases (see \citealt{Boeche2013a} for a discussion), it is useful to compare the two different gradient measurements. 

It is also worth noting that our results on the abundance gradients compare very well with the findings of \citet{Hayden2013}, who use a different set of spectrophotometric distances (Hayden et al., in prep.) for their APOGEE sample.
The gradient measured by Hayden et al., using our adopted vertical and radial ranges, is compatible with the values we obtain in figure \ref{b2}. For stars with $6<R<11$ and $0<z<0.4$ kpc, the authors also obtain a gradient of $\frac{d\mathrm{[M/H]}}{dR}\simeq-0.08$ dex/kpc.
As the authors limit their analysis to smaller distances from the plane, they do not find a positive radial [M/H] gradient at large heights. 

As discussed in Sect. \ref{gradients}, future work using more APOGEE data will certainly help to understand and resolve the discrepancies seen between Figs. \ref{b1}, \ref{b3} and \ref{b2}.

\end{document}